\documentclass[aps,prb,twocolumn,superscriptaddress,floatfix,letterpaper,
nofootinbib]{revtex4-2}

\usepackage{xpatch}

\makeatletter
\patchcmd{\@ssect@ltx}
{\addcontentsline{toc}{#1}{\protect\numberl2409.06701ine{}#8}}
{}
{}
{}
\makeatother

\definecolor{refkey}{rgb}{0,0.7,0}
\definecolor{labelkey}{rgb}{0,0.7,0}

\newcommand{\onebb}{\mathbb{1}}

\newcommand{\id}{\mathrm{id}}

\newcommand{\RZ}{{\mathbb{R}/\mathbb{Z}}}

\newcommand{\cRep}{\mathcal{R}\mathrm{ep}}
\newcommand{\cVec}{\mathcal{V}\mathrm{ec}}

\newcommand{\eGau}{\eG\mathrm{au}}

\usepackage{tikz}

\newcommand{\zb}{
\begin{tikzpicture}[scale=.5,baseline={([yshift=-.5ex]current bounding box.center)}]
\draw (0,0) rectangle (1.2,1.2);
\end{tikzpicture}\,
}
\newcommand{\zbh}{
\begin{tikzpicture}[scale=.5,baseline={([yshift=-.5ex]current bounding box.center)}]
\draw (0,0) rectangle (1.2,1.2);
\draw [thick] (0,.8)--(1.2,.8);
\end{tikzpicture}\,
}
\newcommand{\zbv}{
\begin{tikzpicture}[scale=.5,baseline={([yshift=-.5ex]current bounding box.center)}]
\draw (0,0) rectangle (1.2,1.2);
\draw [thick] (.8,0)--(.8,1.2);
\end{tikzpicture}\,
}
\newcommand{\zbx}{
\begin{tikzpicture}[scale=.5,baseline={([yshift=-.5ex]current bounding box.center)}]
\draw (0,0) rectangle (1.2,1.2);
\draw [thick] (0,.8)--(1.2,.8);
\draw [thick] (.8,0)--(.8,1.2);
\end{tikzpicture}\,
}

\usepackage{braket}

\usepackage{array}
\newcolumntype{P}[1]{>{\centering\arraybackslash}p{#1}}
\newcolumntype{M}[1]{>{\centering\arraybackslash}m{#1}}

\usepackage{hhline}

\begin{document}

\begin{titlepage}

\title{Emergent generalized symmetry and maximal
symmetry-topological-order}

\author{Arkya Chatterjee} 
\affiliation{Department of Physics, Massachusetts Institute of Technology,
Cambridge, Massachusetts 02139, USA}

\author{Wenjie Ji} 
\affiliation{ 
The Division of Physics, Mathematics and Astronomy, Caltech,
1200 E California Blvd, Pasadena CA 91125 
}

\author{Xiao-Gang Wen} 
\affiliation{Department of Physics, Massachusetts Institute of Technology,
Cambridge, Massachusetts 02139, USA}

\begin{abstract} 

A characteristic property of a gapless liquid state is its emergent symmetry
and dual symmetry, associated with the conservation laws of symmetry charges
and symmetry defects respectively. These conservation laws, considered on an
equal footing, can't be described simply by the representation theory of a
group (or a higher group).  They are best described in terms of {\it a
topological order (TO) with gappable boundary in one higher dimension}; we call
this the {\it symTO} of the gapless state. The symTO can thus be considered a
fingerprint of the gapless state. We propose that a largely complete
characterization of a gapless state, up to local-low-energy equivalence, can be
obtained in terms of its {\it maximal} emergent symTO. In this paper, we review
the symmetry/topological-order (Symm/TO) correspondence and propose a
definition of {\it maximal symTO}. We discuss various examples to illustrate
these ideas. We find that the 1+1D Ising critical point has a maximal symTO
described by the 2+1D double-Ising topological order.  We provide a derivation
of this result using symmetry twists in an exactly solvable model of the Ising
critical point.  The critical point in the 3-state Potts model has a maximal
symTO of double (6,5)-minimal-model topological order. As an example of a
noninvertible symmetry in 1+1D, we study the possible gapless states of a
Fibonacci anyon chain with emergent double-Fibonacci symTO.  We find the
Fibonacci-anyon chain without translation symmetry has a critical point with
unbroken double-Fibonacci symTO. In fact, such a critical theory has a maximal
symTO of double (5,4)-minimal-model topological order.  We argue that, in the
presence of translation symmetry, the above critical point becomes a stable
gapless phase with no symmetric relevant operator.

\end{abstract}

\maketitle

\end{titlepage}

\setcounter{tocdepth}{1} 
{\small \tableofcontents }

\section{Introduction}

Systematic understanding of strongly correlated gapless states has been a long
standing challenge in theoretical physics \cite{MS8977,MS8916}.  An example of
a strongly correlated gapless state is the $n+1$D critical point of a
spontaneous symmetry-breaking transition that completely breaks the symmetry
described by a finite group $G$.  It is well known that the critical state has
an unbroken symmetry $G$.  It was pointed out  that the critical state also has
an unbroken dual algebraic $(n-1)$-symmetry $\tl G^{(n-1)}$
\cite{JW191213492,KZ200308898,KZ200514178} (which is a noninvertible higher
symmetry). Symmetry and dual algebraic higher symmetry together form a more
complete characterization of the critical point.  

The combination of symmetry $G$ and dual algebraic $(n-1)$-symmetry $\tl
G^{(n-1)}$ was together referred to as ``categorical symmetry" in
\Rfs{JW191213492,KZ200514178}.  This \emph{categorical symmetry} cannot be
described by group or higher group, in general.  One needs to use a topological
order (TO) with gappable boundary in one higher dimension to describe it, which
leads to Symmetry/Topological-Order (Symm/TO) correspondence.  Such a
topological order is a more precise characterization of \emph{categorical
symmetry}. We will refer to this \emph{topological order with gappable boundary
in one higher dimension} as \emph{symTO}. \footnote{Since the term
``categorical symmetry" has been used by many to refer to noninvertible
symmetry, here we will instead use the term \emph{symTO} to refer to the
concept that was named \emph{categorical symmetry} in
\Rfs{JW191213492,KZ200514178}.}  This symTO point of view amounts to
\begin{enumerate} \item viewing a symmetry-breaking critical point in terms of
both order parameter and disorder parameter on an equal footing \cite{JW191213492};

 \item viewing symmetry in terms of conservation (\ie fusion rings)  of both
symmetry charges and symmetry defects on an equal footing \cite{JW191213492};

 \item viewing a symmetric system by restricting to its symmetric sub-Hilbert
space $\cV_\text{symmetric}$ (\ie assuming all probes to the system also
respect the symmetry) \cite{JW190513279,JW191213492}.

\end{enumerate}

Let us discuss the last point in more detail.  For a lattice system, the total
Hilbert space $\cV_\text{total}$ has a tensor product decomposition
\begin{align}
 \cV_\text{total} = \bigotimes_i \cV_i
\end{align}
where $\cV_i$ is the small Hilbert space on site-$i$.  The presence of tensor
product decomposition implies the absence of noninvertible gravitational
anomaly. (Noninvertible gravitational anomaly was discussed in
\Rf{KW1458,FV14095723,M14107442,KZ150201690,KZ170200673,JW190513279}). On the
other hand, the symmetric sub-Hilbert space $\cV_\text{symmetric}$ does not
have the tensor product decomposition. Thus if we view $\cV_\text{symmetric}$
as the total Hilbert space, the lack of tensor product decomposition will
implies a noninvertible gravitational anomaly \cite{YS13094596}.  This led to
the realization that \cite{JW190513279,JW191213492}
\begin{align}
& \ \ \ \
 \text{a generalized symmetry restricted to } \cV_\text{symmetric} 
\nonumber\\
&= \text{a noninvertible gravitational anomaly}.
\end{align}
Since gravitational anomaly corresponds to topological order in one higher
dimension \cite{KW1458}, we obtain
\begin{align}
& \ \ \ \
 \text{a generalized symmetry restricted to } \cV_\text{symmetric} 
\nonumber\\
&= \text{a topological order in one higher
dimension}. 
\end{align}
\Rf{KZ200514178} introduced the notion of holo-equivalent symmetries: two
symmetries are holo-equivalent if they are equivalent when restricted to their
respected symmetric sub-Hilbert spaces.  For example, 1+1D $\Z_4$ symmetry and
$\Z_2\times\Z_2$ symmetry with a mixed anomaly are holo-equivalent
\cite{CW220303596,ZL220601222}.  Thus \cite{KZ200514178}
\begin{align}
& \ \ \ \
 \text{holo-equivalence class of generalized symmetries} 
\nonumber\\
&= \text{a topological order in one higher
dimension}
\end{align}
where the topological order in one higher dimension is referred to as symTO
(see Section \ref{holoequ} for details).  This is the quickest way to see
Symm/TO correspondence.

Symm/TO correspondence is closely related to \emph{topological Wick rotation}
introduced in \Rf{KZ170501087,KZ190504924,KZ191201760}, which summarizes a
mathematical theory on how bulk can determine boundary.  SymTO (\ie
categorical symmetry in the sense of \Rfs{JW191213492,KZ200514178}) has also
been referred to as symmetry topological field theory (symTFT) in the literature
\cite{AS211202092}.  However, in contrast with symTFT, the notion of symTO
stresses and/or clarifies the following key features:
\begin{enumerate}

\item A symTO has a lattice UV completion. 

\item The lattice model for a symTO does not need to be fine tuned and need not have any symmetry at lattice scale, as long as it has an energy
gap that approaches infinity. 

\item A symTO does not depend on its field theory representation. The
correlation length of the lattice model can be of the same order as the lattice
scale, in which case the continuous coarse-grained fields cannot be defined.
Some times, two different topological field theories (TFT) describe the same
topological order.  In this case, the two different symTFT's correspond to the
same symTO.

\end{enumerate}
For example, a 1+1D $\Z_2$ symmetry is described by a symTFT -- a $U(1)\times
U(1)$ mutual Chern-Simons theory, that has a symmetry that exchanges the two
$U(1)$ gauge fields. This might lead one to think that the $\Z_2$ symmetry also
implies the $e$-$m$ exchange symmetry.  The notion of symTO stresses that there
is no $e$-$m$ exchange symmetry at the UV lattice level. Thus 1+1D $\Z_2$
symmetry does not imply $e$-$m$ exchange symmetry.

In addition to the symmetry $G$ and the dual algebraic $(n-1)$-symmetry $\tl
G^{(n-1)}$, the critical point associated with a symmetry-breaking transition
may have additional emergent symmetries.  Putting all the emergent symmetries
together, we obtain a \emph{maximal symTO}.  The emergent symTO was proposed to
be an essential feature of a critical point. In particular, it was proposed in
\Rfs{JW191213492,KZ200514178} that the emergent maximal symTO may largely
determine the local low energy properties of a strongly correlated gapless
liquid phase.

A general classifying understanding of gapless liquid states and critical
points is a long standing challenge in theoretical physics.  It is well known
that a gapless state can have emergent symmetry.  We now realize that such an
emergent symmetry can be a combination of ordinary symmetry (described by
group), higher-form and higher-group symmetry
\cite{NOc0605316,NOc0702377,KT13094721,GW14125148} (described by higher-group),
anomalous ordinary symmetry \cite{H8035,CGL1314,W1313,KT14030617}, anomalous
higher symmetry
\cite{KT13094721,GW14125148,TK151102929,T171209542,P180201139,DT180210104,BH180309336,ZW180809394,WW181211968,WW181211967,GW181211959,WW181211955,W181202517},
noninvertible 0-symmetry (in 1+1D)
\cite{PZh0011021,CSh0107001,FSh0204148,FSh0607247,FS09095013,DR11070495,BT170402330,CY180204445,TW191202817,KZ191213168,I210315588,Q200509072},
noninvertible higher symmetry (also called algebraic higher symmetry)
\cite{KZ200308898,KZ200514178,HV201200009,KZ211101141,CS211101139,BT220406564,FT220907471},
and/or noninvertible gravitational anomaly
\cite{KW1458,FV14095723,M14107442,KZ150201690,KZ170200673,JW190513279}, which
include anomaly-free/anomalous noninvertible higher symmetry (described by
fusion higher category) \cite{JW191213492,KZ200308898,KZ200514178,FT220907471}.

The term ``noninvertible higher symmetry'' just means ``symmetry beyond group
and higher group'', but what symmetries does it include?  How are they
classified?  There are two approaches to this classification.  In the first
approach,\cite{JW191213492,KZ200308898,KZ200514178} one groups the $n+1$D
noninvertible higher symmetries into holo-equivalence classes and classifies
the holo-equivalence classes using topological orders in one higher dimension
(\ie using braided fusion $n$-categories with trivial center).  This
classification encompasses anomaly-free and anomalous symmetries (defined
below).  

A second approach was used in \Rf{KZ200514178} (see also \Rf{TW191202817} for
1+1D case) to classify all anomaly-free noninvertible higher symmetries, \ie
the symmetries that allow non-degenerate gapped ground state on any closed
space.  It was shown that anomaly-free noninvertible higher symmetries in
$n$-dimensional space are classified by \emph{local fusion $n$-categories}
$\tl\cR$ \cite{KZ200514178}.  \Rf{KZ200514178} also defined and classified
anomalous noninvertible higher symmetries with \emph{invertible} anomalies, in
terms of local fusion $n$-categories $\tl\cR$ and the automorphisms in their
centers $\eZ(\Si\tl\cR)$.  \Rf{FT220907471} used \emph{fusion $n$-categories}
without the \emph{local} condition to classify generic noninvertible higher
symmetries, that can be anomaly-free and anomalous, both invertibly (as
mentioned above) and noninvertibly.  In Section \ref{fusioncat} we will give a
discussion about using fusion $n$-categories to describe
generalized symmetries.  

Since emergent symmetries are so rich, it may be reasonable to conjecture that
a gapless state is largely characterized by its maximal emergent symmetry
insofar as \emph{we may develop a general classifying theory of gapless liquid
states via their maximal emergent symmetries.}

In the next section, we review the unified theory for these 
different emergent symmetries.  In Section III, we propose a definition of
maximal symTO, using the Symm/TO correspondence and the isomorphic holographic
decomposition  introduced in \Rf{KZ150201690} (see Fig.  \ref{CDiso}).  In
section IV, we discuss some simple 1+1D strongly correlated gapless
liquids, and their emergent maximal symTO.  In particular, we compute the
modular invariant partition functions for strongly correlated gapless liquids
for systems with anomaly-free and anomalous $S_3$ symmetries, as well as 
Fibonacci symmetry.  In section V, we present a way to compute the maximal 
symTO of the Ising critical point using symmetry twists. 

\section{Symmetry/Topological-Order (S\lowercase{ymm}/TO) correspondence: a review}

It was proposed in
\Rfs{JW191213492,JW190513279,KZ200308898,KZ200514178,KZ201102859,KZ210703858,KZ220105726,CW220303596,FT220907471}
that all the rich and seemingly very different emergent symmetries have a
unified description in terms of noninvertible gravitational anomaly, or
equivalently \cite{KW1458}, in terms of topological orders $\eM$ in one higher
dimension,\cite{FT220907471} if the symmetries are finite.  Related ideas were
discussed for 1+1D systems, for special models (such as rational conformal
field theory), or in different contexts (such as duality and gauging) in
\Rfs{FSh0204148,FSh0607247,FS09095013,YS13094596,HV14116932,KZ170501087,FT180600008,KZ190504924,KZ191201760,CZ190312334,TW191202817,JW191209391,LB200304328,GK200805960}.
In this section, we mainly consider finite symmetries.

\subsection{From holo-equivalence to homomorphism between quantum field
theories}

\begin{figure}[t]
\begin{center}
\includegraphics[width=0.48\linewidth]{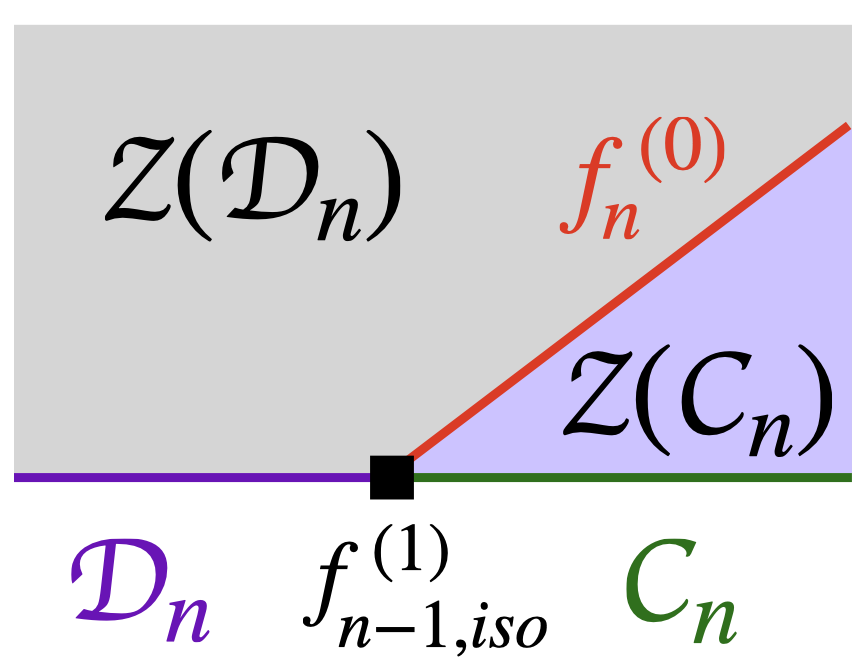}
\hfill
\includegraphics[width=0.48\linewidth]{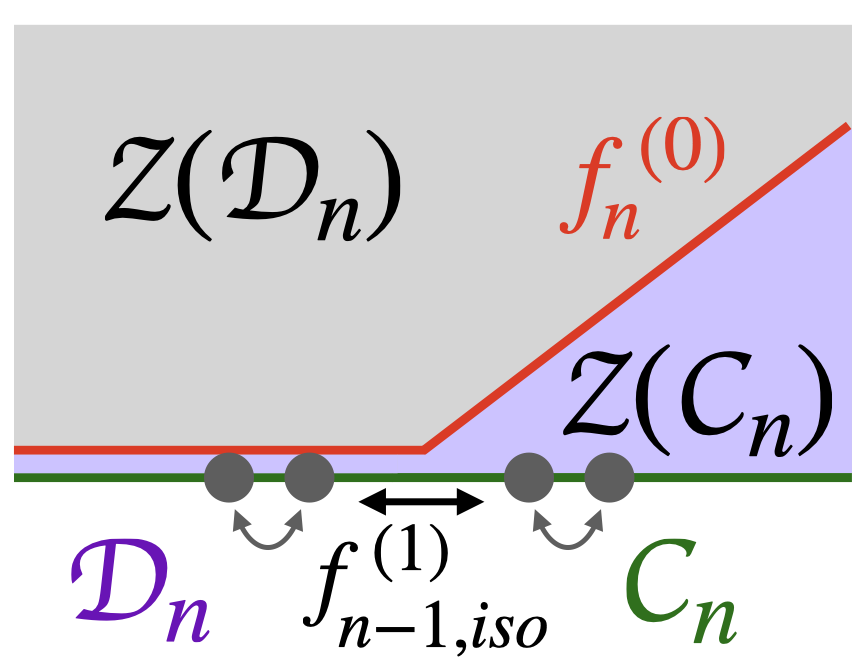}
\end{center}
\caption{An isomorphism $f_{n-1}^{(1)}$ (\ie a transparent domain wall in
spacetime) between two anomalous $n+1$D (gapped or gapless) quantum field
theories, $\cD_n$ and $\cC_n\boxtimes_{\eZ(\cC_n)} f^{(0)}_n $ (cf. eqn. (4.3)
of \Rf{KZ150201690}), describes a local low energy equivalence
(holo-equivalence) of the two quantum field theories.  Here $\eZ$ is the
boundary to bulk function defined in \Rf{KW1458,KZ150201690}.  Such an
equivalence exposes the emergent symmetry described by the symTO $\eZ(\cC_n)$
in quantum field theory $\cD_n$.  Note that the anomaly is given by the
topological order $\eZ(\cD_n)$ in one higher dimension.  } \label{CDiso} 
\end{figure}

In fact, a holographic theory was already developed in \Rf{KZ150201690} (see
Fig. \ref{CDiso}).  However, at that time, it was formulated as a holographic
description of topological orders (\ie gapped quantum liquid phases) and
noninvertible gravitational anomalies.  A few years later, we realized that
noninvertible gravitational anomalies can be viewed as generalized symmetries
\cite{JW190513279,JW191209391}, and the theory developed in \Rf{KZ150201690} is
in fact a unified theory of generalized symmetries.  Such a holographic point
of view was used in \Rf{KZ200308898,KZ200514178} to classify topological phases
and symmetry protected topological (SPT) phases \cite{CGL1204,CGL1314} of
generalized symmetries.

We know that structure preserving map -- homomorphism -- is the single most
important concept in mathematics.  So to have a systematic understanding of
quantum liquid phases (gapped or gapless), \ie to have a systematic
understanding of quantum field theories (which can be strongly coupled without
weakly coupled modes and without Lagrangian description), we introduce
homomorphism between two quantum liquid phases (or two quantum field theories).
Usually, a morphism between two quantum field theories is defined by a domain
wall between them.  However, such a morphism does not preserve the important
structures that we care about in quantum field theories.  

But, what structures do we want to preserve?  Since we want to understand
gapless phases, the structure that we want to preserve is the so-called
\emph{local low energy properties}, which are defined as long range
correlations of local operators (or local symmetric operators for symmetric
systems).  To stress the importance of local low energy properties,
\Rf{KZ200514178} introduced \emph{holo-equivalence} between two quantum field
theories (or two quantum liquids): \frmbox{Two quantum field theories are
\textbf{holo-equivalent} if their corresponding local (symmetric) operators
have the same long range correlations.}  We see that for two quantum field
theories, $QFT$ and $QFT'$, are holo-equivalent, if there exist gapped quantum
field theories, $gapped$ and $gapped'$, such that stacking with $gapped$ and
$gapped'$ make $QFT$ and $QFT'$ identical 
\begin{align} 
QFT\boxtimes gapped = QFT'\boxtimes gapped'.  
\end{align} 
For example $QFT\boxtimes gapped$ and $QFT'\boxtimes gapped'$ have identical
partition function on any large spacetime $M^{n+1}$:
\begin{align} 
Z(QFT\boxtimes gapped, M^{n+1}) =Z(QFT'\boxtimes gapped', M^{n+1}).  
\end{align}

Now we apply the above idea to define holo-equivalence between two anomalous
$n+1$D quantum field theories, $\cC_n$ and $\cD_n$.  The (noninvertible)
gravitational anomalies in $\cC_n$ and $\cD_n$ are described by bulk
topological orders $\eZ(\cC_n)$ and $\eZ(\cD_n)$.  This leads to the setup in
Fig. \ref{CDiso}, which says that $\cC_n$ and $\cD_n$ are holo-equivalent if
they differ by a gapped domain wall $f_n^{(0)}$ between the two bulk
topological orders $\eZ(\cC_n)$ and $\eZ(\cD_n)$.  In other words, $\cC_n$ and
$\cD_n$ are holo-equivalent if $\cC_n$ and
$\cD_n\boxtimes_{\eZ(\cD_n)}f_n^{(0)} $ are isomorphic.  The isomorphism is
given by a transparent domain wall $f_{n-1}^{(1)}$:
\begin{align}
\label{fn1iso}
 f_{n-1}^{(1)}:\ \cD_n \cong \cC_n\boxtimes_{\eZ(\cC_n)}f_n^{(0)}.
\end{align}
We will call such an isomorphism an \emph{isomorphic holographic
decomposition}. This decomposition defines a
homomorphism described by a pair $(f_n^{(0)}, f_{n-1}^{(1)})$
\cite{KZ150201690}:
\begin{align}
\label{CDmorphism}
 (f_n^{(0)}, f_{n-1}^{(1)}):\ \cC_n \to \cD_n.
\end{align}
Such a homomorphism preserves the local low energy properties, and is the
mathematical description of the holo-equivalence.

In this paper, we give a physical description of the isomorphic holographic
decomposition, $f_{n-1}^{(1)}: \cD_n \cong f_{n}^{(0)} \boxtimes_{\eZ(\cC_n)}
\cC_n$: The original (gapped or gapless) theory $\cD_n$ has the same partition
function as the composite theory on any close spacetime manifolds\footnote{The
composite theory $ f_{n}^{(0)} \boxtimes_{\eZ(\cC_n)} \cC_n$ is a slab of bulk
topological order $\eZ(\cC_n)$ with a lower (gapped or gapless) boundary
described by an anomalous theory $\cC_n$ and an upper gapped domain wall
between $\eZ(\cC_n)$ and $\eZ(\cD_n)$ described by $f_{n}^{(0)}$.  The bulk
topological orders, $\eZ(\cC_n)$ and $\eZ(\cD_n)$, and the domain wall
$f_{n}^{(0)}$ are assumed to have a lattice UV completion with an infinite gap.
The lattice UV completion does not need to be fine tuned and may not have any
symmetry.} $ f_{n}^{(0)} \boxtimes_{\eZ(\cC_n)} \cC_n$
\begin{align}
\label{ZZ}
 Z(\cD_n) = Z(f_{n}^{(0)} \boxtimes_{\eZ(\cC_n)} \cC_n),
\end{align}
assuming the bulk $\eZ(\cC_n)$ and the upper boundary $f_{n}^{(0)}$ to have
infinite gap.  The above relation between partition functions is the key, which
physically defines the isomorphic holographic decomposition.  

One can see that, in the composite theory $f_{n}^{(0)} \boxtimes_{\eZ(\cC_n)}
\cC_n$, the lower boundary $\cC_n$ captures the local low energy properties of
the original theory $\cD_n$, while the bulk topological orders $\eZ(\cC_n),\
\eZ(\cD_n)$, and the domain wall $f_{n}^{(0)}$ captures the global low energy
properties, such as ground state degeneracy. In fact, $\eZ(\cC_n),\
\eZ(\cD_n)$, and $f_{n}^{(0)}$ describe an emergent symmetry in the original
theory $\cD_n$ (see Section \ref{iso2symm}). 

When the two quantum field theories, $\cC_n$ and $\cD_n$, are gapped, their
excitations are described by fusion $n$-categories, also denoted as $\cC_n$ and
$\cD_n$ respectively.  In this case, the boundary to bulk functor $\eZ$ becomes
the generalized Drinfeld center that maps a fusion $n$-category to a braided
fusion $n$-category, and the homomorphism \eqref{CDmorphism} becomes a monoidal
functor that preserves the fusion (see Fig. \ref{CDiso}(right)).
\Rf{KZ150201690,KZ170200673} used the homomorphism \eqref{CDmorphism} to show
the boundary to bulk map $\eZ$ corresponds to the mathematical notion of
\emph{center} in a very general setting. This gives rise to the topological
holographic principle: \emph{boundary determines bulk} \cite{KW1458}, but bulk
does not determine boundary.

\subsection{From isomorphic holographic decomposition to emergent symmetry}
\label{iso2symm}

A few years later, in \Rfs{JW190513279,JW191213492}, it was realized that noninvertible
gravitational anomalies can also be viewed as (generalized) symmetries.  Thus,
Fig. \ref{CDiso} actually is a holographic description of symmetry, which leads
to the Symm/TO correspondence.  In other words, the isomorphism \eq{fn1iso} can
be viewed as a decomposition of the anomalous theory $\cD_n$ which reveals the
(generalized) symmetry in $\cD_n$ described by $f_n^{(0)}$ and $\eZ(\cC_n)$.
In fact, $\eZ(\cC_n)$ is the symTO mentioned above.  $f_n^{(0)}$ provides a 
more
detailed description of symmetry than the symTO $\eZ(\cC_n)$ description.

\begin{figure}[t]
\begin{center}
\includegraphics[width=0.45\linewidth]{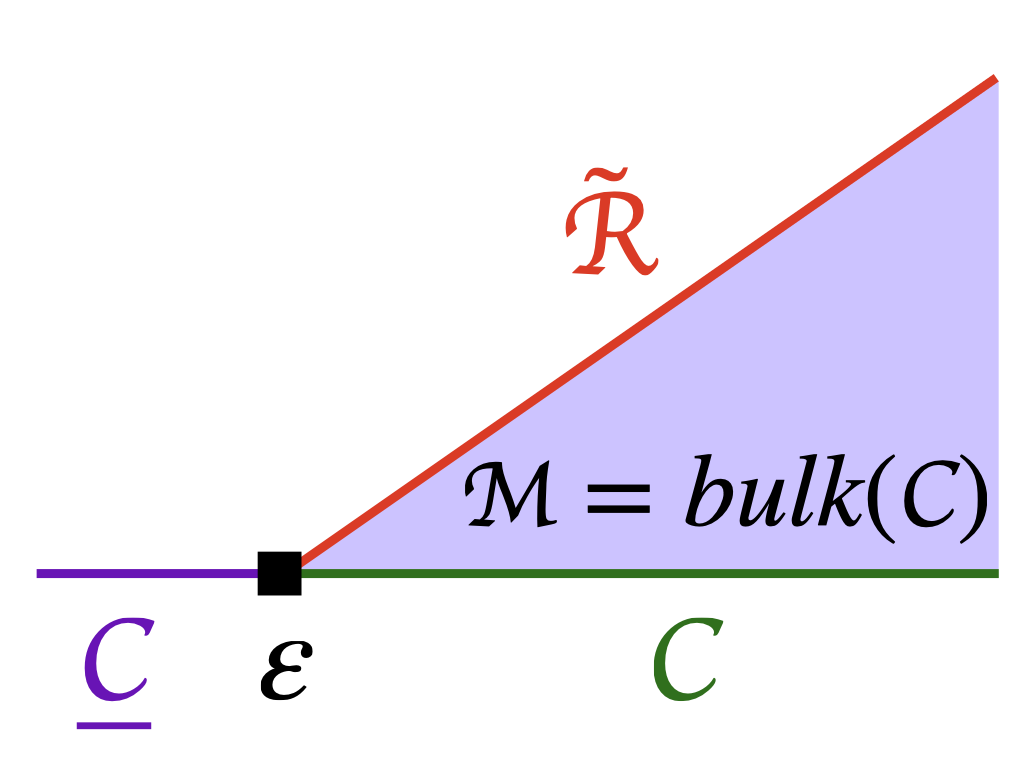}
\end{center}
\caption{A special case of Fig. \ref{CDiso}, where $f^{(0)}_n =\tl \cR$, $\cD_n
= \underline{\cC}$, $\cC_n = \cC$, and $\eZ(\cC_n) = \text{bulk}(\cC)$.
$f^{(1)}_{n-1} =\veps$ is an isomorphism, \ie a transparent domain wall  (cf.
Fig. 24 and Fig. 29 in \Rf{KZ200514178}).  Here, $\tl\cR$ is a fusion higher
category describing the gapped excitations on a gapped boundary of the symTO.
It describes the emergent symmetry in $\underline{\cC}$.  We will refer to such
a symmetry as $\tl\cR$-symmetry.  Also, $\text{bulk}(\cC)$ is the symTO
$\eM$ describing the holo-equivalence class of emergent symmetry $\tl\cR$.  }
\label{CCmorph} 
\end{figure}

\Rf{KZ200514178} used a special case of Fig. \ref{CDiso} with $\eZ(\cD_n) = $
trivial (see Fig.  \ref{CCmorph}) to describe emergent symmetry in $n+1$D
systems with no gravitational anomaly. In this case, the isomorphism
\begin{align}
\label{CCiso}
\veps: \ \underline{\cC} \cong \tl\cR \boxtimes_\eM \cC 
\end{align}
is viewed as a decomposition of the anomaly-free theory $\underline{\cC}$ which
reveals the (generalized) symmetry in $\underline{\cC}$ described by $\tl\cR$
and $\eM$, where $\eM =\eZ(\tl\cR)$.  Here $\tl\cR$ is a fusion $n$-category
that describes the excitations on the upper boundary, and $\eM$ is a braided
fusion $n$-category that describes the excitations in the bulk topological
order (\ie the symTO mentioned before).  In this case, $\tl\cR$ describes the
fusion of the symmetry defects, which in turn describes the generalized
symmetry, and $\eM$ is the symTO of the symmetry.

\subsection{S\lowercase{ym}TOs classify Holo-equivalence classes of symmetries}
\label{holoequ}

\begin{figure}[t]
\begin{center}
 \includegraphics[width=0.6\linewidth]{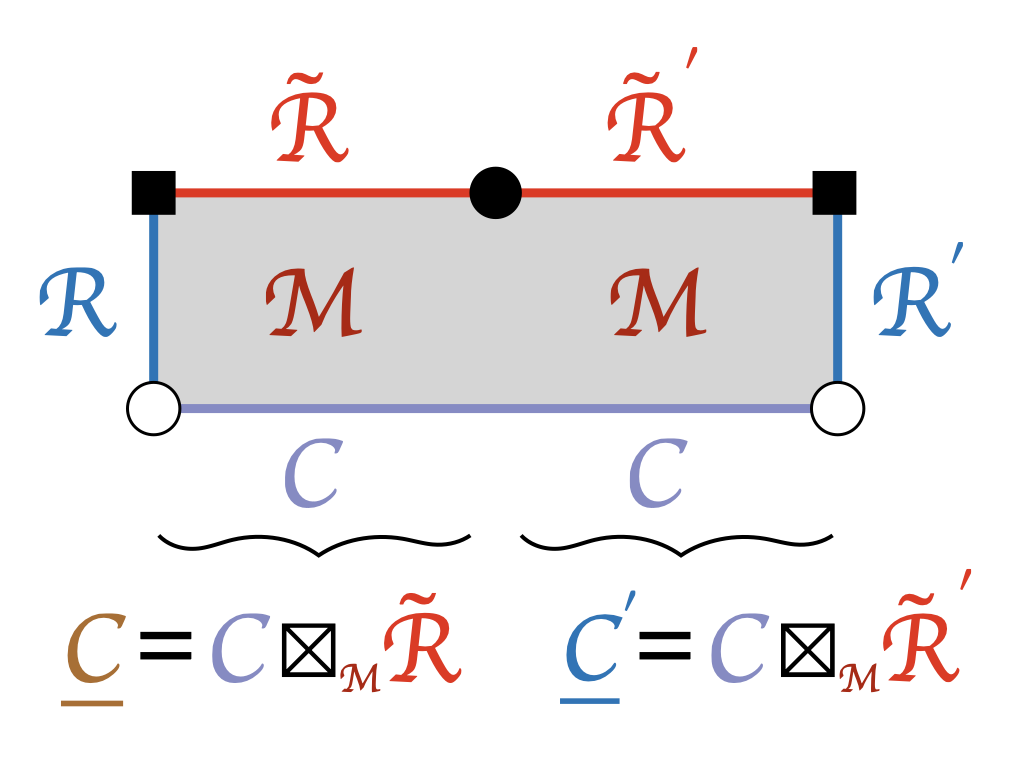}
\end{center}
\caption{Consider two systems, $\underline{\cC}$ and $\underline{\cC'}$, with
holo-equivalent symmetries, $\cR$ and $\cR'$.  After restricting to the
respected symmetric sub-Hilbert spaces, the two systems become identical 
$\cC
=\cC'$ and are described by the same boundary of the symTO $\eM$.  However,
$\underline{\cC}$ and $\underline{\cC'}$ may have different global low energy
properties from different charged sectors: $\cC\boxtimes_\eM \cR \neq
\cC\boxtimes_\eM \cR'$.  } \label{RRsymmC} 
\end{figure}

As an application of the above holographic picture of symmetries, let us define
the notion of holo-equivalence of symmetries \cite{KZ200514178}: \frmbox{Two
$n+1$D generalized symmetries described by fusion $n$-categories $\cR$ and
$\cR'$ are \textbf{holo-equivalent} if they have the same bulk: $\eZ(\cR) =
\eZ(\cR')=\eM$, \ie the same symTO.}  We see that \cite{KZ200514178}
\frmbox{Holo-equivalent classes of symmetries are one-to-one classified by
symTOs (\ie topological order with gappable boundary in one higher dimension).} 

To understand the physical meaning of holo-equivalence of two symmetries, we
note that there are many symmetric systems for each symmetry.  The
holo-equivalence means that there is an one-to-one correspondence between
$\cR$-symmetric systems and $\cR'$-symmetric systems, such that the
corresponding systems, $\underline{\cC}$ and $\underline{\cC}'$, have the same
spectrum after restricting to the respected symmetric sub-Hilbert spaces,
$\cV_\text{symmetric}$ and $\cV_\text{symmetric}'$.  In other words, $\cC =
\cC'$ (see Fig. \ref{RRsymmC}).  In short, two systems with holo-equivalent
symmetries are identical, after restricting to their corresponding symmetric
sub-Hilbert spaces. On the other hand, the two systems may have different
charged sectors.  Since only local dynamics within $\cV_\text{symmetric}$ is
considered here, we may have some unexpected equivalence. For example, 1+1D
$\Z_4$ symmetry is holo-equivalent to 1+1D $\Z_2\times \Z_2$ symmetry with
mixed anomaly \cite{CW220303596,ZL220601222}.  

It is surprising to see that symmetry is so closely related to topological
order in one higher dimension -- symTO.  We know that symmetry constrains local
low energy dynamics.  Similarly, topological order also constrains boundary
local low energy dynamics just as symmetry does.  Thus ``a symmetry is
described by a symTO'' has the following physical meaning
\cite{JW191213492,KZ200514178,CW220303596}:  \frmbox{A \textbf{system}
(described by a Hamiltonian) with a generalized finite symmetry is
\emph{exactly locally reproduced} by a \textbf{boundary} (described by a
boundary Hamiltonian) of the corresponding symTO.} Here, \emph{exactly locally
reproduced} means that the local symmetric operators in the system have a
one-to-one correspondence with the local operators on the boundary of the
topological order.  The corresponding local operators have identical
correlations on the respective ground states, assuming the bulk topological
order has an infinite gap.  

Since the boundary of the symTO $\eM$ in Fig. \ref{CCmorph} exactly simulates
the symmetric system $\underline{\cC}$ after the restriction to the symmetric
sub-Hilbert space $\cV_\text{symmetric}$, the decomposition $\underline{\cC}
\stackrel{\veps}{\cong} \tl\cR \boxtimes_\eM \cC$ implies the following relation
of partition functions
\begin{align}
&\ \ \ \ 
 Z(\underline{\cC} \text{ restricted to } \cV_\text{symmetric})
\nonumber\\
& =
 Z(\cC) = \Tr_{\cV_\text{symmetric}} \ee^{-\bt H_{\underline{\cC}}}
\nonumber\\
&= Z(\cC \text{ boundary of } \eM),
\end{align}
where we have used the fact that the system $\cC$ is nothing but
$\underline{\cC}$ restringing to $\cV_\text{symmetric}$.  The total partition
function of the symmetric system $\underline{\cC}$ also contains contributions
from charged excitations not in $\cV_\text{symmetric}$.  These charged
excitations are not included in $\cC$, but are included in the composition
$\tl\cR \boxtimes_\eM \cC$ \cite{FT220907471}.  Thus the decomposition
$\underline{\cC} \stackrel{\veps}{\cong} \tl\cR \boxtimes_\eM \cC$ also implies
\begin{align}
 Z(\underline{\cC}) = \Tr \ee^{-\bt H_{\underline{\cC}}}
= Z(\tl\cR \boxtimes_\eM \cC),
\end{align}
which is the physical meaning of the isomorphic holographic decomposition
$\veps$.  We will illustrate the above relation through some examples later.
To summarize \frmbox{ The decomposition $\underline{\cC}
\stackrel{\veps}{\cong} \cC \boxtimes_{\eM} \tl\cR$ means that the partition
function of the system $\underline{\cC}$ (gapless or gapped) is the same as the
partition function of the composite system $\cC \boxtimes_{\eM} \tl\cR $ (see
Fig. \ref{CCmorph}), assuming the bulk $\eM$ and the boundary $\tl\cR$ have
infinite energy gap.  } Under such a Symm/TO correspondence, using emergent
symmetry to characterize a gapless state becomes equivalent to using symTO for
that purpose.  \Rf{KZ200514178} concentrated on the pair $(\eM,\cC)$ in Fig.
\ref{CCmorph}, and used the Symm/TO correspondence to classify SPT orders and
symmetry enriched topological (SET) orders for generalized symmetries described
by symTO $\eM$.  Symm/TO correspondence allows us to see some general results,
such as symmetry protected gaplessness (see Section \ref{symmgapless}).  

\subsection{Local fusion higher categories classify anomaly-free
noninvertible higher symmetries } \label{afsymm}

\Rf{KZ200514178} also used local fusion $n$-category $\tl\cR$ to classify
anomaly-free generalized symmetries.  To obtain this result, we first need to
define anomaly-free condition for noninvertible higher symmetries. One
definition was proposed in \Rfs{TW191202817,KZ200514178}: 
\frmbox{\underline{Definition:} A symmetry is
\textbf{anomaly-free} if it allows a gapped non-degenerate ground state for all 
closed spaces of any homotopy type.}  

\begin{figure}[t]
\begin{center}
 \includegraphics[width=0.6\linewidth]{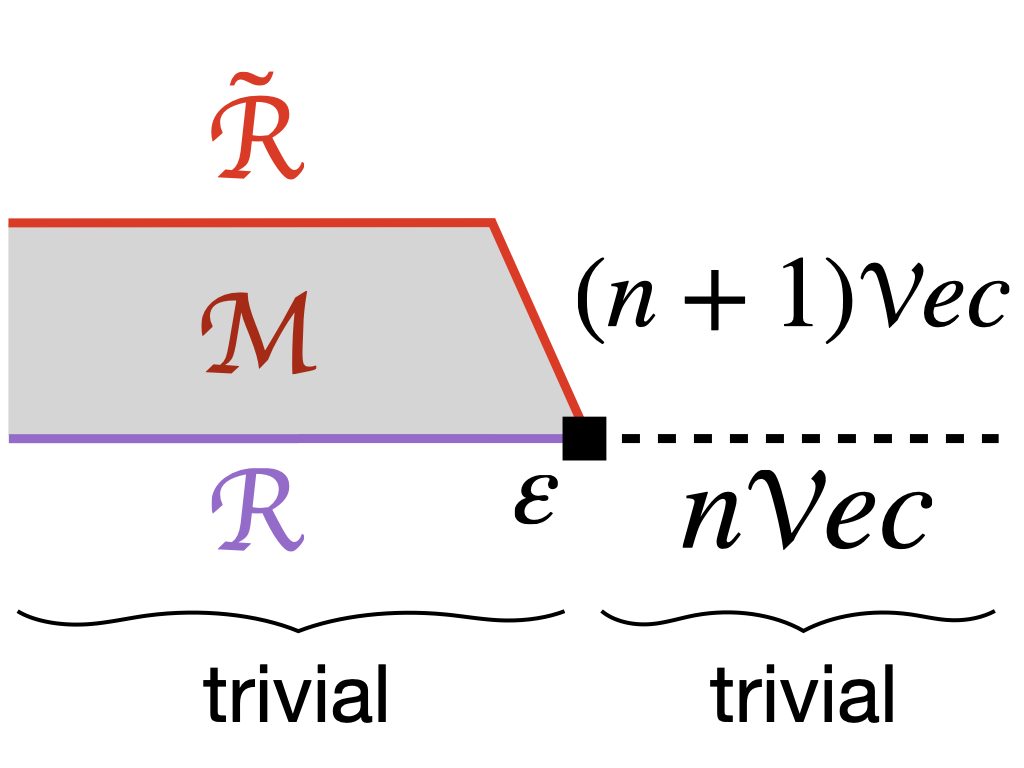}
\end{center}
\caption{An $n+1$D $\tl\cR$-symmetry is anomaly-free if there exists a
fusion $n$-category $\cR$ such that $\cR \boxtimes_{\eM}\tl\cR
\stackrel{\veps}{\cong} n\cVec$ where $\eM=\eZ(\tl\cR)$.  } \label{Rsymm}
\end{figure}

We note that, according to a conjecture in \Rf{KW1458}, an $n+1$D gapped phase
with non-degenerate ground state for all closed spaces of any homotopy types
has a trivial topological order described by $n\cVec$.  Then, from the above
definition, we see that a symmetry described by a fusion $n$-category $\tl\cR$
is anomaly-free if there exists a fusion $n$-category $\cR$ (which is the $\cC$
in Fig.  \ref{CCmorph}), such that $\tl\cR \boxtimes_\eM \cR
\stackrel{\veps}{\cong} n\cVec$ ($n\cVec$ is the $\underline{\cC}$ in Fig.
\ref{CCmorph}).  This isomorphic holographic decomposition is described in Fig.
\ref{Rsymm}, which is equivalent to a monoidal functor (\ie a homomorphism)
\begin{align} 
\label{tR2Vec}
(\cR, \veps): \tl\cR \to n\cVec.  
\end{align} 
Such a monoidal functor to $n\cVec$ is called a fiber functor.
\frmbox{\underline{Definition:} a fusion $n$-category with a monoidal functor
to $n\cVec$ is a \textbf{local} fusion $n$-category.} Thus \frmbox{the most
general $n+1$D anomaly-free generalized symmetries (\ie anomaly-free algebraic
higher symmetries), are classified, with an one-to-one correspondence, by local
fusion $n$-categories $\tl\cR$.  \cite{KZ200514178}} 

We remark that $\tl\cR$ in Fig. \ref{CCmorph} is the fusion $n$-category
describing the fusion of the symmetry defects for both anomaly-free and
anomalous symmetry, while the $\cR$ introduced above is the fusion $n$-category
describing the fusion of the symmetry charges (\ie the excitations above the
gapped non-degenerate ground state introduced above\footnote{The gapped
non-degenerate ground state corresponds to symmetric product state.}) for
anomaly-free symmetry (see Section \ref{commutantPO})
\cite{JW191213492,KZ200514178}.  Clearly, $\cR$ is also a local fusion
$n$-category, satisfying (see Fig.  \ref{Rsymm}) 
\begin{align} 
\label{R2Vec}
(\tl\cR, \veps): \cR \to n\cVec.  
\end{align} 
The above means that an anomaly-free symmetry is completely breakable by
perturbations of local operators, \ie by adding local operators as
perturbations we can break the symmetry $\tl\cR$ to trivial symmetry $n\cVec$.
This is because we can view the homomorphism $\cR \to n\cVec$ as induced by
adding the top morphisms which correspond to local operators.

For example, for a system with $SU(2)$ symmetry, $\cR$ contains spin-$\frac12$
excitations, where the top morphisms in $\cR$ are $SU(2)$ symmetric operators.
After we add generic local operators to top morphisms, the spin-$\frac12$
excitation becomes a direct sum of two trivial excitations $\onebb$ 
\begin{align}
 \text{spin-}\frac12 = \onebb \oplus \onebb.
\end{align}
Thus the fiber functor \eqref{R2Vec} for the symmetry-charge local fusion
$n$-category $\cR$ describes a symmetry breaking process by perturbations of
local operators, that breaks the symmetry completely.  We note that
the above discussion only applies to anomaly-free symmetry.

We like to remark that that there is another definition of anomaly-free
symmetry, based on the obstruction of gauging the full symmetry.  The two
definitions of anomaly-free symmetry are inequivalent \cite{CS230509713}.

\subsection{Holo-equivalence, duality, and generalized gauging}

The isomorphic holographic decomposition in Fig. \ref{CCmorph} defines the
holo-equivalence between different theories.  If we fix $\cC$ and $\eM$ in Fig.
\ref{CCmorph}, but choose different gapped upper boundaries $\tl \cR$ and $\tl
\cR'$, we will obtain different anomaly-free theories, $\underline{\cC}$ and
$\underline{\cC}'$:
\begin{align}
\underline{\cC} & \cong \tl\cR \boxtimes_\eM \cC , 
\nonumber\\
\underline{\cC}' & \cong \tl\cR' \boxtimes_\eM \cC . 
\end{align}
In this case, the two theories, $\underline{\cC}$ and $\underline{\cC}'$, are
holo-equivalent, by definition.  We also say that the two theories,
$\underline{\cC}$ and $\underline{\cC}'$, are dual to each other.  Such a
duality relation includes the well known Kramers-Wannier duality.

Holo-equivalence and gauging finite symmetry are also closely related: Two
theories related by gauging a finite symmetry \cite{BT170402330,T171209542} are
always holo-equivalent.  In fact, we can use this fact to define a generalized
gauging: changing the upper boundary from $\tl \cR$ to $\tl \cR'$ is a
generalized gauging.

Note that the two theories, $\underline{\cC}$ and $\underline{\cC}'$ have
different symmetries: the original theory $\underline{\cC}$ has $\tl
\cR$-symmetry, while the dual theory or the gauged theory $\underline{\cC}'$
has $\tl \cR'$-symmetry.  In fact, symmetry and duality are closely related: if
there is no symmetry, there is no duality.  Or more precisely, the following
three sets have one-to-one correspondences: (1) the set of dual thoeries; (2)
the set of holo-equivalent symmetries; (3) the set of gapped boundary phases of
the symTO.  Thus the isomorphic holographic decomposition gives us a general
theory of duality.

\subsection{Description and classification of gapless and gapped liquid phases}

\begin{figure}[t]
\begin{center}
 \includegraphics[width=0.5\linewidth]{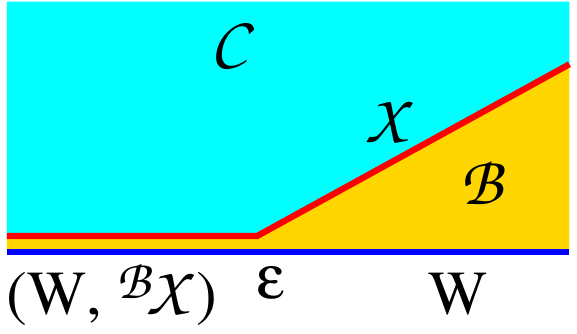}
\end{center} 
\caption{$n+1$D gapless (or gapped) liquid phases, with a gravitational anomaly
described by a bulk topological order $\eC$, are described and classified by
the data $(W, ^\eB\hskip -1mm\cX)$.  Based on the isomorphic holographic
decomposition $\veps$, $W$ can be viewed as an anomalous quantum field theory,
$\eC$ and $\eB$ be viewed as bulk topological orders (\ie braided fusion
$n$-categories) and $\cX$ be viewed as a gapped domain wall (\ie a fusion
$n$-category) between $\eC$ and $\eB$.  } \label{WBXC}
\end{figure}

A unified mathematical description and classification of 1+1D (potentially
anomalous) gapless and gapped phases was given in Theorem 6.7 in
\Rf{KZ190504924} and Theorem 5.9 in \Rf{KZ191201760}, which are summarized
compactly below:\\ 
The gapped/gapless edges of a 2+1D topological order $(\eC,
c)$, where $\eC$ is a modular tensor category and $c$ is the chiral central
charge, are precisely described and classified by pairs $(W, ^\eB\hskip
-1mm\cX)$, where\\
\begin{enumerate}
\item 
$W$ is a vertex operator algebra of central charge $c$ when the edge is chiral;
$W$ is a full field algebra \Rf{HKm0511328} with the chiral central charges
$c_L$ and anti-chiral central charges $c_R$ such that $c = c_L - c_R$ when the
edge is non-chiral; $W$ is the trivial full field algebra (i.e.  $W = \C$) when
the edge is gapped;

\item
$^\eB\hskip -1mm\cX$ is a $\eB$-enriched fusion category defined by the pair
$(\eB, \cX)$ via the canonical construction, where $\eB := \mathrm{Mod}_W$ and
$\cX$ is a closed fusion $\eC$-$\eB$-bimodule (see \Rf{KZ150700503}, Definition
2.6.1).

\end{enumerate}


Such a result has an interpretation in terms of the isomorphic holographic
decomposition introduced in \Rf{KZ150201690} (see Fig. \ref{WBXC}).  Since the
isomorphic holographic decomposition is valid in any dimensions, we expect the
above result of \Rf{KZ190504924} and \Rf{KZ191201760} can be generalized to any dimension (see Section 7 in both \Rf{KZ190504924} and \Rf{KZ191201760}).  In
particular, when the topological order $\eC$ is trivial, the above result
becomes a description and a classification of anomaly-free gapless and gapped
liquid phases.

\begin{figure}[t]
\begin{center}
 \includegraphics[width=0.5\linewidth]{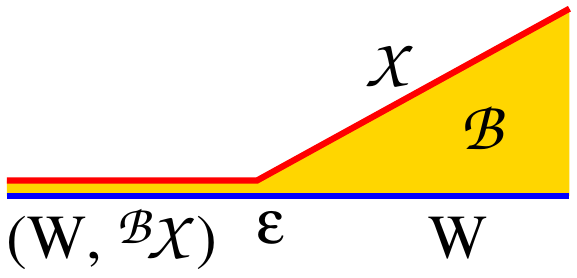}
\end{center}
\caption{ $n+1$D anomaly-free gapless (or gapped) liquid phases (\ie realizable
by lattice models) are described and classified by the data $(W, ^\eB\hskip
-1mm\cX)$, where $W$ can be viewed as an anomalous quantum field theory, $\eB$
be viewed as the bulk topological order $\eB=\eZ(W)$ describing the
non-invertible gravitational anomaly, and $\cX$ be viewed as a gappled boundary
of $\eB$.  } \label{WBX} 
\end{figure}

Since non-invertible gravitational anomalies (\ie bulk topological orders
$\cB$) can be viewed as emergent generalized
symmetries\cite{JW190513279,JW191209391}, we see that the above description and
classification of gapless (and gapped) phases are based on the emergent
generalized symmetries described by the data $^\eB\hskip -1mm\cX$ or
$(\eB,\cX)$, where $\eB=\eZ(\cX)$ is the symTO\cite{JW191209391} and $\cX$ is
the fusion higher category describing the emergent generalized
symmetry\cite{KZ200308898,KZ200514178}.  The data given by the pair $(\eB,\cX)$
is also called  ``topological symmetry'' by \Rf{FT220907471}. Thus the result
of \Rf{KZ170501087,KZ190504924,KZ191201760} and result of \Rf{FT220907471} are
closely related.

\subsection{Fusion $n$-categories classify emergent generalized
symmetries} \label{fusioncat}

\begin{figure}[t]
\begin{center}
\includegraphics[width=0.45\linewidth]{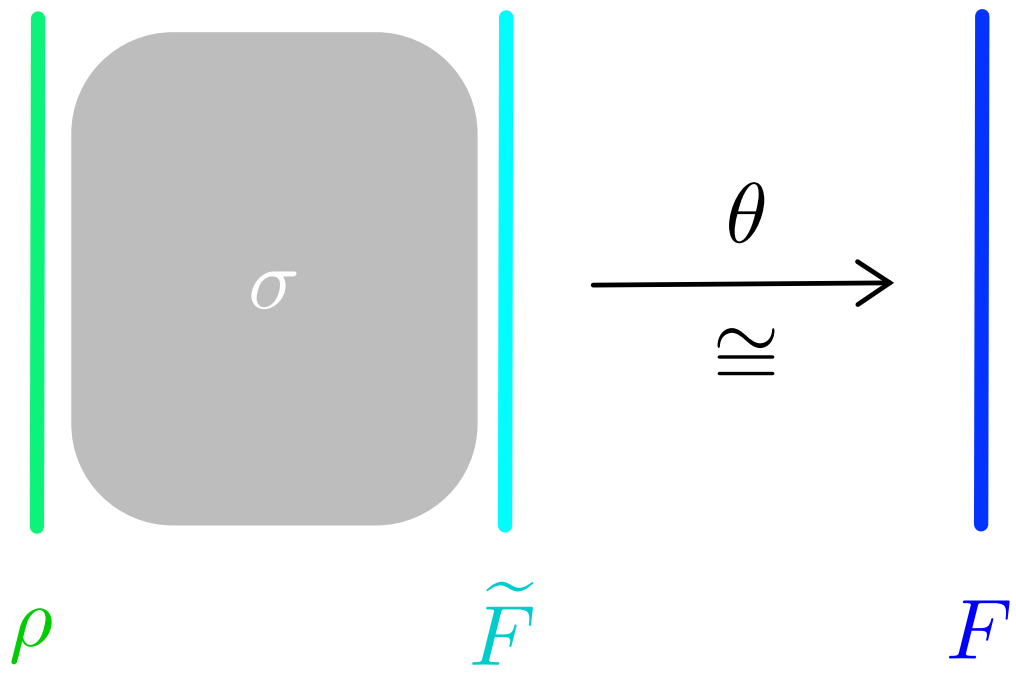}
\hfill
\includegraphics[width=0.45\linewidth]{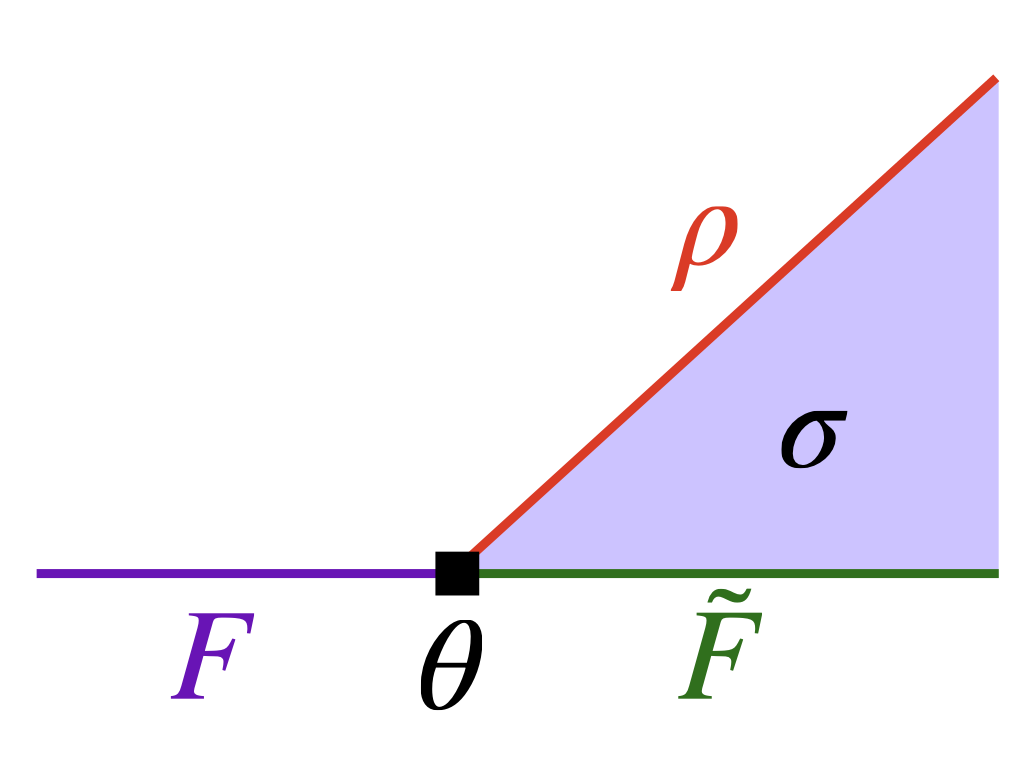}
\end{center}
\caption{ (Left) A pair $(\rho,\si)$ describes a ``topological symmetry'' in an
anomaly-free field theory $F$ (Fig. 1 in \Rf{FT220907471}).  (Right) A
redrawing of (Left) to make the equivalence $\te$ explicit.  }
\label{sandwich} 
\end{figure}

The notion of ``topological symmetry'' discussed in \Rf{FT220907471}
corresponds to a pair $(\rho,\si)$ (see Fig.  \ref{sandwich}(left)), where
$\si$ is the symTO discussed above,  and $\rho$  is a gapped boundary of the
symTO, which is $\tl \cR$  discussed above.  The pair $(\rho,\si)$ describes a
(generalized) symmetry in a quantum field theory $F$ via the equivalence
relation:  $F \stackrel{\te}{\cong} \rho \boxtimes_{\si} \tl F$ where $\tl F$ is
a boundary of $\si$, \ie $\text{bulk}(\tl F)=\si$.  However, to describe the
equivalence $\stackrel{\te}{\cong}$ explicitly, we need to connect the right
part and the left part of Fig.  \ref{sandwich}(left) via an isomorphism (or an
invertible domain wall), and redraw Fig. \ref{sandwich}(left) as Fig.
\ref{sandwich}(right).

Since $\eM$ in Fig. \ref{CCmorph} is determined by $\tl\cR$: $\eM=\eZ(\tl\cR)$
(or $\si$ is determined by $\rho$: $\si=\eZ(\rho)$), we may roughly say that
the most general emergent symmetries in $n+1$D systems $\underline{\cC}$ with
no gravitational anomaly are described and classified by fusion $n$-categories
$\tl\cR$.  The above emergent symmetries include anomalous
symmetries.\footnote{Here, we define anomalous symmetries as symmetries that
are not anomaly-free, along the lines of
\Rfs{TW191202817,KZ230107112,ZC230401262}. Such a definition will include both
invertible and noninvertible anomalies, just like gravitational anomalies
referred to in this paper include both invertible and noninvertible anomalies
\cite{KW1458,FV14095723,M14107442,KZ150201690,KZ170200673,JW190513279}.
Anomalous symmetries defined and classified via the boundary of non-trivial SPT
orders are invertible anomalies\cite{W1313}. Notably, \Rf{KZ200514178} gave a
definition and classification of invertible anomalies for noninvertible
symmetries.} 

However, the pair $(\tl\cR, \eM)$, or the triple $(\tl\cR, \cA, \eM)$ provided
a more complete description of the generalized symmetry.  Here $\cA$ the
condensable algebra that give rise to the gapped boundary $\tl\cR$, which is
the ``glue'' that connect the boundary $\tl\cR$ and the bulk $\eM$.  Some times
different condensable algebras, $\cA$ and $\cA'$, give rise to the same fusion
category $\tl\cR$.  In this case, the different triples, $(\tl\cR, \cA, \eM)$
and $(\tl\cR, \cA', \eM)$, describe different symmetries.  This is why the
triple  $(\tl\cR, \cA, \eM)$ provided a more complete description of the
generalized symmetry.

\subsection{Symmetry protected gaplessness}

\label{symmgapless}

It is well known that perturbative anomalies for continuous symmetries
\cite{H8035} and perturbative gravitational anomalies \cite{AW8469,W8597} imply
gaplessness
\cite{LSM6107,CG8205,W8322,H8353,W9125,Oc9911137,Oc0002392,H0431,GM08120351,CGW1107,L13017355}.
This can be regarded as \emph{perturbative-anomaly protected gaplessness}.
Even global anomalies for discrete symmetries may imply gaplessness
\cite{CLW1141,WS14011142,FO150307292,WS160406807,SV170501557,VP190506969,ET190708204,TV200806638,LZ220403131},
which can be regarded as \emph{anomalous-symmetry protected gaplessness}.
Symmetry fractionalization may also imply gaplessness
\cite{W0213,L160605652,WL191206167}, which can be regarded as
\emph{symmetry-fractionalization protected gaplessness}.  Some noninvertible
symmetry can imply gaplessness as well \cite{AL221214605,LZ230704788}, which
can be regarded as \emph{noninvertible symmetry protected gaplessness}.
Symm/TO correspondence provides a unified point of view to understand these
different kinds of protected gaplessness \cite{CW220506244}.  

In fact there are three types of symmetry protected gaplessness.  For the first
type, we consider states that do not spontaneously break the symTO of the
symmetry.  In this case, we can show the following: \cite{CW220506244}
\frmbox{A state with a non-trivial unbroken symTO must be gapless.} Such a
gapless state corresponds to a $\onebb$-condensed boundary --- which also must
be gapless --- of the corresponding topological order in one higher dimension
(\ie the unbroken symTO).\footnote{For a proof of a special case of this
statement, see \Rf{L190309028}.}  The gaplessness of a $\onebb$-condensed
boundary was shown in a general setting, and was referred to as
\emph{topological Wick rotation}, in \Rfs{KZ170501087,KZ190504924,KZ191201760}.  

For the second type of symmetry protected gaplessness, we consider the states
of a system with an $\tl\cR$-symmetry that do not spontaneously break the
symmetry. Such a symmetric state may be a gapped state and may be a gapless
state, depending on the fusion $n$-category $\tl\cR$.  In the following, we will
try to describe the conditions on  $\tl\cR$, such that an $\tl\cR$  symmetric state
must be gapless.

In order to make the above statement clearer, we need to define the notion of
spontaneous breaking of a generic $\tl\cR$-symmetry. In the holographic picture,
(see Fig. \ref{CCmorph}), the gapped $\tl\cR$ boundary is obtained by condensing
a maximal set of \emph{elementary topological excitations}, called a Lagrangian
condensable algebra, of the bulk $\eM$ corresponding to the charges of the
$\tl\cR$-symmetry.  \frmbox{\underline{Definition:} The $\tl\cR$-symmetry is 
\textbf{spontaneously broken} if one of the elementary topological excitations in
the Lagrangian condensable algebra that produces the $\tl\cR$-boundary condenses
on the lower boundary $\cC$ in Fig.  \ref{CCmorph}.  }

\begin{figure}[t]
\begin{center}
 \includegraphics[width=0.7\linewidth]{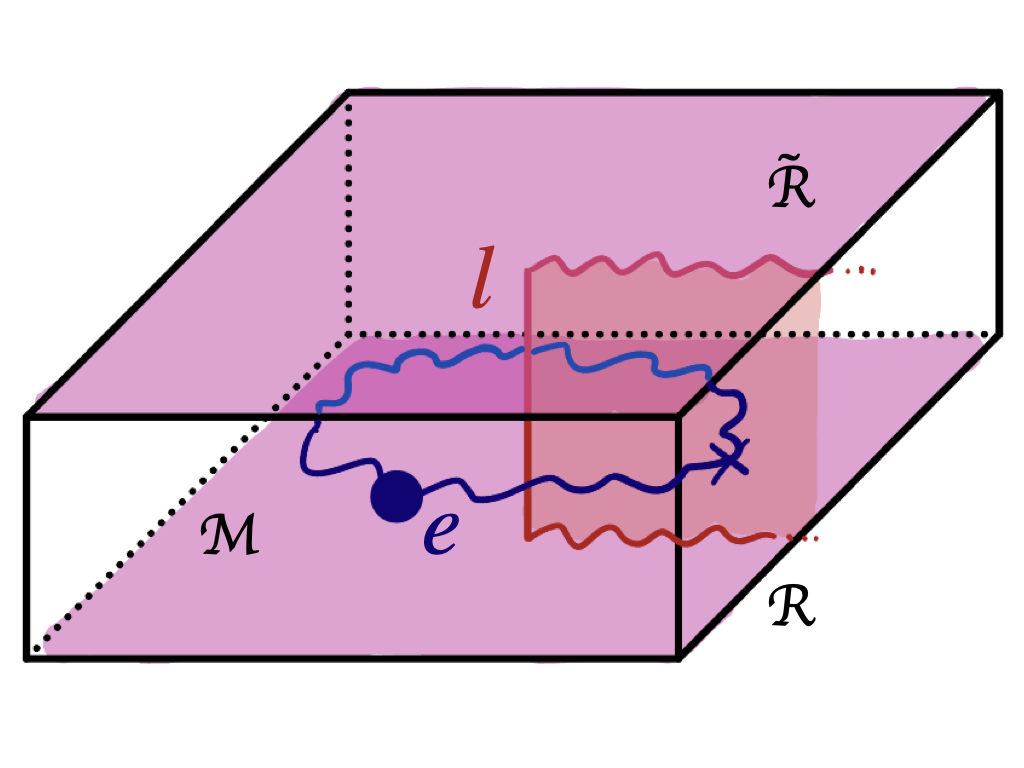}
\end{center}
\caption{A slab of 3+1D topological order $\eM$, with two gapped boundaries
$\tl\cR$ and $\cR$.  The string $l$ condenses on both the boundaries.  The
composite system is denoted as $\cR\boxtimes_\eM \tl\cR$.  } \label{lcond} 
\end{figure}

\begin{figure}[t]
	\begin{center}
		\includegraphics[width=0.7\linewidth]{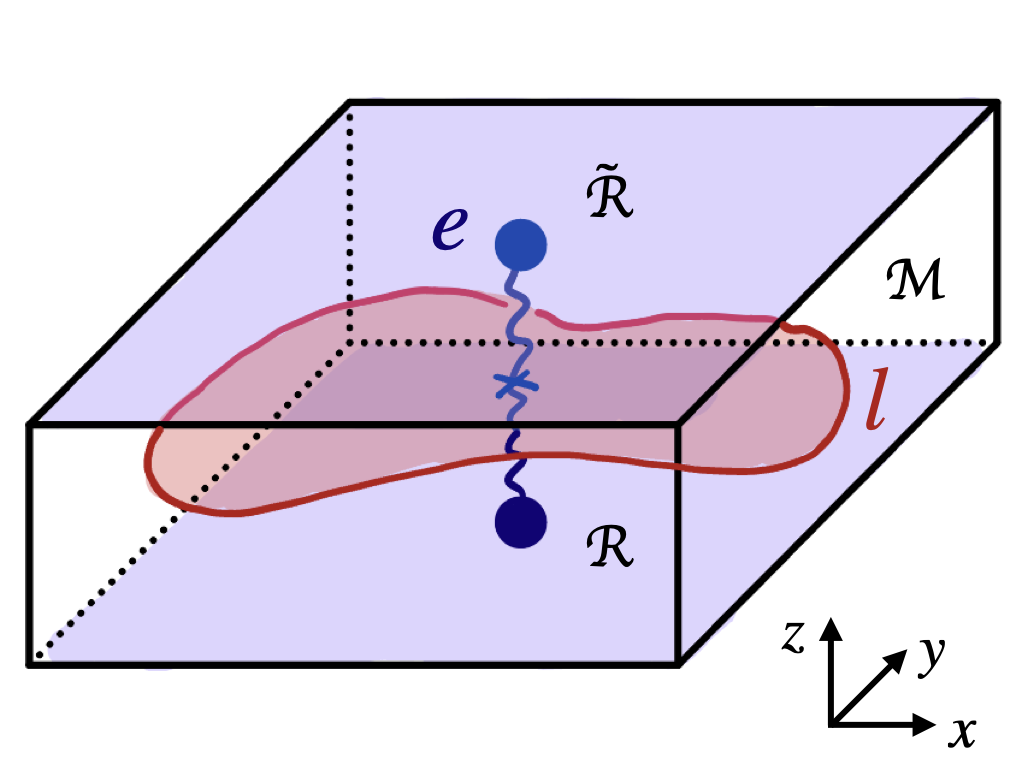}
		
	\end{center}
	\caption{A slab of 3+1D topological order $\eM$, with two gapped boundaries
		$\tl\cR$ and $\cR$.  The particle $e$ condenses on both the boundaries.  The
		composite system is denoted as $\cR\boxtimes_\eM \tl\cR$.  } \label{econd} 
\end{figure}

With this notion of spontaneous breaking of symmetry, we have the following
result: \frmbox{An anomaly-free $\tl\cR$-symmetry allows a gapped state that
does not spontaneously break the $\tl\cR$-symmetry.  In fact, the non-degenerate
gapped state used to define anomaly-free symmetry is one of these symmetric 
gapped
states.  } To show this result, we note that there exists a fusion $n$-category
$\cR$ such that $\cR \boxtimes_{\eZ(\tl\cR)} \tl\cR = n\cVec$, since $\tl\cR$ is
anomaly-free.  Such a state $\cR \boxtimes_{\eZ(\tl\cR)} \tl\cR = n\cVec$ is
actually a state that does not spontaneously break the $\tl\cR$-symmetry, \ie
there is no elementary topological excitation\footnote{The notion elementary
topological excitation was introduced in \Rf{KW1458}.  In general, the
excitations in a topological order can be divided into two classes: elementary
excitation and descendent excitation.  All point-like excitations are
elementary excitations.  A descendent string-like excitation is a phase formed
by  point-like excitations along a string.  A descendent membrane-like
excitation is a phase formed by point-like excitations and string-like
excitations on a membrane.  Thus, descendent excitations are excitations that
can have a boundary.  On the other hand, elementary excitations are excitations
that cannot have a boundary.  The elementary excitations satisfy the principle
of remote detectability \cite{L13017355,KW1458}: a non-trivial  elementary
excitation can always be detected by excitations via remote operations.  } that
condenses on both the $\tl\cR$- and $\cR$-boundaries. Let us argue why this is
the case.

For concreteness, let us consider a symTO in 3 spatial dimensions, \ie a
symmetry of a 2+1D system.  Assume there is a string-like excitation $l$ in
3+1D topological order $\eM$, that condenses on both boundaries (see Fig.
\ref{lcond}). Then, there must be a point-like excitation $e$ in $\eM$ that can
remotely detect $l$ via braiding.  In the dimension-reduced state
$\cR\boxtimes_\eM \tl\cR$, both $l$ and $e$ become point-like excitations which
can remotely detect each other.  Hence the dimension reduced state
$\cR\boxtimes_\eM \tl\cR$ has a non-trivial topological order, which leads to a
contradiction.  

Similarly, let us consider a particle-like excitation $e$ that condenses on both
boundaries. Then we can find a string excitation $l$ in the 3+1D topological
order $\cM$ that has nontrivial braiding with it, due to the remote
detectability principle. Now consider a membrane operator that creates $l$ at
its boundary and take this boundary to infinity in the $ x $ and $ y $
directions (see Fig. \ref{econd}).  Then we take the limit of making the slab
thin along $ z $. This means that the above membrane operator now acts as an
operator $O_l$ with support on all of 2d space in the dimension reduced system.
Now consider a state $\ket{\psi}$ in the ground state subspace of the
dimension-reduced system, and without loss of generality assume it to be in an
eigenstate of $O_l$. Since $e$ is condensed on both boundaries of the original
3+1D system, we can locally create an $e$ particle for vanishing energy cost.
Let us call the new state $\ket{\psi_e}$; this state must also be in the ground
state subspace. Now consider the action of the operator $O_l$ on these two
states. Since $l$ and $e$ braid nontrivially in $\eM$, the action of $O_l$ on
$\ket{\psi}$ and $\ket{\psi_e}$ must differ by the braiding phase.  This then
tells us that $\ket{\psi_e}$ and $\ket{\psi}$ must be linearly independent, and
hence the ground state subspace is degenerate. This tells us that the dimension
reduced system is nontrivial and does not have the structure of $2\cVec$, which leads to a contradiction. 

We see that an anomaly-free $\tl\cR$-symmetry does not have symmetry protected
gaplessness of the second type.  Can we conclude that anomalous 
$\tl\cR$-symmetry
has symmetry protected gaplessness of the second type?  The answer is no.  
Only
some anomalous $\tl\cR$-symmetries have symmetry protected gaplessness of 
the second
type.\cite{ZC230401262}  In particular, if the symTO $\eM = \eZ(\tl\cR)$ has 
only one gapped
boundary $\tl\cR$, then any $\tl\cR$-symmetric state must be gapless, since the 
only gapped boundary is obtained by condensing the $ \tl\cR $-symmetry 
charges and hence necessarily breaks the symmetry.  In
general, we have the following conclusion: \frmbox{Let $\eM= \eZ(\tl\cR)$ be 
the symTO of a $\tl\cR$-symmetry and
let all the gapped boundaries of $\eM$ be described by $\cR_i$,
$i=1,2,\cdots$.  If all $\cR_i$-boundaries share at least one common 
condensation of
an elementary topological excitation with the $\tl\cR$-boundary, then any
$\tl\cR$-symmetric state must be gapless.  }

In the third type of symmetry protected gaplessness, we assume the lattice
system to have an exact symmetry, such as non-on-site $U(1)$ symmetry with
perturbative t' Hooft anomaly.  Then such a lattice system cannot have a 
gapless
phase for any choice of lattice Hamiltonian, as long as the non-on-site lattice
symmetry is not explicitly broken.  An example of such a lattice non-on-site
symmetry is given in \Rf{DW230503024}. It seems that the third type of symmetry
protected gaplessness only appears for non-finite symmetries.

\subsection{Lattice realization of any generalized symmetry as emergent
symmetry} \label{slab}

\begin{figure}[t]
\begin{center}
 \includegraphics[width=0.7\linewidth]{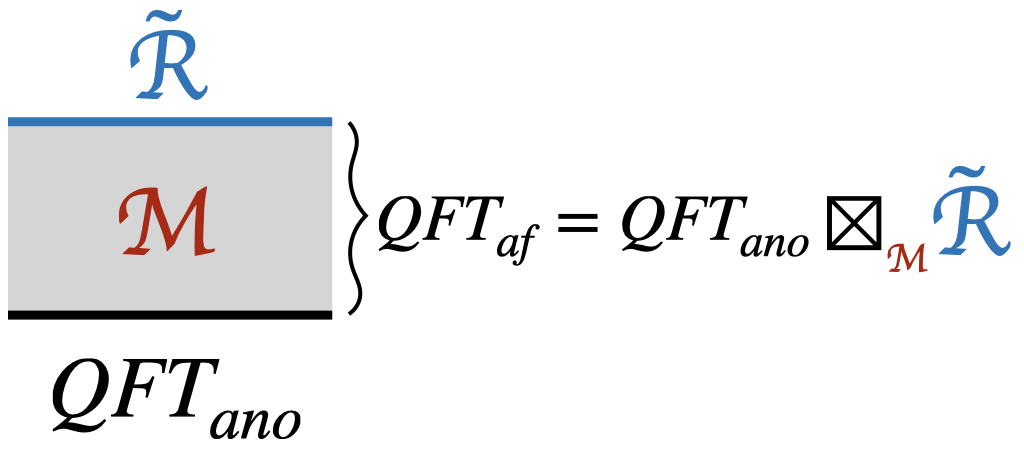}

\end{center}
\caption{The composite system $ QFT_{ano} \boxtimes_{\eM} \tl\cR$, where the
energy gaps of the bulk $\eM$ and the boundary $\tl\cR$ are assumed to be
infinite. We also assume that the thickness of the slab is finite and is much
larger than the correlation length of the bulk $\eM$.  } \label{QFTR} 
\end{figure}

We have mentioned that $n+1$D generalized symmetries are classified by minimal
fusion $n$-categories $\tl\cR$  \cite{KZ200308898,KZ200514178,FT220907471}.  The
isomorphic holographic decomposition in Fig. \ref{CCmorph} also provides a
construction of a lattice model that realized a finite $\tl\cR$-symmetry as an emergent symmetry.

The construction is given as follows.  Let $\eM$ be the braided fusion
$n$-category that is the center of $\tl\cR$.  We also use $\eM$ to denote the
$n+2$D topological order whose excitations are described by $\eM$.  Since $\eM$
has a gapped boundary, the topological order $\eM$ can be realized by a
commuting projector lattice model (such as the Levin-Wen models \cite{LW0510}
and Walker-Wang models \cite{WW1132}).  Let us consider a slab of commuting
projector lattice model realizing $\eM$ (see Fig. \ref{QFTR}).  The upper
boundary is gapped with boundary excitations described by $\tl\cR$.  The lower
boundary $QFT_\text{ano}$ can be gapped or gapless. We assume that the 
bulk $\eM$
and the upper boundary $\tl\cR$ have a large energy gap.  Below that energy
gap, all the low energy excitations are on the lower boundary $QFT_\text{ano}$.
The low energy effective theory of the slab is given by $QFT_\text{af} =
QFT_\text{ano} \boxtimes_{\eM} \tl\cR$.  We see that the low energy effective
theory $QFT_\text{af}$ has an emergent $\tl\cR$-symmetry below the bulk
energy gap.

\subsection{Generalized symmetry as algebra of local symmetric operators and
its transparent patch operators}

\label{commutantPO}

\begin{figure}[t]
\begin{center}
 \includegraphics[width=0.5\linewidth]{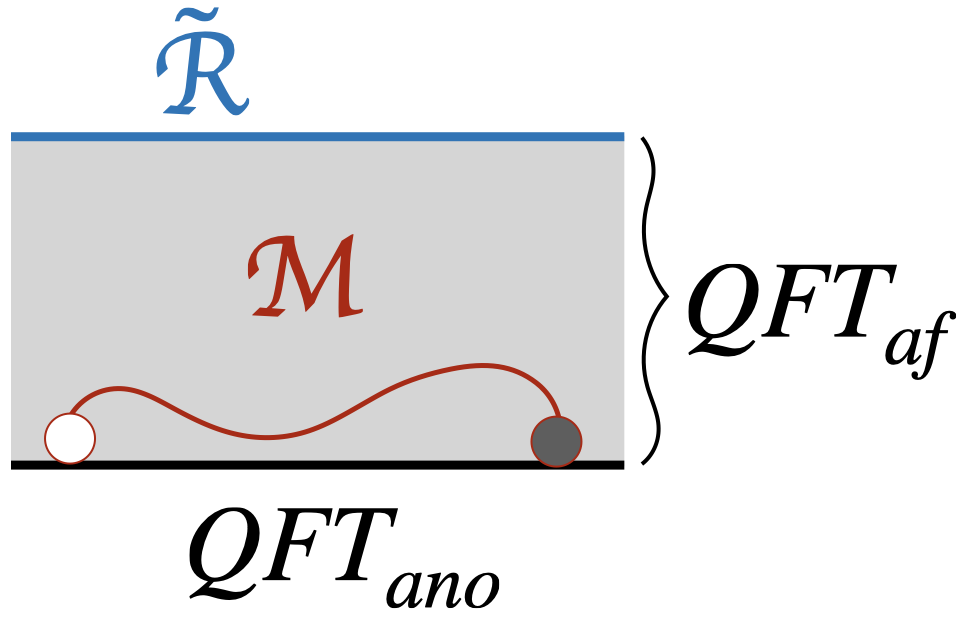}

\end{center}
\caption{A 1-dimensional transparent patch operator corresponds to a topological
string operator in the bulk that creates a pair of anyon-anti-anyon $a,\bar a$.
If $a$ has a quantum dimension $d_a >1$ (\ie if $a$ is a non-Abelian anyon),
the symmetry generated by the transparent patch operator will be a noninvertible
symmetry.  If $a$ has non-trivial statistics, the symmetry generated by the
transparent patch operator will be anomalous \cite{W181202517}.  } \label{QFTRpc} 
\end{figure}

In its most general setting, a generalized symmetry is defined by the algebra
$\sA^\text{symm}$ of some local operators.  
\begin{align}
 \sA^\text{symm} = \{\text{some local operators}\} .
\end{align}
The local operators in such a collection, by definition, are called \emph{local
symmetric operators} (LSOs), \ie $LSO \in \sA^\text{symm}$.  The choice of such
a collection of ``some local operators'' defines the symmetry.  For a special
case, the algebra $\sA$ of all local operators
\begin{align}
 \sA = \{\text{all local operators} \} 
\end{align}
describes a trivial symmetry (\ie no symmetry).

Usually, we describe a symmetry not by the algebra of LSOs, but by the algebra
of commutant  operators, \ie operators that commute with all the LSOs.  The
commutant  operators are the usual symmetry transformations.  When symmetry is
defined in such a general way, the symmetry transformations may not form a
group, which leads to non-invertible symmetry.

However, these commutant  operators are defined with a macroscopically large
support.  Since our systems of interest are local, it is desirable to describe
symmetry using only the information in a local patch.  This motivates us to
introduce transparent patch operators \cite{CW220303596} (which is called
quantum currents in \Rf{LZ230512917}).  If space is a manifold, then such a
patch is a sub-manifold that may have a lower dimensionality and a boundary.  A
transparent patch operator is defined as an operator that is a sum of products
of local symmetric operators on the patch.  We also require the transparent
patch operators to commute with all the LSOs, as long as the LSOs are far away
from the boundary of the patch.  

It was shown, through some simple examples \cite{CW220303596}, that the algebra
of the transparent patch operators encodes a non-degenerate braided fusion
$n$-category $\eM$, which gives rise to the symTO of the symmetry.

Two different but \emph{equivalent} algebras of LSOs can describe the same
symmetry.  We define the equivalence as follows:\\
(1) Two algebras of LSOs, $\sA_1^\text{symm}$ and $\sA_2^\text{symm}$ are
equivalent (denoted as $\sA_1^\text{symm} \cong \sA_2^\text{symm}$ ) if there
exist an unitary operator $U_\text{local}$ that maps all the local operators to
local operators and maps $\sA_1^\text{symm}$ to $\sA_2^\text{symm}$:
\begin{align}
 U_\text{local} \sA_1^\text{symm} U_\text{local}^\dag =
 \sA_2^\text{symm} .
\end{align}
(2) Two  algebras of LSOs, $\sA_1^\text{symm}$ and $\sA_2^\text{symm}$ are
equivalent if there exist $\sA_1$ and $\sA_2$, such that
\begin{align}
 \sA_1^\text{symm} \sA_1
 \cong \sA_2^\text{symm} \sA_2
\end{align}
The equivalence classes of the above equivalence relation define the notion of
symmetry (which include generalized symmetries).

If we replace (1) by\\
(1') Two algebras of LSOs, $\sA_1^\text{symm}$ and $\sA_2^\text{symm}$ are
equivalent if there exist an unitary operator $U_\text{non-local}$ that maps
$\sA_1^\text{symm}$ to $\sA_2^\text{symm}$:
\begin{align}
 U_\text{non-local} \sA_1^\text{symm} U_\text{non-local}^\dag =
 \sA_2^\text{symm} .
\end{align}
The equivalence classes of the equivalence relations (1') and (2) define the
notion of holo-equivalent symmetries.  We remark that $U_\text{non-local}$ only
maps the LSOs to LSOs.  $U_\text{non-local}$ may map a local non-symmetric
operator to non-local operators.

It turns out that holo-equivalence classes of $n+1$D symmetries are 1-to-1
classified by symTO's (\ie non-degenerate braided fusion $n$-categories)
\cite{JW191213492,KZ200514178}.  On the other hand, the equivalence classes of
$n+1$D symmetries are classified by fusion $n$-categories (see Section
\ref{fusioncat}).

In our lattice realization of $\tl\cR$-symmetry Fig. \ref{QFTR}, what
are the transparent patch operators that describe the $\tl\cR$-symmetry?  It 
turns out that a $k$-dimensional transparent patch operator is
given by a $k$-brane topological operator in the bulk
\cite{JW191213492,CW220303596}, that creates a $k-1$-dimensional topological
excitation on its boundary (see Fig.  \ref{QFTRpc}).  (A $k$-brane
topological operator corresponds to a $k+1$-dimensional topological defect in
spacetime.)

\begin{table*}[tb] 
\caption{The classification of 2+1D topological orders (up to $E(8)$ invertible
topological order) for bosonic systems with no symmetry, up to 11 types of
anyons (the first row).  This leads to a classification of 2+1D symTOs (the
second row), which classify all the 1+1D generalized global symmetries up to
holo-equivalence.  Such a classification includes all finite-group symmetries
with potential anomalies (the third row).  It also includes beyond-group
symmetries.  } \label{toptable} \centering \setlength{\tabcolsep}{5pt}
\begin{tabular}{|c | c|c|c|c|c|c|c|c|c|c|c|}
\hline
\# of anyon (symmetry charges/defects) types  
& 1 & 2 &  3 &  4 &  5 &  6 &  7 &  8 &  9 & 10 & 11\\
\hline
\hline
\# of 2+1D topological orders 
& 1 & 4 & 12 & 18 & 10 & 50 & 28 & 64 & 81 & 76 & 44\\
\hline
\# of symTOs (TOs with gappable boundary) 
& 1 & 0 & 0 & 3 & 0 & 0 & 0 & 6 & 6 & 3 & 0\\
\hline
\# of finite-group symmetries (with anomaly $\om$) 
& 1 & 0 & 0 & 2$_{\Z_2^\om}$ & 0 & 0 & 0 & 6$_{S_3^\om}$ & 3$_{\Z_3^\om}$ & 0 & 0\\
\hline
\end{tabular}
\end{table*}

After the upper boundary $\tl\cR$ is specified, the  transparent patch operators
can be divided into patch charge operators, patch symmetry operators, and their
combinations.  A transparent patch operator is a patch charge operator if the
topological excitation created on the boundary condenses on the upper boundary
$\tl\cR$.  The boundary of a patch charge operator corresponds to symmetry
charges.  A transparent patch operator is a patch symmetry operator (or a
combination of a patch symmetry operator and a patch charge operator) if the
topological excitation created on the boundary does not condenses on the upper
boundary $\tl\cR$.  The boundary of patch symmetry operator corresponds to
symmetry defects, and the patch symmetry operator performs the symmetry
transformation within the patch \cite{JW191213492,CW220303596}.  This is why
the excitations on the upper boundary $\tl\cR$ correspond to symmetry defects.
\frmbox{The fusion $n$-category $\tl\cR$ describes the fusion of symmetry
defects of the $\tl\cR$-symmetry.}

\subsection{Classification of 1+1D symTOs and generalized symmetries}

We have seen that the emergent symmetries from lattice systems can be
generalized symmetries that go beyond the group and higher group 
descriptions.  Such
emergent generalized symmetries (up to holo-equivalence) are fully described
and classified by symTOs.  Recently, 2+1D topological orders (up to $E_8$
invertible topological orders) with 11 or less anyon types were classified (see
Table \ref{toptable}) \cite{NW230809670}.  The symTOs correspond to such
topological orders with gappable boundary, which have also been classified 
(see Table
\ref{toptable}).  This leads to a classification of 1+1D generalized
symmetries.  

For example, there are three symTOs describing three holo-equivalent classes of 1+1D symmetries with 4
types of symmetry charges/defects \cite{NW230809670}: 
\begin{enumerate}

\item
$\Z_2$ topological order
(\ie $\Z_2$ gauge theory) in 2+1D, 
$\eGau_{\Z_2}$.
Its holo-equivalent class of symmetries
contains two symmetries:
$\cVec_{\Z_2}$-symmetry and
$\cRep_{\Z_2}$-symmetry,
since the symTO $\eGau_{\Z_2}$
is the Drinfeld center of two
fusion categories
$\cVec_{\Z_2}$ and $\cRep_{\Z_2}$:
\begin{align}
 \eGau_{\Z_2} = \eZ(\cVec_{\Z_2}) = \eZ(\cRep_{\Z_2}).
\end{align}

When $\tl\cR = \cVec_{\Z_2}$ in Fig. \ref{CCmorph}, the corresponding
$\cVec_{\Z_2}$-symmetry is nothing but the ordinary $\Z_2$ symmetry,
since the $\Z_2$ symmetry defects form the fusion category $\cVec_{\Z_2}$.
When $\tl\cR = \cRep_{\Z_2}$ in Fig. \ref{CCmorph}, the corresponding
$\cRep_{\Z_2}$-symmetry is $\tl\Z_2$, the dual of $\Z_2$ symmetry
discussed in \Rf{JW191213492}, since the $\tl\Z_2$ symmetry defects form the
fusion category $\cRep_{\Z_2}$.

We remark that, for Abelian group $G$, the $\cVec_{G}$-symmetry (\ie the ordinary $G$-symmetry) and the $\cRep_{G}$-symmetry
(\ie the dual of $G$-symmetry, $\tl G$) are isomorphic.  For non-Abelian group
$G$, the $\cVec_{G}$-symmetry and the $\cRep_{G}$-symmetry
are not isomorphic.  The $\cVec_{G}$-symmetry is the ordinary
$G$-symmetry, while the $\cRep_{G}$-symmetry is the dual of
$G$-symmetry, which is a noninvertible algebraic symmetry
\cite{TW191202817,JW191213492,KZ200514178}.  We also remark that the
$\cVec_{G}$-symmetry and the $\cRep_{G}$-symmetry (or more
generally, all the symmetries in the same holo-equivalence class) are
holo-equivalent \cite{KZ200514178,CW220303596}, in the sense that they provide
the same constrain on the dynamics of the systems within the symmetric
sub-Hilbert space $\cV_\text{symmetric}$.

\item 
Double-semion topological order, denoted by
$\eM_\text{dSem}$.  Its holo-equivalent class of symmetries contains just one
symmetry: $\tl\cR_\text{Sem}$-symmetry, since the symTO
$\eM_\text{dSem}$ is the Drinfeld center of just one fusion category
$\tl\cR_\text{Sem}$:
\begin{align}
 \eM_\text{dSem} = \eZ(\tl\cR_\text{Sem}).
\end{align}
Here, $\tl\cR_\text{Sem}$ is the fusion category formed by anyons in
$\eM_\text{Sem}$, and $\eM_\text{Sem}$ is the braided fusion category formed by
a single semion.  We denote such a relation as $\tl\cR_\text{Sem} \leftarrow
\eM_\text{Sem}$.  The $\tl\cR_\text{Sem}$-symmetry is nothing but the
anomalous $\Z_2$ symmetry.

\item 
Double-Fibonacci topological order, denoted by
$\eM_\text{dFib}$.  Its holo-equivalent class of symmetries contains just one
symmetry: $\tl\cR_\text{Fib}$-symmetry, since the symTO
$\eM_\text{dFib}$ is the Drinfeld center of just one fusion category
$\tl\cR_\text{Fib}$:
\begin{align}
\label{Rfib}
\eM_\text{dFib}  = \eZ(\tl\cR_\text{Fib}).
\end{align}
Here, $\tl\cR_\text{Fib} \leftarrow \eM_\text{Fib} $ is the fusion category
formed by anyons in the braided fusion category $\eM_\text{Fib}$ of a single
Fibonacci anyon.  See Section \ref{FibSymm} for a lattice realization and a
discussion of gapless states with the $\tl\cR_\text{Fib}$-symmetry.

The $\tl\cR_\text{Fib}$-symmetry is a noninvertible symmetry.  Since
$\tl\cR_\text{Fib}$ is not a local fusion category, $\tl\cR_\text{Fib}$-symmetry
is an anomalous noninvertible symmetry.  In fact, the anomaly of the
noninvertible $\tl\cR_\text{Fib}$-symmetry is itself not invertible, i.e. the
symmetry of the system cannot be made anomaly-free by stacking it with other
such anomalous $\tl\cR_\text{Fib}$-symmetric systems.  
Also, $\tl\cR_\text{Fib}$-symmetric state must be gapless.

\end{enumerate}

There are six symTOs and holo-equivalent classes of 1+1D symmetries with
9 types of symmetry charges/defects \cite{NW230809670}: 
\begin{enumerate}

\item
$\Z_3$ gauge theory $\eGau_{\Z_3}$.  Its holo-equivalence
class contains two symmetries: $\Z_3$ symmetry and dual of $\Z_3$ symmetry,
$\tl\Z_3$. 

\item
Dijkgraaf-Witten $\Z_3$ gauge theory $\eGau^{(1)}_{\Z_3}$.
Its holo-equivalence class contains one symmetry: anomalous $\Z^{(1)}_3$
symmetry.

\item
Another Dijkgraaf-Witten $\Z_3$ gauge theory
$\eGau^{(2)}_{\Z_3}$.  Its holo-equivalence class contains one symmetry:
another anomalous $\Z^{(2)}_3$ symmetry.

\item
Double-Ising topological order
$\eM_\text{dIs}$.
Its holo-equivalence class contains one symmetry:
$\tl\cR_\text{Is}$-symmetry, where
$\tl\cR_\text{Is} \leftarrow \eM_\text{Is}$ and
$\eM_\text{Is}$ is the rank-3 Ising topological order with central charge $c=\frac12$.
see Section \ref{dIsSymmTO} for a realization of this symmetry.

\item
Twisted-double-Ising topological order
$\eM_\text{twdIs}$.  Its holo-equivalence class contains one symmetry:
$\tl\cR_\text{twIs}$-symmetry, where $\tl\cR_\text{twIs} \leftarrow
\eM_\text{twIs}$ and $\eM_\text{twIs}$ is the rank-3 twisted Ising topological
order with central charge $c=\frac32$.  Note that the Ising topological order
$\eM_\text{Is}$ contains a fermion $\psi$.  If $\psi$'s condense into a filling
fraction $\nu=1$ integer quantum Hall state, the Ising topological order
$\eM_\text{Is}$ will change into the twisted Ising topological order
$\eM_\text{twIs}$. 

\item
$PSU(2)_5$ topological order $\eZ(\tl\cR_{PSU(2)_5})$.
$\tl\cR_{PSU(2)_5}$ is a fusion category $\tl\cR_{PSU(2)_5} \leftarrow
\eM_{PSU(2)_5}$.  $\eM_{PSU(2)_5}$ is a factor of $\eM_{SU(2)_5}$, the modular
tensor category of $SU(2)_5$ Kac-Moody algebra
\begin{align}
 \eM_{SU(2)_5} = \eM_{PSU(2)_5}\boxtimes \eM_\text{Sem} .
\end{align}
The holo-equivalence class of the symTO $\eZ(\tl\cR_{PSU(2)_5})$ contains
one symmetry: $\tl\cR_{PSU(2)_5}$-symmetry, which is an
anomalous noninvertible symmetry.\cite{KZ200514178}

\end{enumerate}

\section{Definition of maximal \lowercase{sym}TO}

We have mentioned that a state with a non-trivial unbroken symTO $\eM$ must
be gapless
\cite{L190309028,KZ190504924,KZ191201760,JW191213492,KZ200514178,CW220506244}.
Such a state corresponds to a $\onebb$-condensed boundary of the corresponding
symTO $\eM$.\cite{KZ190504924,KZ191201760,CW220506244}  Thus the gaplessness
of the state is directly associated with the non-trivialness of unbroken symTO\
$\eM$.  This supports the idea that a gapless state is characterized by its
symTO $\eM$. 

However, the correspondence between the gapless state and its symTO $\eM$ is
not one-to-one.  This is because the same topological order $\eM$ can have many
different $\onebb$-condensed boundaries, which leads to many different gapless
states with the same symTO.  These gapless states have varying numbers of
gapless excitations.  A gapless state with more gapless excitations may have a
large emergent symmetry, \ie may be a $\onebb$-condensed boundary of a larger
symTO.  This leads to the notion of maximal symTO $\eM_\text{max}$ for the
gapless state: it is the largest topological order in one higher dimension
which has a $\onebb$-condensed boundary that has the same number of gapless
excitations as that of the gapless state.

\begin{figure}[t]
\begin{center}
 \includegraphics[width=0.7\linewidth]{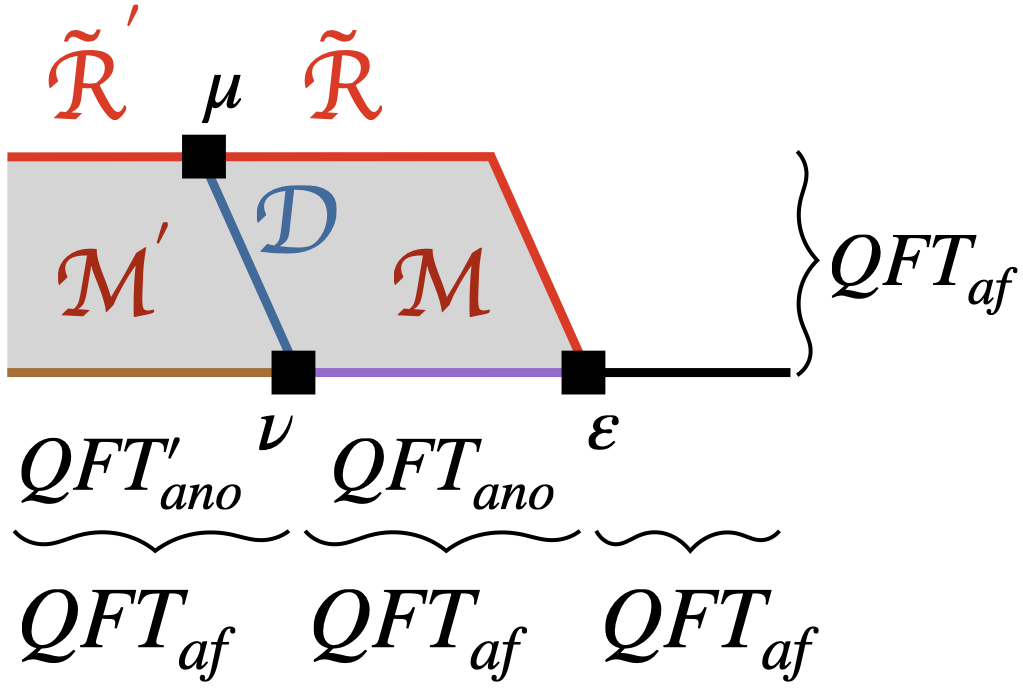}

\end{center}
\caption{ Decompositions of an anomaly-free quantum field theory $QFT_{af}$
expose the possible emergent symmetries in the quantum field theory.  The bulk
topological orders $\eM,\eM'$ are the revealed emergent symTOs.  The gapped
boundaries of $\eM,\eM'$, $\tl\cR,\tl\cR'$, describe the revealed emergent
symmetries $\cR,\cR'$.  The gapless anomalous field theories, $QFT_{ano}$ and
$QFT_{ano}'$, are $\onebb$-condensed boundaries of $\eM,\eM'$.  } \label{QFTQFT} 
\end{figure}

In this section, we will define the notion of maximal symTO, using Symm/TO
correspondence.  Consider an $n+1$D anomaly-free gapless conformal field
theory,\footnote{In this paper, ``field theory'' is defined as a ground state
along with all its low energy excitations.  A field theory is anomaly-free if
it can be realized by a lattice model.  A ``conformal field theory'' is gapless
with a linear dispersion with a single velocity.} $QFT_{af}$, which is
described by a single-component partition function $Z^{af}$.  If $QFT_{af}$ has
some (emergent) symmetry, then we can decompose it as a stacking in Figs.
\ref{QFTR} and \ref{QFTQFT}:\cite{KZ200514178,FT220907471}
\begin{align}
 QFT_{af} = QFT_{ano} \boxtimes_{\eM} \tl\cR ,
\end{align}
to expose the (emergent) symmetry.  Here $QFT_{ano}$ is an anomalous field
theory described by a multi-component partition function
$Z^{ano}_a$.\cite{JW190513279,JW191209391}  $\eM$ is the exposed emergent symTO
and $\tl\cR$ is a gapped boundary of $\eM$ describing the exposed emergent
symmetry.  The above decomposition has a meaning that the invariant partition
function $Z^{af}$ can be constructed from the multi-component partition
function $Z^{ano}_a$, and the data describing the gapped bulk $\eM$ and the
gapped boundary $\tl\cR$.  See next section for a detailed construction for 1+1D
field theory.  

The above decomposition is not unique. For a given $QFT_{af}$, we may have
several decompositions (see Fig.  \ref{QFTQFT}):
\begin{align}
 QFT_{af} = 
QFT_{ano} \boxtimes_{\eM} \tl\cR 
=QFT_{ano}' \boxtimes_{\eM'} \tl\cR' .
\end{align}
Such different decompositions reveal different parts of the (emergent) symmetry
of $QFT_{af}$. Here we have assumed that the bulks $\eM$, $\eM'$ and the
boundaries $\tl\cR$, $\tl\cR'$ have infinite energy gap and all the excitations
on the boundaries $QFT_{ano}$ and $QFT_{ano}'$ have zero or finite energy gaps. These
excitations, along with the possible degenerate ground states (\ie the global
excitations) from $\eM$, $\eM'$, $\tl\cR$, and $\tl\cR'$, give rise to the finite
energy excitations of $QFT_{af}$.  In other words, we have assumed that
$QFT_{ano}$ is a $\onebb$-condensed boundary of $\eM$ or a nearly
$\onebb$-condensed boundary of $\eM$. By ``nearly $\onebb$-condensed'', we mean
that if $QFT_{ano}$ is induced by some non-trivial condensations, these
condensations are assumed to be weak and only lead to small energy gaps.
Similarly, we have assumed that $QFT_{ano}'$ is a $\onebb$-condensed boundary of
$\eM'$ or a nearly $\onebb$-condensed boundary of $\eM'$.
The decomposition that gives rise to the \emph{largest} $\eM$ reveals the
maximal symTO in the state $QFT_{af}$.

In making the above statement, we define a symTO $\cM$ to be \emph{larger} than symTO $\eM'$ if $\eM$ has a larger total quantum dimension.  Let us review the notion of quantum dimension of topological orders.
A topological order $\eM$ can have two kinds of excitations: elementary
excitations and descendent excitations.\cite{KW1458} The descendent excitations
are formed by the condensation of elementary excitations.  Let us label all the
elementary excitations by $a$. Each elementary excitation has a quantum
dimension $d_a$ describing the number of its internal degrees of freedom. For
example, a spin-$\frac12$ particle has a quantum dimension $d=2$. The total quantum
dimension of $ \eM $ is $D^2 = \sum_{a\in \eM} d_a^2$.  So the maximal
symTO corresponds to the topological order $\eM$ with the largest total
quantum dimension.  This leads to a definition of  maximal symTO for an
anomaly-free gapless state.

Using Fig. \ref{QFTQFT}, we can give another definition of ``larger'' symTO,
which is consistent with the above definition based on total quantum dimension.
Since the symTOs are all in trivial Witt class (\ie have gappable
boundaries), there is a gapped domain wall $\cD$ between the two symTOs,
$\eM$ and $\eM'$.  The gapped domain wall $\cD$ can be viewed as a gapped
boundary of $\eM\boxtimes \overline{\eM}'$.  In other words, $\cD$ corresponds
to a Lagrangian condensable algebra $\cA = \oplus_{a,\bar b} A_{a,\bar b}
a\otimes \bar b$ in $\eM\boxtimes \overline{\eM}'$, where $a$ are elementary
excitations in $\eM$ and $\bar b$ are elementary excitations in
$\overline{\eM}'$.  Now we can define that \frmbox{$\eM'$ is \textbf{larger} than 
$\eM$
if there exists a domain wall $\cD$, such that $A_{a,\bar b} =\Big|_{\bar b
=\onebb} \del_{a,\onebb}$.} In other words, $\cD$ can be viewed as a
$\onebb$-condensed boundary of $\eM$.  This is equivalent to say that $\eM$ is
induced from $\eM'$ by condensing excitations in a condensable algebra
$\cA_{\eM'} = \oplus_{\bar b} A_{a=\onebb,\bar b} b$ in $\eM'$:\cite{K13078244,CW220506244}
\begin{align}
 \eM = \eM'_{/\cA_{\eM'}}.
\end{align}

Being able to compare some topological orders may help us define maximal
symTO.  However, among all possible emergent symTOs of a given system,
the limit of increasingly large symTOs may not be unique.  So in general,
we will define maximal symTO formally: \frmbox{The \textbf{maximal symTO}
of a given system is defined formally as the collection of all  possible
emergent symTOs of that system. } If the limit of increasingly large 
symTOs is unique, we can use that limit to represent the  maximal
symTO.  Otherwise, we need to use the whole collection of all  possible
emergent symTOs to represent the maximal symTO.

We also note that the three gapless states $QFT_{af}$, $QFT_{ano}$,
$QFT_{ano}'$, are \emph{local low energy equivalent} since they only differ by
stacking gapped states with large energy gaps.  Here, we propose that
\frmbox{the local-low-energy-equivalence classes of gapless liquid states are
largely characterized by their emergent maximal symTOs.  } In \Rf{W0213}, the
notion of projective symmetry group (PSG) was introduced to characterize
gapless states (as well as gapped states).  The maximal symTO is a much more
improved version of PSG and can characterize a gapless state much more
completely.

\section{Examples and constructions of maximal \lowercase{sym}TOs}

In this section, we give some 1+1D examples of the Symm/TO correspondence described
above. Anomaly-free 1+1D gapless states (\ie CFTs) are described by
modular invariant partition functions:
\begin{align}
 Z^{af}(\tau+1)= Z^{af}(-1/\tau)=Z^{af}(\tau).
\end{align}
It is well known that a 1+1D rational CFT is closely related to a 2+1D
topological quantum field theory \cite{MS8977,W8951,FSh0204148,FSh0607247,FS09095013}.  Here, we will consider a
different problem: we note that a 1+1D rational CFT can be related to many
different  2+1D topological orders (\ie we can focus on different emergent
symTOs for a 1+1D rational CFT).  We want to identify the maximal emergent
symTO of a given CFT, and propose that the maximal emergent symTO largely
determines the CFT.  We also consider a related problem: given a symTO, what
is the minimal CFT that has the symTO unbroken?

\subsection{1+1D Ising critical point}

\subsubsection{Modular invariant and modular covariant partition functions }

As an example, let's consider the 1+1D CFT describing the $\Z_2$ symmetry breaking transition,
denoted as $Is_{af}$, which has the following modular invariant partition function on
a ring-like space \cite{CFT12}:
\begin{align}
 Z^{af}_{Is}(\tau,\bar \tau) =
|\chi^\text{Is}_0(\tau)|^2+|\chi^\text{Is}_\frac12(\tau)|^2 +
|\chi^\text{Is}_{\frac1{16}}(\tau)|^2
\end{align}
where $\chi^\text{Is}_h(\tau)$ are the conformal characters of Ising CFT (the
(4,3) minimal model), and the subscript $h$ is the scaling dimension of the
corresponding primary fields.

\begin{figure}[t]
\begin{center}
 \includegraphics[width=0.5\linewidth]{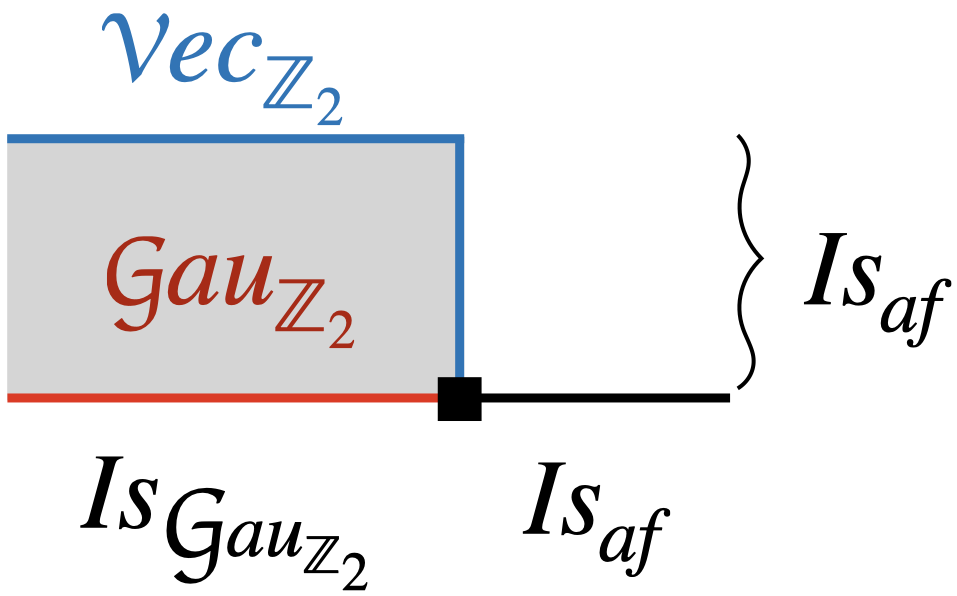}

\end{center}
\caption{A decomposition of an anomaly-free Ising critical point $Is_{af}$
exposes an emergent $\cVec_{\Z_2}$-symmetry (which is the
ordinary symmetry $\Z_2$), as well as the emergent symTO: $\eM =\eGau_{\Z_2}$.
The gapless boundary $Is_{\eGau_{\Z_2}}$
is a $\onebb$-condensed boundary of $\eGau_{\Z_2}$.
} \label{IsingZ2}
\end{figure}

To reveal the emergent symTO, following \Rf{JW191213492}, we
restrict to the sub-Hilbert space of $\Z_2$ invariant states.  Restricting to
the symmetric Hilbert space converts the symmetry to the noninvertible
gravitational anomaly, since the symmetric sub-Hilbert space 
$\cV_\text{symmetric}$
does not have the tensor product decomposition $\cV_\text{symmetric} \neq \otimes
\cV_i$, where $\cV_i$ is local Hilbert space on a site $i$.  The partition
function in symmetric sub-Hilbert space is given by
\begin{align}
 Z(\tau,\bar \tau) =
|\chi^\text{Is}_0(\tau)|^2+|\chi^\text{Is}_\frac12(\tau)|^2 
\end{align}
which is not modular invariant, but it is part of 4-component partition
function \cite{JW190513279}:
\begin{align}
\label{Z2cri}
\begin{pmatrix}
 Z^{\eGau_{\Z_2}}_{\onebb\text{-cnd};\onebb}(\tau,\bar \tau)\\
 Z^{\eGau_{\Z_2}}_{\onebb\text{-cnd};e}(\tau,\bar \tau)\\
 Z^{\eGau_{\Z_2}}_{\onebb\text{-cnd};m}(\tau,\bar \tau)\\
 Z^{\eGau_{\Z_2}}_{\onebb\text{-cnd};f}(\tau,\bar \tau)\\
\end{pmatrix} 
&=
\begin{pmatrix}
|\chi^\text{Is}_0(\tau)|^2+|\chi^\text{Is}_\frac12(\tau)|^2\\
|\chi^\text{Is}_{\frac1{16}}(\tau)|^2\\
|\chi^\text{Is}_{\frac1{16}}(\tau)|^2\\
\chi^\text{Is}_0(\tau) \bar \chi^\text{Is}_\frac12(\tau) +\chi^\text{Is}_\frac12(\tau) \bar \chi^\text{Is}_0(\tau) \\
\end{pmatrix} ,
\end{align}
which is modular covariant 
\cite{CZ190312334,JW190513279,JW191213492}
\begin{align}
\label{STtrans}
 Z_a(\tau+1) = T_{ab} Z_b(\tau), \ \ \ \
 Z_a(-1/\tau) = S_{ab} Z_b(\tau),
\end{align} 
with the $S,T$ matrices are given by
\begin{align}
\label{Z2STmat}
  T^{\eGau_{\Z_2}}&=
\begin{pmatrix}
    1&0&0&0\\
    0&1&0&0\\
    0&0&1&0\\
    0&0&0&-1
  \end{pmatrix} ,
&
  S^{\eGau_{\Z_2}}&=\frac12 \begin{pmatrix}
    1&1&1&1\\
    1&1&-1&-1\\
    1&-1&1&-1\\
    1&-1&-1&1
  \end{pmatrix}
\end{align}
The above four-component partition function is the partition function of an
anomalous CFT, denoted as $Is_{\eGau_{\Z_2}}$, which can be viewed as a
gapless boundary of a 2+1D $\Z_2$ topological order
$\eGau_{\Z_2}$ (see Fig. \ref{IsingZ2}).  The $\Z_2$ bulk
topological order $\eGau_{\Z_2}$ has 4 types of excitations $\onebb,e,m,f$, where
$\onebb,e,m$ are bosons and $f$ is a fermion.  $e,m,f$ have $\pi$ mutual
statistics between them.  They have the following fusion rule
\begin{align}
 e\otimes e =
 m\otimes m =
 f\otimes f = \onebb,\ \ \ 
 e\otimes m = f.
\end{align}
The above fusion rule implies the mod-2 conservation of $e$-particles,
$m$-particles, and $f$-particles, which correspond to the three $\Z_2$ 
symmetries:
the $\Z_2^m$ symmetry (\ie the $\Z_2$ symmetry) where $e,f$ carry its charge;
the $\Z_2^e$ symmetry (\ie the dual $\tl\Z_2$ symmetry) where $m,f$ carry its
charge; the $\Z_2^f$ symmetry where $e,m$ carry its charge.

The $\Z_2$ topological order $\eGau_{\Z_2}$, the symTO of the $\Z_2$ symmetry,
is characterized by the $S,T$ matrices in \eqn{Z2STmat}.  We see that the $S,T$
matrices for the topological order in one higher dimension constrain the
partition function of 1+1D CFT via the modular covariance condition
\eq{STtrans}.  This is how a symTO largely determines a gapless state.

The above results can also be obtained within 1+1D CFT, if we consider the
following four partition functions with $\Z_2$ twisted boundary conditions
\cite{CY180204445}, 
\begin{align}\label{Z2bc}
&Z_{++}=\zb,\quad Z_{+-}=\zbh,\nonumber \\
&Z_{-+}= \zbv, \quad Z_{--}= \zbx ,
\end{align}
where the vertical direction is the time direction.  We find
\begin{align}
\label{IsingZ1}
 Z_{++}(\tau) &= |\chi^\text{Is}_0|^2+|\chi^\text{Is}_\frac{1}{2}|^2 +|\chi^\text{Is}_\frac{1}{16}|^2
  \\ 
 Z_{+-}(\tau) &= |\chi^\text{Is}_0|^2+|\chi^\text{Is}_\frac{1}{2}|^2 -|\chi^\text{Is}_\frac{1}{16}|^2
 \nonumber \\ 
Z_{-+}(\tau) &= 
|\chi^\text{Is}_\frac{1}{16}|^2
+\chi^\text{Is}_0 \bar \chi^\text{Is}_\frac{1}{2} + \chi^\text{Is}_\frac12 \bar \chi^\text{Is}_0
 \nonumber \\ 
Z_{--}(\tau) &= |\chi^\text{Is}_\frac{1}{16}|^2 -
\chi^\text{Is}_0 \bar \chi^\text{Is}_\frac{1}{2} - \chi^\text{Is}_\frac12 \bar \chi^\text{Is}_0
\nonumber 
\end{align}

In the $G$-symmetry-twist basis of 
partition functions, the $S$ and $T$ matrix for modular transformation is
\begin{equation}\label{ZpropG}
\begin{split}
 Z_{g',h'}(-1/\tau) &=  S_{(g',h'),(g,h)} Z_{g,h}(\tau),\\
 Z_{g',h'}(\tau+1) &=  T_{(g',h'),(g,h)} Z_{g,h}(\tau),\\
 Z_{g',h'}(\tau) &=  R_{(g',h'),(g,h)}(u) Z_{g,h}(\tau),\\
S_{(g',h'),(g,h)} &= \del_{(g',h'),(h^{-1},g)},\\
T_{(g',h'),(g,h)} &= \del_{(g',h'),(g,hg)},\\
R_{(g',h'),(g,h)}(u) &= \del_{(g',h'),(ugu^{-1},uhu^{-1})}
,
\end{split}
\end{equation}
where 
\begin{align}
g,h,g',h'\in G, \ \ \ gh=hg,\ \ \ g'h'=h'g',
\end{align}
describe the symmetry twists of the symmetry group $G$.
For $G=\Z_2=\{+,-\}$, we find
\begin{align}
\label{Z2ST1}
S=\begin{pmatrix}
1 & 0 & 0 & 0 \\
0 & 0 & 1 & 0 \\
0 & 1 & 0 & 0 \\
0 & 0 & 0 & 1
\end{pmatrix}
,\ \
T=\begin{pmatrix}
1 & 0 & 0 & 0 \\
0 & 1 & 0 & 0 \\
0 & 0 & 0 & 1 \\
0 & 0 & 1 & 0
\end{pmatrix}
,\ \ R=1 .
\end{align}
This way, we can obtain modular covariant multi-component partition functions
for 1+1D CFTs with symmetry.  Including the symmetry twist and considering
modular covariant multi-component partition functions is a way to expose the
symmetry in a CFT.  The modular invariant single component partition function
corresponds to a point of view of ignoring the symmetry.

We note that each component of the partition function is a polynomial of
$q=\ee^{\ii 2\pi \tau}$ and $\bar q$, times a factor $q^{-\frac c {24} + h}
\bar q^{-\frac {\bar c} {24} + \bar h}$.  Here $c,\bar c$ are the central
charges for the right movers and left movers, and $h,\bar h$ are the right and
left scaling dimensions of the primary fields of the corresponding sector.  We
can choose a different basis where the expansion coefficients are all
non-negative integers.  Such a basis is the so-called quasiparticle basis:
\begin{align}
Z^{\eGau_{\Z_2}}_{\onebb\text{-cnd};\onebb} =&\frac{Z_{++}+Z_{+-}}{2},\ \
Z^{\eGau_{\Z_2}}_{\onebb\text{-cnd};e} =\frac{Z_{++}-Z_{+-}}{2}
\\
Z^{\eGau_{\Z_2}}_{\onebb\text{-cnd};m} =&\frac{Z_{-+}+Z_{--}}{2},\ \
Z^{\eGau_{\Z_2}}_{\onebb\text{-cnd};f} =\frac{Z_{-+}-Z_{--}}{2}.
\nonumber 
\end{align}
The partition functions in the quasiparticle basis\cite{JW190513279} are given
by \eqn{Z2cri}, and transform as \eqn{STtrans}, with $S,T$ given by
\eqn{Z2STmat}.  Note that $T$ is always diagonal in the quasiparticle basis.  

This example demonstrates how to convert a symmetry to a noninvertible
gravitational anomaly (characterized by the $S,T$ matrices for the topological
order in one higher dimension).  We can view a global symmetry as a
noninvertible gravitational anomaly, \ie as a topological order in one higher
dimension.  Viewing the CFT at the $\Z_2$ symmetry breaking transition as the
gapless boundary of 2+1D $\Z_2$ topological order, not only allows us to see 
the
$\Z_2$ symmetry, it also allows us to see two additional symmetries, $\tl\Z_2$
and $\Z_2^f$.

\subsubsection{The decomposition in terms of partition functions }

Using this explicit example, we can explain the decomposition
(see Fig. \ref{IsingZ2}) 
\begin{align}
 Is_{af} 
= Is_{\eGau_{\Z_2}} \boxtimes_{\eM} \tl\cR
= Is_{\eGau_{\Z_2}} \boxtimes_{\eGau_{\Z_2}} \cVec_{\Z_2}
\end{align}
in more detail.  The decomposition reveals an emergent $\Z_2$ symmetry which is
described by the fusion 1-category $\tl\cR=\cVec_{\Z_2}$ (formed by the $\Z_2$
symmetry defect $m$) \cite{FT180600008,JW191213492}. 

A gapped boundary is described by $\tau$ independent multi-component
partition function $Z^\text{gapped}_a$ which is modular covariant
\begin{align}
 Z^\text{gapped}_a = T_{ab} Z^\text{gapped}_b, \ \ \ \
 Z^\text{gapped}_a = S_{ab} Z^\text{gapped}_b,
\end{align} 
The gapped boundary $\tl\cR=\cVec_{\Z_2}$ in Fig. \ref{IsingZ2} is an
$e$-condensed boundary (so that the boundary excitations are given by $m$'s).
Such a $\cVec_{\Z_2}$-boundary is described by the following constant
multi-component partition function 
\begin{align}
\label{VecZ2}
\begin{pmatrix}
 Z^{\eGau_{\Z_2}}_{e\text{-cnd};\onebb}\\
 Z^{\eGau_{\Z_2}}_{e\text{-cnd};e}\\
 Z^{\eGau_{\Z_2}}_{e\text{-cnd};m}\\
 Z^{\eGau_{\Z_2}}_{e\text{-cnd};f}\\
\end{pmatrix} 
&=
\begin{pmatrix}
1\\
1\\
0\\
0\\
\end{pmatrix} ,
\end{align}
where $Z^{\eGau_{\Z_2}}_{e\text{-cnd};e}=1$ indicates the $e$-condensation, and
$Z^{\eGau_{\Z_2}}_{e\text{-cnd};\onebb}=1$ indicates the $\onebb$-condensation on
the boundary.  In fact the trivial particle $\onebb$ always condenses on the
boundary and $Z^{\eGau_{\Z_2}}_{e\text{-cnd};\onebb}$ is always a positive
integer, describing the ground state degeneracy of the boundary.  Now, the
formal decomposition $ Is_{af} = Is_{\eGau_{\Z_2}} \boxtimes_{\eGau_{\Z_2}}
\cVec_{\Z_2} $ have an explicit meaning
\begin{align}
Z^{af}_{Is} (\tau,\bar\tau) =
\sum_{a=\{\onebb,e,m,f\}}
Z^{\eGau_{\Z_2}}_{\onebb\text{-cnd};a} (\tau,\bar\tau) 
\left (Z^{\eGau_{\Z_2}}_{e\text{-cnd};a} \right )^*
. 
\end{align}

\begin{figure}[t]
\begin{center}
 \includegraphics[width=0.5\linewidth]{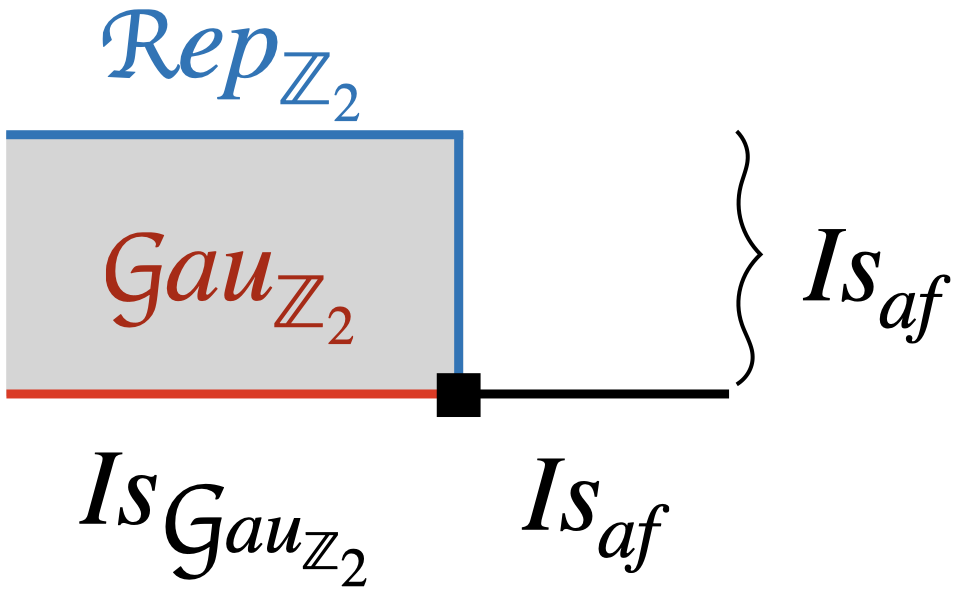}

\end{center}
\caption{ The second decomposition of an anomaly-free Ising critical point
$Is_{af}$ exposes an emergent $\cRep_{\Z_2}$-symmetry (which is
the dual-symmetry $\tl\Z_2$), as well as the emergent symTO: $\eM
=\eGau_{\Z_2}$.  
The gapless boundary $Is_{\eGau_{\Z_2}}$
is a $\onebb$-condensed boundary of $\eGau_{\Z_2}$.
} \label{IsingtZ2}
\end{figure}

The Ising critical point $Is_{af}$ has another decomposition
(see Fig. \ref{IsingtZ2}) 
\begin{align}
 Is_{af} 
= Is_{\eGau_{\Z_2}} \boxtimes_{\eM} \tl\cR
= Is_{\eGau_{\Z_2}} \boxtimes_{\eGau_{\Z_2}} \cRep_{\Z_2}
\end{align}
This second decomposition reveals an emergent $\tl\Z_2$ symmetry which is the
dual of the $\Z_2$ symmetry and is described by the fusion 1-category
$\tl\cR=\cRep_{\Z_2}$ (formed by the $\Z_2$ charges $e$).

The gapped boundary $\tl\cR=\cRep_{\Z_2}$ in Fig. \ref{IsingtZ2} is an
$m$-condensed boundary (so that the boundary excitations are given by $e$'s).
Such a $\cRep_{\Z_2}$-boundary is described by the following constant
multi-component partition function 
\begin{align}
\label{RepZ2}
\begin{pmatrix}
 Z^{\eGau_{\Z_2}}_{m\text{-cnd};\onebb}\\
 Z^{\eGau_{\Z_2}}_{m\text{-cnd};e}\\
 Z^{\eGau_{\Z_2}}_{m\text{-cnd};m}\\
 Z^{\eGau_{\Z_2}}_{m\text{-cnd};f}\\
\end{pmatrix} 
&=
\begin{pmatrix}
1\\
0\\
1\\
0\\
\end{pmatrix} ,
\end{align}
where $Z^{\eGau_{\Z_2}}_{m\text{-cnd};m}=1$ indicates the $m$-condensation on
the boundary.  The formal decomposition $ Is_{af} = Is_{\eGau_{\Z_2}}
\boxtimes_{\eGau_{\Z_2}} \cRep_{\Z_2} $ implies the following relation
between partition functions:
\begin{align}
Z^{af}_{Is} (\tau,\bar\tau) =
\sum_{a=\{\onebb,e,m,f\}}
Z^{\eGau_{\Z_2}}_{\onebb\text{-cnd};a} (\tau,\bar\tau) 
\left (Z^{\eGau_{\Z_2}}_{m\text{-cnd};a}\right )^* . 
\end{align}

From the above discussion, we see that for each gapped boundary, we can
construct a modular invariant partition function.  This relation between
modular invariant partition functions and gapped boundaries (\ie Lagrangian
condensable algebras) was noticed before, in \Rfs{KR08073356,M09092537}, 
but was very mysterious at the time. Now, within the framework of
Symm/TO correspondence, it becomes very natural.

\subsubsection{Maximal symTO}
\label{dIsSymmTO}

\begin{figure}[t]
\begin{center}
 \includegraphics[width=0.5\linewidth]{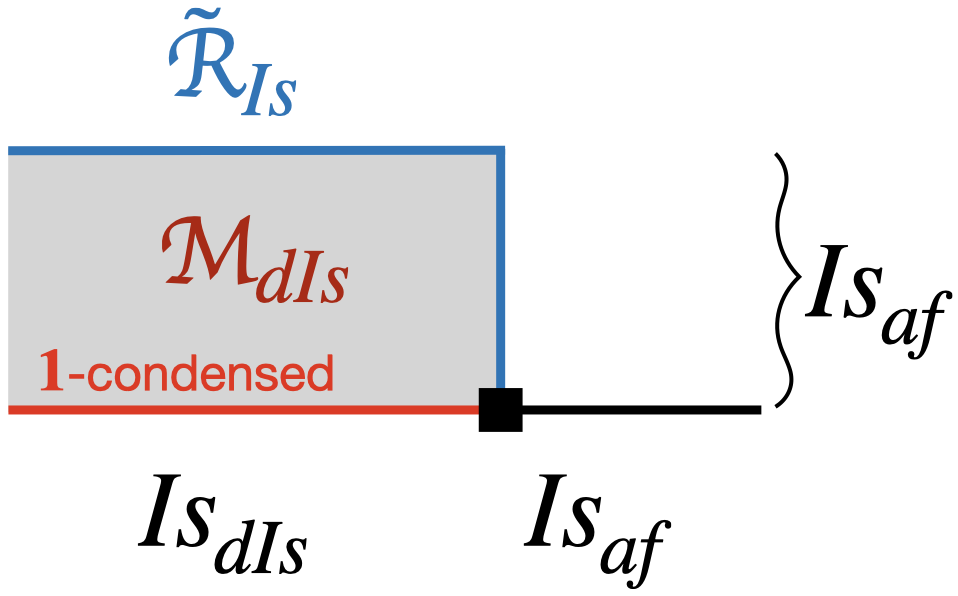}
\end{center}
\caption{ The third decomposition of an anomaly-free Ising critical point
$Is_{af}$ exposes the maximal emergent symTO: $\eM =\eM_\text{dIs}$, as well as
the emergent $\tl\cR_\text{Is}$-symmetry.  The $\tl\cR_\text{Is}$-symmetry is a
noninvertible symmetry with noninvertible anomaly.} \label{CFTdIs}
\end{figure}

The above two decompositions only reveal emergent $\Z_2$ symmetry or dual
$\Z_2$ symmetry. Their associated symTO is $\eM = \eGau_{\Z_2}$. However,
$\eGau_{\Z_2}$ is not the maximal symTO.  The Ising critical point also has an
$\Z_2^{em}$ symmetry that exchange $e$ and $m$, which is not included in the
symTO $\eGau_{\Z_2}$.  To reveal the emergent maximal symTO, we need to
consider another decomposition \cite{HV14116932,CZ190312334,JW191213492}
\begin{align}
  Is_{af} = Is_\text{dIs} \boxtimes_{\eM_\text{dIs}} \tl\cR_\text{Is} .
\end{align}
Here $\eM_\text{dIs} = \eM_\text{Is} \boxtimes \bar \eM_\text{Is} $ is the 1+2D
double-Ising topological order, which has 9 anyons labeled by $(h,\bar h)$,
$h,\bar h=0,\frac12, \frac1{16}$.  $h=0,\frac12, \frac1{16}$ correspond to the
three anyons $\onebb,\psi,\si$ in the Ising topological order $\eM_\text{Is}$:
\begin{align}
\begin{matrix}
\text{anyons}: & \onebb &  \psi  & \si     \\
d_a: &  1 & 1 & \sqrt2  \\
h_a: & 0 & \frac12 & \frac1{16} \\
\end{matrix}
\end{align}
where $d_a$ is the quantum dimension and $h_a$ is the topological spin of the
corresponding anyon.

$Is_\text{dIs}$ is an anomalous CFT (a $\onebb$-condensed
boundary of $\eM_\text{dIs}$) which is described
by the following multi-component partition function, one component
for each anyon $(h,\bar h)$:
\begin{align}
 Z^{\eM_\text{dIs}}_{\onebb\text{-cnd};(h,\bar h)}(\tau,\bar \tau) =
\chi^\text{Is}_{h}(\tau)
\bar \chi^\text{Is}_{\bar h}(\bar \tau),\ \
h,\bar h\in \left \{0,\frac12, \frac1{16}\right \}
.
\end{align}
$\tl\cR_\text{Is}$ is the gapped boundary of $\eM_\text{dIs}$ which is described
by the following modular covariant multi-component constant partition function:
\begin{align}
 Z^{\eM_\text{dIs}}_{\tl\cR_\text{Is};(0,0)} &=
 Z^{\eM_\text{dIs}}_{\tl\cR_\text{Is};(\frac12,\frac12)} =
 Z^{\eM_\text{dIs}}_{\tl\cR_\text{Is};(\frac1{16},\frac1{16})} =1,
\nonumber\\
\text{ others } &= 0.
\end{align}
In fact, $\tl\cR_\text{Is}$ is the fusion 1-category formed by $\onebb,\psi,\si$.
The relation between partition functions
\begin{align}
Z^{af}_{Is} (\tau,\bar\tau) =
\sum_{h,\bar h}
Z^{\eM_\text{dIs}}_{\onebb\text{-cnd};(h,\bar h)} (\tau,\bar\tau) 
\left (Z^{\eM_\text{dIs}}_{\tl\cR_\text{Is};(h,\bar h)}\right )^* 
\end{align}
confirms the decomposition $  Is_{af} = Is_\text{dIs}
\boxtimes_{\eM_\text{dIs}} \tl\cR_\text{Is} $.

Let us note that $\tl\cR_\text{Is}$ is not a local fusion 1-category, since
there is no \emph{dual} local fusion 1-category $\tl\cR$ that satisfies
$\tl\cR_\text{Is} \boxtimes_{\eM_\text{dIs}} \tl\cR = \cVec$.  $\si$ in
$\tl\cR_\text{Is}$ having a non-integral quantum dimension $\sqrt 2$ also
implies that $\tl\cR_\text{Is}$ is not a local fusion
1-category.\cite{TW191202817} Therefore,  $\tl\cR_\text{Is}$ does not describe
an anomaly-free noninvertible symmetry.  Thus, \frmbox{the decomposition $
Is_{af} = Is_\text{dIs} \boxtimes_{\eM_\text{dIs}} \tl\cR_\text{Is} $ reveals an
emergence of $\tl\cR_\text{Is}$-symmetry and $ \eM_\text{dIs} $ symTO, for the
Ising critical point.  However, there is no emergent  anomaly-free noninvertible
symmetry for this emergent symTO.  } This is an interesting example, where
anomaly-free symmetry can no longer properly describe the emergent symmetry,
even after including those that are beyond group and higher group.  One
may use such an anomalous symmetry to properly describe the emergent symmetry.
Note that the concept of anomaly for noninvertible symmetry is quite subtle. 
In \Rf{KZ200514178}, the authors provided a definition of an \emph{invertible anomaly}, 
according to which the $\tl\cR_\text{Is}$-symmetry above is beyond such 
invertible anomaly. This example demonstrates the richness of emergent 
symmetries, which can be noninvertible with noninvertible anomalies.  On the 
other hand, symTO is a simple, unified, and systematic way to describe the most 
general emergent symmetry.  

Even though the $\tl\cR_\text{Is}$-symmetry has a noninvertible anomaly, one 
can still consistently describe its symmetry transformations. The associated 
symmetry defects are
described by the fusion category $\tl\cR_\text{Is}$.  Let us discuss these 
symmetry transformations in the
slab model (see Fig. \ref{CFTdIs}) discussed in Section \ref{slab}, which
realizes the emergent $\tl\cR_\text{Is}$-symmetry.  The symmetry transformations
are given by string operators in the bulk that create a pair of
anyon-anti-anyons.  There are nine such string operators which are denoted as
$O_\text{str}(a,a^*)$, corresponding to the nine types of anyons of the 
double-Ising topological order: $a = $ $\onebb$, $\psi$, $\si$, $\bar\psi$, 
$\bar\si$,
$\psi\bar\psi$, $\psi\bar\si$, $\si\bar\psi$, $\si\bar\si$.

However, the string operators $O_\text{str}(\onebb,\onebb)$,
$O_\text{str}(\psi\bar\psi,\psi\bar\psi)$,
$O_\text{str}(\si\bar\si,\si\bar\si)$, do not generate symmetry
transformations; they correspond to patch charge operators (in the language of 
\Rf{CW220303596}).  This is
because the anyons they create, $\onebb$, $\psi\bar\psi$, and $\si\bar\si$,
condense on the $\tl\cR_\text{Is}$ boundary.  

The non-trivial patch symmetry operators are given by $O_\text{str}(\psi,\psi)$
and $O_\text{str}(\si,\si)$, which correspond to the $\Z_2^f$ symmetry and the
$\Z_2^{em}$ symmetry mentioned before.  Since $\psi$ and $\si$ are not bosons,
the symmetry generated by $O_\text{str}(\psi,\psi)$ and $O_\text{str}(\si,\si)$
are anomalous.  The symmetry $\Z_2^f$ generated by
$O_\text{str}(\psi,\psi)$ is invertible since $\psi\otimes \psi = \onebb$.  The
symmetry $\Z_2^{em}$ generated by $O_\text{str}(\si,\si)$ is noninvertible
since $\si\otimes \si = \onebb\oplus \psi$.

Let us note that the Ising model at the critical coupling
\begin{align}
\label{HIsingC}
 H = -\sum_i (Z_iZ_{i+1} + X_i)
\end{align}
and the closely related Majorana fermion model \eqref{HMajC} realize the
$\Z_2^f$ and $\Z_2^{em}$ symmetries, but only in the low energy limit.
The $\Z_2^{em}$ is realized via lattice
translation in the Majorana model \eqref{HMajC}.  This is different from the
slab lattice model Fig. \ref{CFTdIs}, where the $\Z_2^{em}$ symmetry
transformation does not involve translation.  So the  slab lattice model Fig.
\ref{CFTdIs} is quite different from the model \eqref{HIsingC}.

\subsection{1+1D critical points for models with $G$ symmetry or
dual $\tl G$ symmetry}

\subsubsection{Two 1+1D lattice models with group-like symmetry $G$ and
algebraic symmetry $\tl G$}

We consider two 1+1D lattice models on a ring, where lattice sites are labeled
by $i$, the links labeled by $ij$.  In the first model, the physical degrees of
freedom live on the vertices and are labeled by group elements $g$ of a finite
group $G$.  The many-body Hilbert space is spanned in the following local basis
\begin{align}
|\{g_i\}\>, \ \ \ g_i \in G.
\end{align}
The Hamiltonian is given by
\begin{align}
\label{HG}
 H_G = - J \sum_{i} f(g_ig_{i+1}^{-1}) - \sum_i \sum_{h \in G} L_h(i),
\end{align}
where $ f(g)$ is a positive function that is peaked at $ g = \id$.  Also,
the operator $L_h(i)$ is given by
\begin{align}
 L_h(i) |g_1,\cdots,g_i,\cdots,g_N\>= |g_1,\cdots,hg_i,\cdots,g_N\>.
\end{align}
The Hamiltonian $H_G$ has an on-site  $G$ symmetry
\begin{align}
\label{Uh}
 U_h H_G = H_G U_h, \ \ \ 
U_h =\prod_i L_h(i).
\end{align}
We see that when $J \gg 1$, $H_G$ is in the symmetry breaking phase, and when
$J \ll 1$, $H_G$ is in the symmetric phase.

The second bosonic lattice model has
degrees of freedom living on the links.  On an oriented link $ij$ pointing from
$i$-site to $j$ site, the degrees of freedom are labeled by $g_{ij} \in G$. The
many-body Hilbert space has the following local basis
\begin{align}
|\{g_{ij}\}\>, \ \ \ g_{ij} \in G.
\end{align} Here, $g_{ij}$'s on links with
opposite orientations satisfy 
\begin{align}
 g_{ij}=g_{ji}^{-1}.
\end{align}
The second model is related to the first model.  A state
$|g_1,\cdots,g_i,\cdots,g_N\>$ in the first model is mapped to a state
$|\cdots,g_{i,i+1},\cdots\>$ in the second model where
$g_{i,i+1}=g_ig_{i+1}^{-1}$.

This connection allows us to design the Hamiltonian of the second model as
\begin{align}
\label{HtG}
 H_{\tl G} = &- J \sum_{i} f(g_{i,i+1}) 
- \sum_i \sum_{h \in G} Q_h(i) ,
\end{align}
where the star term $Q_h(i)$ acts on the two links $(i,i+1)$ and $(i-1,i)$:
\begin{align}
\label{Qh}
&\ \ \ \
 Q_h(i) |\cdots,g_{i-1,i} , g_{i,i+1},\cdots\> 
\nonumber\\
& = |\cdots,
g_{i-1,i}h^{-1}, 
hg_{i,i+1} ,
\cdots\>.
\end{align}
The second model has an algebraic symmetry, denoted as $\tl G$
\cite{KZ200514178}, 
\begin{align}\label{Wq}
 W_q H_{\tl G} = H_{\tl G} W_q  ,~~~
W_q = \Tr \prod_{i} R_q(g_{i,i+1}),
\end{align}
where $R_q$ is an irreducible
representation (irrep) of $G$.  We see that the algebraic symmetry $\tl G$
is generated by the Wilson loop operators $W_q$, for
all irrep $q$.  We note that the algebraic symmetry
$\tl G$ is different from the usual symmetry characterized by a group $G$,
when $G$ is non-Abelian. However, when $G$ is Abelian, the  dual symmetry 
$\tl G$ reduces to a group-like symmetry, and is isomorphic to $G$.  

\begin{table*}[t]
\caption{The point-like excitations and their fusion rules in 2+1D $\eGau_{S_3}$
topological order (\ie $S_3$ gauge theory with charge excitations).  The $S_3$
group is generated by $(1,2)$ and $(1,2,3)$.  Here $\bm 1$ is the trivial
excitation.  $a_1$ and $a_2$ are pure $S_3$ charge excitations, where $a_1$
corresponds to the 1-dimensional sign irreducible representation (irrep), and 
$a_2$ the 2-dimensional
irrep of $S_3$.  $b$ and $c$ are pure $S_3$ flux excitations, where
$b$ corresponds to the conjugacy class $\{(1,2,3),(1,3,2)\}$, and $c$ conjugacy
class $\{(1,2),(2,3),(1,3)\}$.  $b_1$, $b_2$, and $c_1$ are charge-flux bound
states.  $d,s$ are the quantum dimension and the topological spin of an
excitation.  } \label{S3FusionRules} 
\setlength\extrarowheight{-2pt}
\setlength{\tabcolsep}{2pt}
\centering
\begin{tabular}{|c | c|c|c|c|c|c|c|c|}
\hline
$d,s$ & $1,0$ & $1,0$ & $2,0$ & $2,0$ & $2,\frac13$ & $2,-\frac13$ & $3,0$ & $3,\frac12$\\
\hline
$\otimes$ & $\bm 1$ &  $a_1$  & $a_2$ &  $b$  &  $b_1$  &  $b_2$  & $c$  & $c_1$    \\
\hline
$ \bm 1 $ & $\bm 1$ & $a_1$ & $a_2$   & $b$  & $b_1$  &  $b_2$          & $c$  & $c_1$    \\
$a_1$ & $a_1$ & $\bm 1$ & $a_2$	  & $b$  &$b_1$  & $b_2$      &  $c_1$  & $c$  \\
$a_2$ & $a_2$ & $a_2$   & $\bm 1\oplus a_1\oplus a_2$      & $b_1\oplus b_2$    & $b\oplus b_2$ & $b\oplus b_1$      & $c\oplus c_1$  & $c\oplus c_1$\\
$b$  & $b$  & $b$ & $b_1\oplus b_2$     & $\bm 1\oplus a_1\oplus b$ & $b_2\oplus a_2$  & $b_1\oplus a_2$     & $c\oplus c_1$   & $c\oplus c_1$  \\
$b_1$  & $b_1$   & $b_1$ & $b\oplus b_2$      & $b_2\oplus a_2$ & $\bm 1\oplus a_1\oplus b_1$  & $b\oplus a_2$      & $c\oplus c_1$ & $c\oplus c_1$ \\
$b_2$  & $b_2$  & $b_2$  & $b\oplus b_1$    & $b_1\oplus a_2$ & $b\oplus a_2$  & $\bm 1\oplus a_1\oplus b_2$      & $c\oplus c_1$  & $c\oplus c_1$ \\
$c$ & $c$ & $c_1$ & $c\oplus c_1$     & $c\oplus c_1$   & $c\oplus c_1$  & $c\oplus c_1$     & $\bm 1\oplus a_2\oplus b\oplus b_1\oplus b_2$ & $a_1 \oplus a_2\oplus b\oplus b_1\oplus b_2$  \\
$c_1$ & $c_1$ & $c$  & $c\oplus c_1$     & $c\oplus c_1$ & $c\oplus c_1$  & $c\oplus c_1$     & $a_1 \oplus a_2\oplus b\oplus b_1\oplus b_2$  & $\bm 1\oplus a_2\oplus b\oplus b_1\oplus b_2$ \\
\hline
\end{tabular}
\end{table*}

\subsubsection{Critical points and their holographic picture}

Let us assume that for an appropriately chosen function $f(g)$, the model 
$H_G$ has a
continuous spontaneous symmetry breaking transition at $J=J_c$.  Due to the
duality, the model $H_{\tl G}$ also has a continuous transition at $J=J_c$.
What are the partition functions for these two critical points?

\begin{figure}[t]
\begin{center}
 \includegraphics[width=0.5\linewidth]{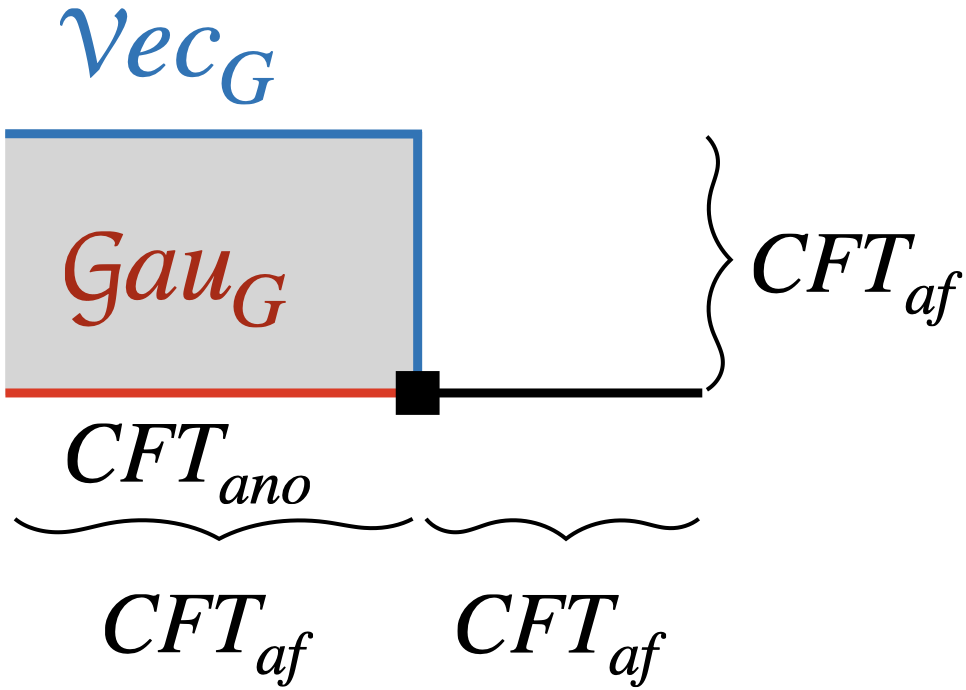}

\end{center}
\caption{ A decomposition of an anomaly-free critical point $CFT_{af}$ of model
$H_G$ \eq{HG}, exposes an emergent symmetry $G$, as well as the emergent symTO:
$\eM =\eGau_{G}$. The symmetry $G$ is described by $\tl\cR =\cVec_{G}$ for the
fusion of the symmetry defects.  } \label{CFTCFTG} 
\end{figure}

\begin{figure}[t]
\begin{center}
 \includegraphics[width=0.5\linewidth]{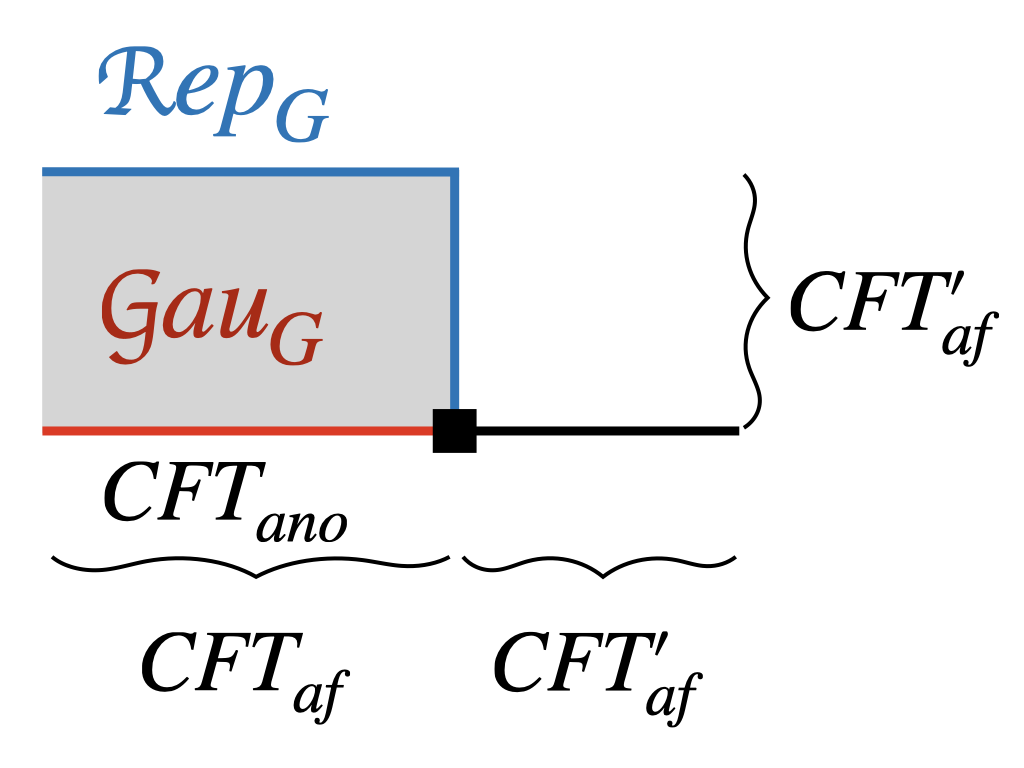}

\end{center}
\caption{ A decomposition of an anomaly-free critical point $CFT_{af}'$ of
model $H_{\tl G}$ \eq{HtG}, exposes an emergent algebraic symmetry $\tl G$, as
well as the emergent symTO: $\eM =\eGau_{G}$. The algebraic symmetry $\tl G$ is
described by $\tl\cR =\cRep_{G}$ for the fusion of the symmetry defects.}
\label{CFTCFTtG} 
\end{figure}

Both the symmetry $G$ and the dual algebraic symmetry $\tl G$ have the same
symTO, given by 2+1D $G$-gauge theory $\eGau_G$ (see Table \ref{S3FusionRules}
for $G=S_3$).\cite{JW191213492,KZ200514178} Therefore the critical point in
model $H_G$, denoted as $CFT_{af}$, is given by Fig.  \ref{CFTCFTG}.  This is
because the symmetry $G$ is the $\cVec_G$-symmetry, which leads to
Fig.  \ref{CFTCFTG}.  On the other hand, the critical point in model $H_{\tl
G}$, denoted as $CFT'_{af}$, is given by Fig.  \ref{CFTCFTtG}.  This is because
the algebraic symmetry $\tl G$ is the $\cRep_G$-symmetry, which leads
to Fig.  \ref{CFTCFTtG}.

The anomalous gapless boundary $CFT_{ano}$ in Fig.  \ref{CFTCFTG} and
\ref{CFTCFTtG} is a $\onebb$-condensed boundary of $\eGau_G$. For $G=S_3$, it is
described by the following multi component partition function, labeled by the
anyons in $\eGau_{S_3}$:
\begingroup
\allowdisplaybreaks
\begin{align}
\label{S3bdy}
 Z^{\eGau_{S_3}}_{\onebb\text{-cnd};\onebb} &=  |\chi^{m6}_{0}|^2 +  |\chi^{m6}_{3}|^2 +  |\chi^{m6}_{\frac{2}{5}}|^2 +  |\chi^{m6}_{\frac{7}{5}}|^2 
 \nonumber \\ 
Z^{\eGau_{S_3}}_{\onebb\text{-cnd};a_1} &=  \chi^{m6}_{0} \bar\chi^{m6}_{3} +  \chi^{m6}_{3} \bar\chi^{m6}_{0} +  \chi^{m6}_{\frac{2}{5}} \bar\chi^{m6}_{\frac{7}{5}} +  \chi^{m6}_{\frac{7}{5}} \bar\chi^{m6}_{\frac{2}{5}} 
 \nonumber \\ 
Z^{\eGau_{S_3}}_{\onebb\text{-cnd};a_2} &=  |\chi^{m6}_{\frac{2}{3}}|^2 +  |\chi^{m6}_{\frac{1}{15}}|^2 
 \nonumber \\ 
Z^{\eGau_{S_3}}_{\onebb\text{-cnd};b} &=  |\chi^{m6}_{\frac{2}{3}}|^2 +  |\chi^{m6}_{\frac{1}{15}}|^2 
\\ 
Z^{\eGau_{S_3}}_{\onebb\text{-cnd};b_1} &=  \chi^{m6}_{0} \bar\chi^{m6}_{\frac{2}{3}} +  \chi^{m6}_{3} \bar\chi^{m6}_{\frac{2}{3}} +  \chi^{m6}_{\frac{2}{5}} \bar\chi^{m6}_{\frac{1}{15}} +  \chi^{m6}_{\frac{7}{5}} \bar\chi^{m6}_{\frac{1}{15}} 
 \nonumber \\ 
Z^{\eGau_{S_3}}_{\onebb\text{-cnd};b_2} &=  \chi^{m6}_{\frac{2}{3}} \bar\chi^{m6}_{0} +  \chi^{m6}_{\frac{2}{3}} \bar\chi^{m6}_{3} +  \chi^{m6}_{\frac{1}{15}} \bar\chi^{m6}_{\frac{2}{5}} +  \chi^{m6}_{\frac{1}{15}} \bar\chi^{m6}_{\frac{7}{5}} 
 \nonumber \\ 
Z^{\eGau_{S_3}}_{\onebb\text{-cnd};c} &=  |\chi^{m6}_{\frac{1}{8}}|^2 +  |\chi^{m6}_{\frac{13}{8}}|^2 +  |\chi^{m6}_{\frac{1}{40}}|^2 +  |\chi^{m6}_{\frac{21}{40}}|^2 
 \nonumber \\ 
Z^{\eGau_{S_3}}_{\onebb\text{-cnd};c_1} &=  \chi^{m6}_{\frac{1}{8}} \bar\chi^{m6}_{\frac{13}{8}} +  \chi^{m6}_{\frac{13}{8}} \bar\chi^{m6}_{\frac{1}{8}} +  \chi^{m6}_{\frac{1}{40}} \bar\chi^{m6}_{\frac{21}{40}} +  \chi^{m6}_{\frac{21}{40}} \bar\chi^{m6}_{\frac{1}{40}} .
\nonumber 
\end{align}
\endgroup
where $ \chi^{m6}_{h}(\tau)$ is the conformal character of $(6,5)$ minimal
model, and $h$ is the scaling dimension of the corresponding primary field.
The $(6,5)$ minimal model has a central charge $c=\frac45$. We can see that 
the above
boundary is $\onebb$-condensed, because only the $\onebb$-component of the
partition function contains the conformal character $|\chi^{m6}_{0}|^2$ for the
identity primary field.\cite{CW220506244}  

The gapped boundary $\cVec_{S_3}$ in Fig.  \ref{CFTCFTG} is induced by
condensing the condensable algebra $\cA_c = \onebb \oplus a_1 \oplus 2 a_2$ 
formed
by $S_3$ charges.  It is described by the following multi component partition
function:
\begingroup
\allowdisplaybreaks
\begin{align}
\label{VecS3}
 Z^{\eGau_{S_3}}_{\cVec_{S_3};\onebb} &=  1,
 \nonumber \\ 
Z^{\eGau_{S_3}}_{\cVec_{S_3};a_1} &=  1,
 \nonumber \\ 
Z^{\eGau_{S_3}}_{\cVec_{S_3};a_2} &=  2,\text{ and}
 \\ 
Z^{\eGau_{S_3}}_{\cVec_{S_3};\al} &= 0, \text{ for }\al=b,b_1,b_2,c,c_1.\nonumber
\end{align}
\endgroup
The gapped boundary $\cRep_{S_3}$ in Fig.  \ref{CFTCFTtG} is induced by
condensing the condensable algebra $\cA_f = \onebb \oplus b \oplus c$ formed
by $S_3$ flux.  It is described by the following multi component partition
function:
\begingroup
\allowdisplaybreaks
\begin{align}
\label{RepS3}
 Z^{\eGau_{S_3}}_{\cRep_{S_3};\onebb} &=  1,
 \nonumber \\ 
Z^{\eGau_{S_3}}_{\cRep_{S_3};b} &= 1,
\\  
Z^{\eGau_{S_3}}_{\cRep_{S_3};c} &=  1, \text{ and}
 \nonumber \\ 
Z^{\eGau_{S_3}}_{\cRep_{S_3};\al} &=  0, \text{ for }\al=a_1,a_2,b_1,b_2,c_1.\nonumber
\end{align}
\endgroup

The critical point in the $G$-symmetric model $H_G$ is given by $CFT_{af}$ in
Fig. \ref{CFTCFTG} via the decomposition $CFT_{af} = CFT_{ano}
\boxtimes_{\eGau_{S_3}} \cVec_{S_3}$.  Thus the modular invariant partition
function for $CFT_{af}$ is given by
\begin{align}
& Z_{af} = \sum_\al
Z^{\eGau_{S_3}}_{\onebb\text{-cnd};\al}(\tau,\bar\tau)
\left (Z^{\eGau_{S_3}}_{\cVec_{S_3};\al}\right )^*  
\\
 &=  |\chi^{m6}_{0} +  \chi^{m6}_{3}|^2 +  |\chi^{m6}_{\frac{2}{5}} +  \chi^{m6}_{\frac{7}{5}}|^2 
+
2 |\chi^{m6}_{\frac{2}{3}}|^2 + 2 |\chi^{m6}_{\frac{1}{15}}|^2 
.
\nonumber
\end{align}

The critical point in the $\tl G$-symmetric model $H_{\tl G}$ 
is given by $CFT'_{af}$ in
Fig. \ref{CFTCFTtG} via the decomposition $CFT'_{af} = CFT_{ano}
\boxtimes_{\eGau_{S_3}} \cRep_{S_3}$.  Thus the modular invariant partition
function for $CFT'_{af}$ is given by
\begin{align}
& Z_{af}' = \sum_\al
Z^{\eGau_{S_3}}_{\onebb\text{-cnd};\al}(\tau,\bar\tau)
\left (Z^{\eGau_{S_3}}_{\cRep_{S_3};\al}  \right )^*
\\
 &=  |\chi^{m6}_{0}|^2 +  |\chi^{m6}_{3}|^2 
+  |\chi^{m6}_{\frac{2}{5}}|^2 +  |\chi^{m6}_{\frac{7}{5}}|^2 
+ |\chi^{m6}_{\frac{2}{3}}|^2 +  |\chi^{m6}_{\frac{1}{15}}|^2 
\nonumber\\
&\ \ \ \
+ |\chi^{m6}_{\frac{1}{8}}|^2 +  |\chi^{m6}_{\frac{13}{8}}|^2 
+ |\chi^{m6}_{\frac{1}{40}}|^2 +  |\chi^{m6}_{\frac{21}{40}}|^2 
.
\nonumber
\end{align}

Through the above examples, we see that the holographic picture of emergent
symmetry Fig.  \ref{CCmorph} can give rise to concrete partition functions for
the critical point in the models $H_G$ and $H_{\tl G}$.  The different
choices of the gapped boundary $\tl\cR$ give rise to lattice
models with different dual symmetries on the other boundary (cf. Fig. 
\ref{QFTRpc}).  
Although the partition functions for the model $H_G$ and model $H_{\tl
G}$ are different, the partition function for the $G$-symmetric sub-Hilbert
space of model $H_G$ and the partition function for the $\tl G$-symmetric
sub-Hilbert space of the model $H_{\tl G}$ are the same, and both are given by
the $\onebb$-component of the multi-component partition function 
\begin{align}
Z^{\eGau_{S_3}}_{\onebb\text{-cnd};\onebb} =  |\chi^{m6}_{0}|^2 + |\chi^{m6}_{3}|^2
+  |\chi^{m6}_{\frac{2}{5}}|^2 +  |\chi^{m6}_{\frac{7}{5}}|^2 .  
\end{align}
This implies
that the model $H_G$ and the model $H_{\tl G}$ are identical within the
respective symmetric sub-Hilbert space.  In other words, the model $H_G$ and
the model $H_{\tl G}$ are holo-equivalent (\ie  local low energy equivalent).

In addition to $\cA_c =\onebb\oplus a_1\oplus 2 a_2$, $\cA_f =\onebb\oplus b\oplus
c$, the $S_3$-gauge theory $\eGau_{S_3}$ also has two other Lagrangian
condensable algebras: $\tl\cA_c =\onebb\oplus a_1\oplus 2 b$, $\tl\cA_f =\onebb\oplus
a_2\oplus c$.  We note that the $S_3$-gauge theory $\eGau_{S_3}$ has an
automorphism that exchanges $a_2$ and $b$.  The condensable algebras $\tl\cA_c
$, $\tl\cA_f $ are generated from $\cA_c $, $\cA_f $ through the automorphism.
Thus, we denote the boundary induced by $\tl\cA_c $-condensation as $\tl\cVec_G$,
and the boundary induced by $\tl\cA_f $-condensation as $\tl\cRep_G$.  
Replacing
the gapped boundaries $\cVec_G$ and $\cRep_G$ in Fig. \ref{CFTCFTG} and
\ref{CFTCFTtG} by $\tl\cVec_G$ and $\tl\cRep_G$ will give us two other lattice
models, denoted as $\tl H_G$ and $\tl H_{\tl G}$.  All the four lattice models
$H_G$,  $H_{\tl G}$, $\tl H_G$, and $\tl H_{\tl G}$ are local low energy
equivalent.

Because the two boundaries
$\cVec_G$ and $\tl\cVec_G$ are related by an automorphism, we believe that 
we can
choose a proper lattice regularization such that $H_G$ and $\tl H_G$ have the 
same
form, \ie the lattice model is self-dual under the $a_2 \leftrightarrow b$
exchange.  Similarly, we believe that we can choose a proper lattice 
regularization
such that $H_{\tl G}$ and $\tl H_{\tl G}$ have the same form. This is analogous 
to the Kramers-Wannier self-duality of the Ising model.

\subsubsection{Maximal symTO}

The emergent symTO $\eGau_{S_3}$ for the critical points of the four models,
$H_G$,  $H_{\tl G}$, $\tl H_G$ and $\tl H_{\tl G}$, is not the maximal symTO.
From the expression of partition function for the critical point, we see that
the maximal symTO is given by the double $(6,5)$-minimal model: $\eM_{dm6}
=\eM_{m6} \boxtimes \bar \eM_{m6}$, where $\eM_{m6}$ the topological order of
single  $(6,5)$-minimal model with the following set of anyons:
\begin{widetext}
\begin{align}
\begin{matrix}
\text{anyons}\ (s,r): & (1,1) & (2,1) & (3,1) & (4,1) & (5,1) & (1,2) & (2,2) & (3,2) & (4,2) & (5,2) & \\
d_{(s,r)}: &  
1 & \sqrt{3} & 2 & \sqrt{3} & 1 & \frac{1+\sqrt{5}}{2} & 
\frac{\sqrt {15}+\sqrt{3}}{2} & 1+\sqrt{5} 
& \frac{\sqrt{15}+\sqrt{3}}{2} & \frac{1+\sqrt{5}}{2} 
\\
h_{(s,r)}: 
& 0  & \frac{ 1}{8 } & \frac{ 2}{3 } & \frac{ 13}{8 } &  3 & \frac{ 2}{5 } & \frac{ 1}{40 } & \frac{ 1}{15 } & \frac{ 21}{40 } & \frac{ 7}{5} \\
\end{matrix}
\end{align}
\end{widetext}
where we label anyons by
$(s,r)$, $s=1,2,3,4,5$ and $r=1,2$.
In general, for the $(p,q)$-minimal model, the prime fields are labeled by $(s,r)$
with $1\leq s \leq p-1$, $1\leq r \leq q-1$, and the identification $(s,r) =
(q-r,p-s)$.  The scaling dimensions of the corresponding primary fields are
given by
\begin{equation}
 h_{s,r} = \frac{(pr-qs)^2-(p-q)^2}{4pq} .
\end{equation}
The fusion rule is given by
\begin{align}
&\ \ \ \
 (s_1+1,r_1+1) \otimes
 (s_2+1,r_2+1) 
\\
&=
\bigoplus_{r_3 \stackrel{2}{=} |r_1-r_2|}^{\text{min}(r_1+r_2,2q-r_1-r_2-4)}
\
\bigoplus_{s_3 \stackrel{2}{=} |s_1-s_2|}^{\text{min}(s_1+s_2,2p-s_1-s_2-4)}
\hskip -10mm (s_3+1,r_3+1)
\nonumber 
\end{align}
We see that the fusion rule has a $\Z_2\times\Z_2$ grading if both $p$ and $q$
are even, and a $\Z_2$ grading if one of $p$ and $q$ is even.  In particular,
the unitary minimal models all have a $\Z_2$ grading 
(see Appendix \ref{grading}).

Using the conformal characters of the $(6,5)$-minimal model,
$\chi^{m6}_{s,r}(\tau) = \chi^{m6}_{h_{s,r}}(\tau)$, we can construct
two modular invariant partition functions
\begin{align}
 Z_{af}' &= \sum_{s,r} |\chi^{m6}_{s,r}(\tau)|^2
\\
 &=  |\chi^{m6}_{0}|^2 
+ |\chi^{m6}_{\frac{1}{8}}|^2 
+ |\chi^{m6}_{\frac{2}{3}}|^2 
+  |\chi^{m6}_{\frac{13}{8}}|^2 
+  |\chi^{m6}_{3}|^2 
\nonumber\\
&\ \ \ \
+  |\chi^{m6}_{\frac{2}{5}}|^2 
+ |\chi^{m6}_{\frac{1}{40}}|^2 
+  |\chi^{m6}_{\frac{1}{15}}|^2 
+  |\chi^{m6}_{\frac{21}{40}}|^2
+  |\chi^{m6}_{\frac{7}{5}}|^2 
,
\nonumber 
\end{align}
and
\begin{align}
& Z_{af} = 
\sum_{s=\text{odd},r} |\chi^{m6}_{s,r}(\tau)|^2
+\sum_{s=\text{odd},r} \chi^{m6}_{s,r}(\tau) \bar \chi^{m6}_{6-s,r}(\bar \tau)
\nonumber\\
& =
  |\chi^{m6}_{0}|^2 
+  |\chi^{m6}_{\frac{2}{3}}|^2 
+  |\chi^{m6}_{3}|^2 
+  |\chi^{m6}_{\frac{2}{5}}|^2 
+  |\chi^{m6}_{\frac{1}{15}}|^2 
+  |\chi^{m6}_{\frac{7}{5}}|^2 
\nonumber\\
&\ \ \ \
+  \chi^{m6}_{0} \bar \chi^{m6}_{3} 
+  |\chi^{m6}_{\frac{2}{3}} |^2
+  \chi^{m6}_{3} \bar \chi^{m6}_{0}
+  \chi^{m6}_{\frac{2}{5}} \bar \chi^{m6}_{\frac{7}{5}}
\nonumber\\
&\ \ \ \
+  |\chi^{m6}_{\frac{1}{15}}|^2
+  \chi^{m6}_{\frac{7}{5}} \bar \chi^{m6}_{\frac{2}{5}}
\\
 &=  |\chi^{m6}_{0} +  \chi^{m6}_{3}|^2 +  |\chi^{m6}_{\frac{2}{5}} 
+  \chi^{m6}_{\frac{7}{5}}|^2 
+ 2 |\chi^{m6}_{\frac{2}{3}}|^2 + 2 |\chi^{m6}_{\frac{1}{15}}|^2
.
\nonumber 
\end{align}
These modular invariant partition functions happen to describe the
critical points of the model $H_{\tl G}$ and the model $H_G$, respectively.  

The partition function $Z_{af}'$ is obtained by choosing the gapped boundary
$\tl\cR$ in Fig.  \ref{CCmorph} to be described by the following constant
multi-component partition function
\begin{align}
 Z^{\eM_{dm6}}_{h,h'} &= \del_{h,h'}, \ \
h,h' \in \left \{  0  , \frac{ 1}{8 } , \frac{ 2}{3 } , \frac{ 13}{8 } ,  3 , \frac{ 2}{5 } , \frac{ 1}{40 } , \frac{ 1}{15 } , \frac{ 21}{40 } , \frac{ 7}{5} \right \}
\end{align}
The partition function $Z_{af}$ is obtained by choosing the gapped boundary
$\tl\cR$ to be described by
\begin{align}
 Z^{\eM_{dm6}}_{0,0} 
&=Z^{\eM_{dm6}}_{3,3} 
=Z^{\eM_{dm6}}_{0,3} 
=Z^{\eM_{dm6}}_{3,0} 
=Z^{\eM_{dm6}}_{\frac25,\frac25} 
\nonumber\\
&=Z^{\eM_{dm6}}_{\frac75,\frac75} 
=Z^{\eM_{dm6}}_{\frac25,\frac75} 
=Z^{\eM_{dm6}}_{\frac75,\frac25} =1,
\nonumber\\
Z^{\eM_{dm6}}_{\frac23,\frac23} &=
Z^{\eM_{dm6}}_{\frac1{15},\frac1{15}} = 2, \ \ 
\text{and other } Z^{\eM_{dm6}}_{h,h'} = 0.
\end{align}
The above two gapped boundaries are not described by local fusion 1-category.
Thus, the critical points in the four models,  $H_G$,  $H_{\tl G}$, $\tl H_G$ and
$\tl H_{\tl G}$, have the same emergent maximal symTO $\eM_{dm6} =\eM_{m6}
\boxtimes \bar \eM_{m6}$, without the associated emergent anomaly-free
symmetry.

\subsection{Gapless states with anomalous $S_3$ symmetry}

In this subsection, we consider 1+1D gapless states with anomalous $S_3$
symmetry.  The 1+1D anomalous $S_3$ symmetries are classified by
$H^3(S_3;\RZ)=\Z_3\times\Z_2 \cong \Z_6$.\cite{CGL1314}  We label these
anomalies by $S_3^{(m)}$, $m\in \{0,1,2,3,4,5\}$.  The  symTO for an
anomalous $S_3^{(m)}$ symmetry is given by a topological order
$\eGau_{S_3}^{(m)}$ that is described in the IR limit by the 2+1D
Dijkgraaf-Witten gauge theory\cite{DW9093} coupled to gauge charges. Note that the
time reversal conjugate of an anomalous symmetry $S_3^{(m)}$ is
another anomalous symmetry $S_3^{(-m \text{ mod } 6)}$, so we only need to focus on half of these six possible anomalies. 

In the following, we study the
modular invariant partition function for the gapless states that have the above symmetries, along with the corresponding emergent maximal symTOs.

\subsubsection{Anomalous $S_3^{(1)}$ symmetry}

A gapless state for a lattice system with anomalous $S_3^{(1)}$ symmetry
has the following decomposition
\begin{align}\label{S31_decomp}
 CFT_{af} = CFT_{ano}\boxtimes_{\eGau_{S_3}^{(1)}} \tl\cR ,
\end{align}
where the symTO $\eGau_{S_3}^{(1)}$ (\ie 
the 2+1D $\eGau_{S_3}^{(1)}$ topological order) 
has anyons given by
\begin{align}
\begin{matrix}
\text{anyons}: & \onebb &  a_1  & a_2 &  b  &  b_1  &  b_2  & c  & c_1    \\
d_a: &  1 & 1 & 2 & 2 & 2 & 2 & 3 & 3 \\
s_a: & 0 & 0 & 0 & \frac19 & \frac49 & \frac79 & \frac14 & \frac34 \\
\end{matrix}
\end{align}
where anyon $a_1,a_2$ carry the $S_3$-charges.  In fact $a_1$ carries the
non-trivial 1-dimensional irrep of $S_3$, and $a_2$ carries the
2-dimensional irrep of $S_3$.

If the gapless state does not break the symTO $\eGau_{S_3}^{(1)}$, then
$CFT_{ano}$ in the decomposition is given by a $\onebb$-condensed boundary of
$\eGau_{S_3}^{(1)}$, 
as discussed in section VI.A of \Rf{CW220506244}.
This gapless state  is described by a $so(9)_2 \times
u(1)_2 \times \overline{u(1)_2}\times\overline{E(8)_1}$ chiral CFT with central charge $(c, \bar
c)=(9,9)$.  
To describe the anomalous symmetry $S_3^{(1)}$, we need to choose $\tl\cR$ in
the decomposition \eqref{S31_decomp} as the gapped boundary of $\eGau_{S_3}^{(1)}$ obtained from the
condensation of all the $S_3$-charges. In other words, $\tl\cR$ is a $\onebb\oplus
a_1\oplus 2 a_2$-condensed boundary of $\eGau_{S_3}^{(1)}$, described by the
following multi-component partition function:\cite{CW220506244}
\begingroup
\allowdisplaybreaks
\begin{align}\label{Rtbdy}
 Z_{\onebb\oplus a_1\oplus 2a_2\text{-cnd};\onebb}^{\eGau_{S_3}^{(1)}} &= 1,
 \nonumber \\ 
Z_{\onebb\oplus a_1\oplus 2a_2\text{-cnd};a_1}^{\eGau_{S_3}^{(1)}} &= 1,
 \nonumber \\ 
Z_{\onebb\oplus a_1\oplus 2a_2\text{-cnd};a_2}^{\eGau_{S_3}^{(1)}} &= 2,
 \  \text{and}\nonumber \\ 
Z_{\onebb\oplus a_1\oplus 2a_2\text{-cnd};\al}^{\eGau_{S_3}^{(1)}} &= 0,
 \  \text{for $\al=b,b_1,b_2,c,c_1$}.
 \end{align}
\endgroup
From the decomposition \eqref{S31_decomp}, we find the modular invariant partition function of the gapless
state that does not break the symTO $\eGau_{S_3}^{(1)}$:
\footnote{In the above, we have used an abbreviated notation where
	$\chi^{CFT_1\times CFT_2 \times \cdots}_{a_1,h_1;a_2,h_2;\cdots}$ is the product of the conformal characters of $CFT_i$ associated with the primary fields labeled by $a_i$ whose scaling dimensions are $h_i$}
\begin{align}
\label{ZafS31}
& Z_{af} = \sum_\al
Z^{\eGau_{S_3}^{(1)}}_{\onebb\text{-cnd};\al}(\tau,\bar\tau)
\left (Z^{\eGau_{S_3}^{(1)}}_{\onebb\oplus a_1\oplus 2a_2\text{-cnd};\al}\right 
)^*  
\\
 &=  
 \chi^{so(9)_2 \times u(1)_2 \times \overline{u(1)_2}\times\overline{E(8)_1}}_{1,0; 1,0; 1,0} +  \chi^{so(9)_2 \times u(1)_2 \times \overline{u(1)_2}\times\overline{E(8)_1}}_{2,1; 2,\frac{1}{4}; 2,-\frac{1}{4}} 
\nonumber\\
&
+ \chi^{so(9)_2 \times u(1)_2 \times \overline{u(1)_2}\times\overline{E(8)_1}}_{1,0; 2,\frac{1}{4}; 2,-\frac{1}{4}} +  \chi^{so(9)_2 \times u(1)_2 \times \overline{u(1)_2}\times\overline{E(8)_1}}_{2,1; 1,0; 1,0} 
\nonumber\\
&
+ 2\chi^{so(9)_2 \times u(1)_2 \times \overline{u(1)_2}\times\overline{E(8)_1}}_{6,1; 1,0; 1,0} +  2\chi^{so(9)_2 \times u(1)_2 \times \overline{u(1)_2}\times\overline{E(8)_1}}_{6,1; 2,\frac{1}{4}; 2,-\frac{1}{4}} 
,
\nonumber 
\end{align}
where 
$Z^{\eGau_{S_3}^{(1)}}_{\onebb\text{-cnd};\al}(\tau,\bar\tau)$
is given in section VI.A of \Rf{CW220506244}.
In the above, we have used an abbreviated notation, where
$\chi^{CFT_1\times CFT_2 \times \cdots}_{a_1,h_1;a_2,h_2;\cdots}$ is product of
conformal characters of $CFT_i$ for the primary fields labeled by $a_i$ with
scaling dimension $h_i$.  For example,
\begin{align}
\label{cterm}
&\ \ \ \
  \chi^{so(9)_2 \times u(1)_2 \times \overline{u(1)_2}\times\overline{E(8)_1}}_{2,1; 2,\frac{1}{4}; 2,-\frac{1}{4}} 
\nonumber\\
&= 
  \chi^{so(9)_2 }_{2,1}(\tau) 
  \chi^{u(1)_2 }_{ 2,\frac{1}{4} }(\tau) 
  \chi^{\overline{u(1)_2}}_{ 2,-\frac{1}{4}} (\bar \tau)
  \chi^{\overline{E(8)_1}}(\bar \tau),
\end{align}
where $\chi^{so(9)_2 }_{2,1}(\tau)$ is the conformal character of $so(9)_2$
CFT, for the second primary field with scaling dimension $h=1$; $\chi^{u(1)_2
}_{2,\frac{1}{4}}(\tau)$ is the conformal character of $u(1)_2$ CFT, for the
second primary field with scaling dimension $h=\frac14$;
$\chi^{\overline{u(1)_2} }_{2,-\frac{1}{4}}(\bar\tau)$ is the conformal
character of $\overline{u(1)_2}$ CFT, for the second primary field with scaling
dimension $h=\frac14$; $\chi^{\bar E(8)_1 }$ is the conformal character of
$\overline{E(8)_1}$ CFT (the complex conjugate of $E(8)_1$ Kac-Moody algebra).
The $\overline{E(8)_1}$ CFT has only one primary field (the identity), whose
index is suppressed.

From the above result, we see that such a gapless state has a symTO\
larger than $\eGau_{S_3}^{(1)}$
\begin{align}
\eM_\text{larger}=\eM_{so(9)_2} \times \eM_{(2,-2,0)}\times \eM_{\overline{E(8)_1}},
\end{align}
where $ \eM_{so(9)_2}$ is the 2+1D topological order described by $so(9)_2$ Chern-Simons theory, $ \eM_{\overline{E(8)_1}}$ is the 2+1D topological
order described by the time-reversal conjugate of $E(8)_1$ Chern-Simons
theory, and $\eM_{(2,-2,0)}$ is the 2+1D Abelian topological order described by the
$K$-matrix $\begin{pmatrix} 2 &0 \\ 0 &-2\\ \end{pmatrix}$.

We notice that the conformal character $\chi^{u(1)_2}_{2;\frac14}$ is also
contained in the $u(1)_{2n^2}$ CFT, $n\in \Z$.  Therefore,
the gapless state  \eq{ZafS31} has an even larger symTO\
\begin{align}
\eM=\eM_{so(9)_2}\times\eM_{(2n^2,-2n^2,0)}\times\eM_{\overline{E(8)_1}},
\end{align}
where $\eM_{(2n^2,-2n^2,0)}$, $ n\in \Z$, is the 2+1D Abelian topological order
described by the $K$-matrix $\begin{pmatrix} 2n^2 &0 \\ 0 &-2n^2\\
\end{pmatrix}$.  
When $n\to \infty$, the total quantum dimension of the symTO also
approaches $\infty$.  Thus the maximal symTO for the gapless state
\eq{ZafS31} contains, at least, the symTO for a
continuous $U(1)$ symmetry, denoted as $\eGau_{U(1)}$. Here, $\eGau_{U(1)}$ is an appropriately generalized braided fusion category with infinite
objects. In other words, the gapless state \eq{ZafS31} has an emergent $U(1)$
symmetry.

\subsubsection{Anomalous $S_3^{(2)}$ symmetry}

Similarly, a gapless state for a lattice system with anomalous $S_3^{(2)}$
symmetry has the decomposition
\begin{align}\label{S32_decomp}
 CFT_{af} = CFT_{ano}\boxtimes_{\eGau_{S_3}^{(2)}} \tl\cR ,
\end{align}
where the symTO is the 2+1D topological order
$\eGau_{S_3}^{(2)}$, which has anyons given by
\begin{align}
\begin{matrix}
\text{anyons}: & \onebb &  a_1  & a_2 &  b  &  b_1  &  b_2  & c  & c_1    \\
d_a: &  1 & 1 & 2 & 2 & 2 & 2 & 3 & 3 \\
s_a: & 0 & 0 & 0 & \frac29 & \frac59 & \frac89 & 0 & \frac12 \\
\end{matrix}.
\end{align}

If the gapless state does not break the symTO $\eGau_{S_3}^{(2)}$, then
$CFT_{ano}$ in the decomposition is given by a $\onebb$-condensed boundary of
$\eGau_{S_3}^{(2)}$, as discussed in section VI.B of \Rf{CW220506244}.
This gapless state is described by a $E(8)_1 \times \overline{so(9)_2} $
chiral CFT with central charge $(c, \bar c)=(8,8)$.  
As in the previous example, to describe the anomalous symmetry $S_3^{(2)}$, we need to choose $\tl\cR$ in
the decomposition \eqref{S32_decomp} as the gapped boundary of $\eGau_{S_3}^{(2)}$ obtained from the
condensation of all the $S_3$-charges. This is given by the partition function in \eqn{Rtbdy}.
From the decomposition $CFT_{af} = CFT_{ano}\boxtimes_{\eGau_{S_3}^{(2)}}
\tl\cR$, we find the modular invariant partition function of the gapless
state that leaves the symTO $\eGau_{S_3}^{(2)}$ unbroken:
\begin{align}
\label{ZafS32}
Z_{af} &= \sum_\al
Z^{\eGau_{S_3}^{(2)}}_{\onebb\text{-cnd};\al}(\tau,\bar\tau)
\left (Z^{\eGau_{S_3}^{(2)}}_{\onebb\oplus a_1\oplus 2a_2\text{-cnd};\al}\right 
)^*  
\\
 &=  
\chi^{{E(8)_1\times \overline{so(9)_2}}}_{1,0} 
+ \chi^{{E(8)_1\times \overline{so(9)_2}}}_{2,-1} 
+ 2\chi^{{E(8)_1\times \overline{so(9)_2}}}_{6,-1} 
.
\nonumber 
\end{align}
From this result, we see that such a gapless state has a symTO\
larger than $\eGau_{S_3}^{(2)}$
\begin{align}
\eM_\text{larger}=\eM_{E(8)_1} \times \eM_{\overline{so(9)_2}} .
\end{align}
This symTO is still not the maximal symTO since this gapless state
contains many emergent $U(1)$ symmetries.

\subsubsection{Anomalous $S_3^{(3)}$ symmetry}

Last, we consider a gapless state for a lattice system with anomalous
$S_3^{(3)}$ symmetry, which has the decomposition
\begin{align}\label{S33_decomp}
 CFT_{af} = CFT_{ano}\boxtimes_{\eGau_{S_3}^{(3)}} \tl\cR ,
\end{align}
where the symTO is the 2+1D topological order $\eGau_{S_3}^{(3)}$, which
has anyons given by
\begin{align}
\begin{matrix}
\text{anyons}: & \onebb &  a_1  & a_2 &  b  &  b_1  &  b_2  & c  & c_1    \\
d_a: &  1 & 1 & 2 & 2 & 2 & 2 & 3 & 3 \\
s_a: & 0 & 0 & 0 & 0 & \frac13 & \frac23 & \frac14 & \frac34 \\
\end{matrix}.
\end{align}

If the gapless state does not break the symTO $\eGau_{S_3}^{(3)}$, then
$CFT_{ano}$ in the decomposition is given by a $\onebb$-condensed boundary of
$\eGau_{S_3}^{(3)}$, as discussed in section VI.C of \Rf{CW220506244}. 
This gapless state  is described by a $m6 \times u(1)_2 \times \overline{m6}
\times \overline{u(1)_2} $ CFT with central charge $(c, \bar c)=(\frac95,\frac95)$. 
To describe the anomalous symmetry $S_3^{(3)}$, we choose $\tl\cR$ in
the decomposition \eqref{S33_decomp} as the gapped boundary of $\eGau_{S_3}^{(3)}$ obtained from the
condensation of all the $S_3$-charges, described by \eqn{Rtbdy}.
From the decomposition \eqref{Rtbdy}, we find the modular invariant partition function of the gapless
state that does not break the symTO $\eGau_{S_3}^{(3)}$:
\begingroup
\allowdisplaybreaks
\begin{align}
\label{ZafS33}
 Z_{af} &= \sum_\al
Z^{\eGau_{S_3}^{(3)}}_{\onebb\text{-cnd};\al}(\tau,\bar\tau)
\left (Z^{\eGau_{S_3}^{(3)}}_{\onebb\oplus a_1\oplus 2a_2\text{-cnd};\al}\right 
)^*  
\\
 &=  
\chi^{m6 \times u(1)_2 \times \overline{m6} \times \overline{u(1)_2}}_{1,0; 1,0; 1,0; 1,0} +  \chi^{m6 \times u(1)_2 \times \overline{m6} \times \overline{u(1)_2}}_{1,0; 2,\frac{1}{4}; 5,-3; 2,-\frac{1}{4}} 
\nonumber\\&\ \ \ \
+  \chi^{m6 \times u(1)_2 \times \overline{m6} \times \overline{u(1)_2}}_{5,3; 1,0; 5,-3; 1,0} +  \chi^{m6 \times u(1)_2 \times \overline{m6} \times \overline{u(1)_2}}_{5,3; 2,\frac{1}{4}; 1,0; 2,-\frac{1}{4}} 
\nonumber\\ &\ \ \ \ 
+  \chi^{m6 \times u(1)_2 \times \overline{m6} \times \overline{u(1)_2}}_{6,\frac{2}{5}; 1,0; 6,-\frac{2}{5}; 1,0} +  \chi^{m6 \times u(1)_2 \times \overline{m6} \times \overline{u(1)_2}}_{6,\frac{2}{5}; 2,\frac{1}{4}; 10,-\frac{7}{5}; 2,-\frac{1}{4}} 
\nonumber\\ &\ \ \ \ 
+  \chi^{m6 \times u(1)_2 \times \overline{m6} \times \overline{u(1)_2}}_{10,\frac{7}{5}; 1,0; 10,-\frac{7}{5}; 1,0} +  \chi^{m6 \times u(1)_2 \times \overline{m6} \times \overline{u(1)_2}}_{10,\frac{7}{5}; 2,\frac{1}{4}; 6,-\frac{2}{5}; 2,-\frac{1}{4}} 
 \nonumber \\ &\ \ \ \
+ \chi^{m6 \times u(1)_2 \times \overline{m6} \times \overline{u(1)_2}}_{1,0; 1,0; 5,-3; 1,0} +  \chi^{m6 \times u(1)_2 \times \overline{m6} \times \overline{u(1)_2}}_{1,0; 2,\frac{1}{4}; 1,0; 2,-\frac{1}{4}} 
\nonumber\\ &\ \ \ \
+  \chi^{m6 \times u(1)_2 \times \overline{m6} \times \overline{u(1)_2}}_{5,3; 1,0; 1,0; 1,0} +  \chi^{m6 \times u(1)_2 \times \overline{m6} \times \overline{u(1)_2}}_{5,3; 2,\frac{1}{4}; 5,-3; 2,-\frac{1}{4}}
\nonumber\\ & \ \ \ \
 +  \chi^{m6 \times u(1)_2 \times \overline{m6} \times \overline{u(1)_2}}_{6,\frac{2}{5}; 1,0; 10,-\frac{7}{5}; 1,0} +  \chi^{m6 \times u(1)_2 \times \overline{m6} \times \overline{u(1)_2}}_{6,\frac{2}{5}; 2,\frac{1}{4}; 6,-\frac{2}{5}; 2,-\frac{1}{4}} 
\nonumber\\ & \ \ \ \
+  \chi^{m6 \times u(1)_2 \times \overline{m6} \times \overline{u(1)_2}}_{10,\frac{7}{5}; 1,0; 6,-\frac{2}{5}; 1,0} +  \chi^{m6 \times u(1)_2 \times \overline{m6} \times \overline{u(1)_2}}_{10,\frac{7}{5}; 2,\frac{1}{4}; 10,-\frac{7}{5}; 2,-\frac{1}{4}} 
 \nonumber \\ & \ \ \ \
+2  \chi^{m6 \times u(1)_2 \times \overline{m6} \times \overline{u(1)_2}}_{3,\frac{2}{3}; 1,0; 3,-\frac{2}{3}; 1,0} + 2 \chi^{m6 \times u(1)_2 \times \overline{m6} \times \overline{u(1)_2}}_{3,\frac{2}{3}; 2,\frac{1}{4}; 3,-\frac{2}{3}; 2,-\frac{1}{4}} 
\nonumber\\ &\ \ \ \
+ 2 \chi^{m6 \times u(1)_2 \times \overline{m6} \times \overline{u(1)_2}}_{8,\frac{1}{15}; 1,0; 8,-\frac{1}{15}; 1,0} 
+ 2 \chi^{m6 \times u(1)_2 \times \overline{m6} \times \overline{u(1)_2}}_{8,\frac{1}{15}; 2,\frac{1}{4}; 8,-\frac{1}{15}; 2,-\frac{1}{4}} 
.
\nonumber 
\end{align}
\endgroup
From this result, we see that this gapless state has a symTO\
larger than $\eGau_{S_3}^{(3)}$
\begin{align}
\eM=\eM_{m6} \times \eM_{(2n^2,-2n^2,0)} \times \eM_{\overline{m6}} ,
\ \
n\in \Z.
\end{align}
where $\eM_{m6}$ is the 2+1D topological order that has a boundary given by 
the
$(6,5)$ minimal model. Again we see that the maximal symTO for the gapless
state \eq{ZafS33} contains, at least, the symTO $\eGau_{U(1)}$ for a
continuous $U(1)$ symmetry.  In other words, the gapless state \eq{ZafS33} has
an emergent $U(1)$ symmetry.

\subsection{Gapless states with noninvertible $\tl\cR_\text{Fib}$-symmetry}
\label{FibSymm}

\begin{figure}[t]
\begin{center}
\includegraphics[width=0.7\linewidth]{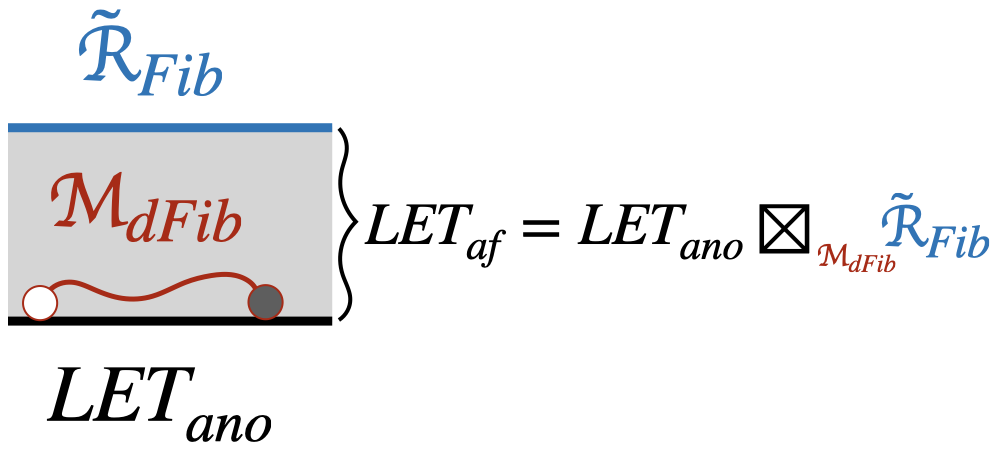}
\end{center}
\caption{ A 1+1D lattice model with emergent $\tl\cR_\text{Fib}$-symmetry at low energies.  The 1+1D lattice model is constructed from a slab of
2+1D lattice.  In the bulk, we have a lattice Hamiltonian that
realizes a double-Fibonacci topological order $\eM_\text{dFib}$ \cite{LW0510}
with a large energy gap.  The top boundary $\tl\cR = \tl\cR_\text{Fib}$ is a 
gapped
boundary of $\eM_\text{dFib}$ with a large energy gap.  The lower boundary is
described by an anomalous low energy theory LET$_{ano}$.  The low energy theory
LET$_{af}$ of the slab has an emergent $\tl\cR_\text{Fib}$-symmetry
below the energy gaps of the bulk and the top boundary.  } \label{emFib} 
\end{figure}

Table \ref{toptable} reveals that the simplest 1+1D noninvertible symmetry is
the $\tl\cR_\text{Fib}$-symmetry which has 4 types of symmetry
charges/defects (see \eqn{Rfib}).  The symTO of $\tl\cR_\text{Fib}$-symmetry is the 2+1D double-Fibonacci topological order $\eM_\text{dFib}$ given
by
\begin{align}
 \eM_\text{dFib} = \eM_\text{Fib} \boxtimes \overline{\eM}_\text{Fib},
\end{align}
where $\eM_\text{Fib}$ is the 2+1D Fibonacci topological order.
$\eM_\text{Fib}$ has 2 types of anyons $\onebb,\phi$ and 
$\eM_\text{dFib}$ has 4 types of anyons: 
\begin{align}
\begin{matrix}
\text{anyons}: & \onebb &  \phi  & \bar \phi &   \phi \bar \phi   \\
d_a: & 1 & \frac{1+\sqrt 5}{2} & \frac{1+\sqrt 5}{2} & \frac{3+\sqrt 5}{2}  \\
s_a: &  0 & \frac25 & \frac35 & 0 \\
\end{matrix}.
\end{align}

The $\tl\cR_\text{Fib}$-symmetry can be a low energy emergent symmetry in a 1+1D
lattice model.  Such a 1+1D lattice model can be constructed from a slab of 2+1D
lattice (see Fig. \ref{emFib}), where in the bulk, we have a commuting-projector
Hamiltonian that realizes a double-Fibonacci topological order
$\eM_\text{dFib}$ \cite{LW0510} with a large energy gap.  The double-Fibonacci
topological order has only one type of gapped boundary obtained by condensing
$\onebb\oplus \phi\bar\phi$.  The top boundary of the slab in Fig. \ref{emFib}
is this gapped boundary, again with a large energy gap.  This gapped boundary can be
described by the vector-valued partition function
$Z_{\tl\cR_\text{Fib}}^{\eM_\text{dFib}}=(1,0,0,1)$ where the components are indexed by the set of anyons, $\{\onebb,   \phi ,  \bar \phi ,   \phi \bar \phi \}$.
The lower boundary is described by an
anomalous low energy theory LET$_{ano}$.  The low energy theory LET$_{af}$ of
the full slab has an emergent $\tl\cR_\text{Fib}$-symmetry below the energy
gaps of the bulk and the gapped top boundary.  

The symmetry transformation of the $\tl\cR_\text{Fib}$-symmetry in the slab
model Fig. \ref{emFib} is given by the patch symmetry operator
$O_\text{str}(\phi,\phi)$ on local patches. This is the string operator that 
creates a pair of $\phi$ anyons in the bulk.\footnote{Note that
the string operator $O_\text{str}(\bar\phi,\bar\phi)$ generates the same
symmetry since $\phi\bar\phi$ condenses on the upper boundary
$\tl\cR_\text{Fib}$.}  
This symmetry is noninvertible since the $\phi$
anyon is non-Abelian.  The string operator
$O_\text{str}(\phi\bar\phi,\phi\bar\phi)$ is a patch charge operator, which
creates a pair of charges $\phi\bar\phi$ of the  $\tl\cR_\text{Fib}$-symmetry
and does not generate any symmetry transformation.

The emergent $\tl\cR_\text{Fib}$-symmetry can also be realized by a
Fibonacci-anyon chain in the 2+1D Fibonacci topological order $\eM_\text{Fib}$; this goes by the name of ``golden chain" \cite{FFc0612341}.  The
topological symmetry of the Fibonacci-anyon chain discussed in \Rf{FFc0612341}
is nothing but the symmetry generated by $O_\text{str}(\phi,\phi)$, \ie the
topological symmetry is the $\tl\cR_\text{Fib}$-symmetry.

\begin{figure}[t]
\begin{center}
\includegraphics[height=2.3in]{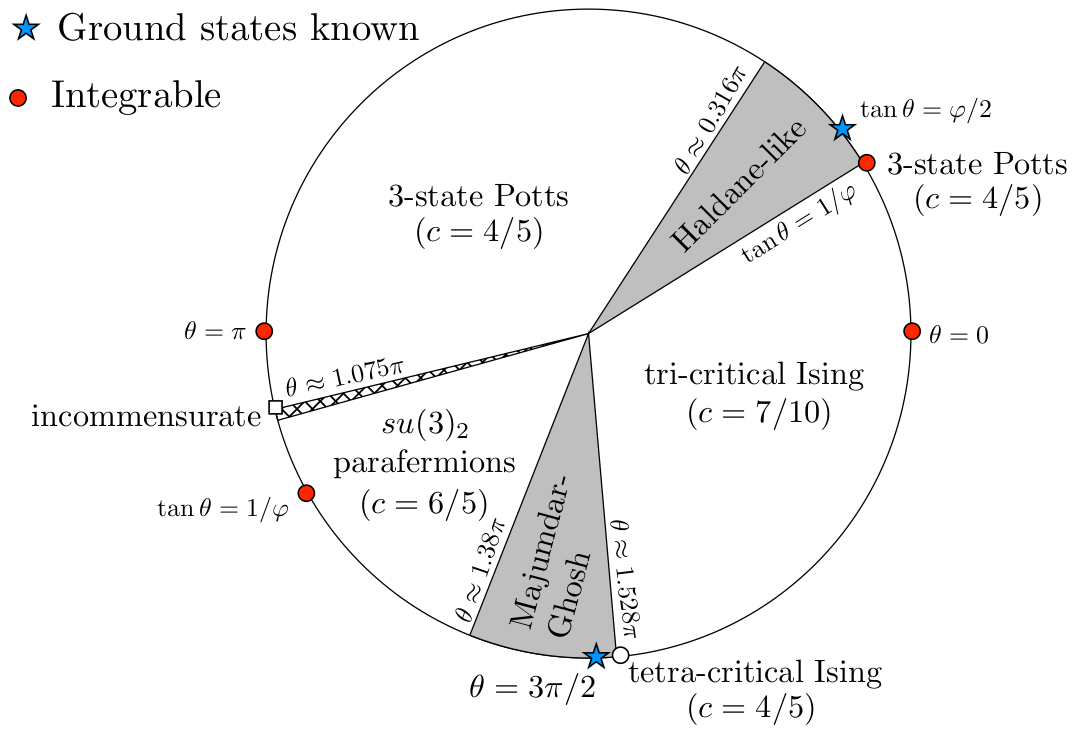} 
\end{center}

\caption{The phase diagram of Fibonacci-anyon chain adapted from
\Rf{TT08014602,KA11100719}, with pairwise fusion term $J_2 = \cos\te$ and
three-particle fusion term $J_3 = \sin \te$.  Here $\vphi=\frac{1+\sqrt 5}{2}$.
The shaded parts are gapped phases.} \label{GoldenChainPhases}
\end{figure}

Let us study the gapless states that have the unbroken double-Fibonacci symTO
$\eM_\text{dFib}$.  Such gapless states correspond to $\onebb$-condensed
boundaries of symTO $\eM_\text{dFib}$. One such boundary
of $\eM_\text{dFib}$ is described by the following vector-valued partition
function
\begingroup
\allowdisplaybreaks
\begin{align}
\label{Zm5m5}
 Z_{\onebb\text{-cnd};\onebb}^{\eM_\text{dFib}} &= \chi^{m5 \times \overline{m5}}_{1,0; 1,0} +  \chi^{m5 \times \overline{m5}}_{4,\frac{3}{2}; 4,-\frac{3}{2}} +  \chi^{m5 \times \overline{m5}}_{5,\frac{7}{16}; 5,-\frac{7}{16}} 
 \nonumber \\ 
Z_{\onebb\text{-cnd};\phi}^{\eM_\text{dFib}} &= \chi^{m5 \times \overline{m5}}_{1,0; 3,-\frac{3}{5}} +  \chi^{m5 \times \overline{m5}}_{4,\frac{3}{2}; 2,-\frac{1}{10}} +  \chi^{m5 \times \overline{m5}}_{5,\frac{7}{16}; 6,-\frac{3}{80}} 
 \nonumber \\ 
Z_{\onebb\text{-cnd};\bar \phi}^{\eM_\text{dFib}} &= \chi^{m5 \times \overline{m5}}_{2,\frac{1}{10}; 4,-\frac{3}{2}} +  \chi^{m5 \times \overline{m5}}_{3,\frac{3}{5}; 1,0} +  \chi^{m5 \times \overline{m5}}_{6,\frac{3}{80}; 5,-\frac{7}{16}} 
 \nonumber \\ 
Z_{\onebb\text{-cnd};\phi \bar\phi}^{\eM_\text{dFib}} &= \chi^{m5 \times \overline{m5}}_{2,\frac{1}{10}; 2,-\frac{1}{10}} +  \chi^{m5 \times \overline{m5}}_{3,\frac{3}{5}; 3,-\frac{3}{5}} +  \chi^{m5 \times \overline{m5}}_{6,\frac{3}{80}; 6,-\frac{3}{80}} 
 \end{align}
\endgroup
This gapless state is described by $(5,4)$-minimal model $m5\times \overline{m5}$
for the right- and left-movers.  Its modular invariant partition function is
\begin{align}
\label{ZafFibMM5}
 Z_{af} &= 
\sum_\al
Z^{\eM_\text{dFib}}_{\onebb\text{-cnd};\al}(\tau,\bar\tau)
\left (Z^{\eM_\text{dFib}}_{\tl\cR_\text{Fib};\al}\right )^*  
\nonumber\\
&=
\chi^{m5 \times \overline{m5}}_{1,0; 1,0} +  \chi^{m5 \times \overline{m5}}_{4,\frac{3}{2}; 4,-\frac{3}{2}} +  \chi^{m5 \times \overline{m5}}_{5,\frac{7}{16}; 5,-\frac{7}{16}} 
\nonumber\\
& \ \ \ \  
+
\chi^{m5 \times \overline{m5}}_{2,\frac{1}{10}; 2,-\frac{1}{10}} +  \chi^{m5 \times \overline{m5}}_{3,\frac{3}{5}; 3,-\frac{3}{5}} +  \chi^{m5 \times \overline{m5}}_{6,\frac{3}{80}; 6,-\frac{3}{80}} , 
\end{align}
which is the partition function of the Ising tricritical point.  So we will refer to this gapless state
as the tricritical Ising CFT. It has a central charge $(c,\bar
c)=(\frac7{10},\frac7{10})$ and one symmetric relevant operator of dimension
$(h,\bar h)= (\frac7{16},\frac7{16})$, as one can see from
$Z_{\onebb\text{-cnd};\onebb}^{\eM_\text{dFib}} $ in \eqn{ZafFibMM5}.

From the form of the partition function \eqn{Zm5m5}, we see that the
tricritical Ising gapless state has double-Fibonacci topological
order $\eM_\text{dFib}$ as its symTO. This gapless state also has an emergent 
maximal symTO of double-(5,4)-minimal-model.

\begin{figure}[t]
\begin{center}
\includegraphics[height=2.3in]{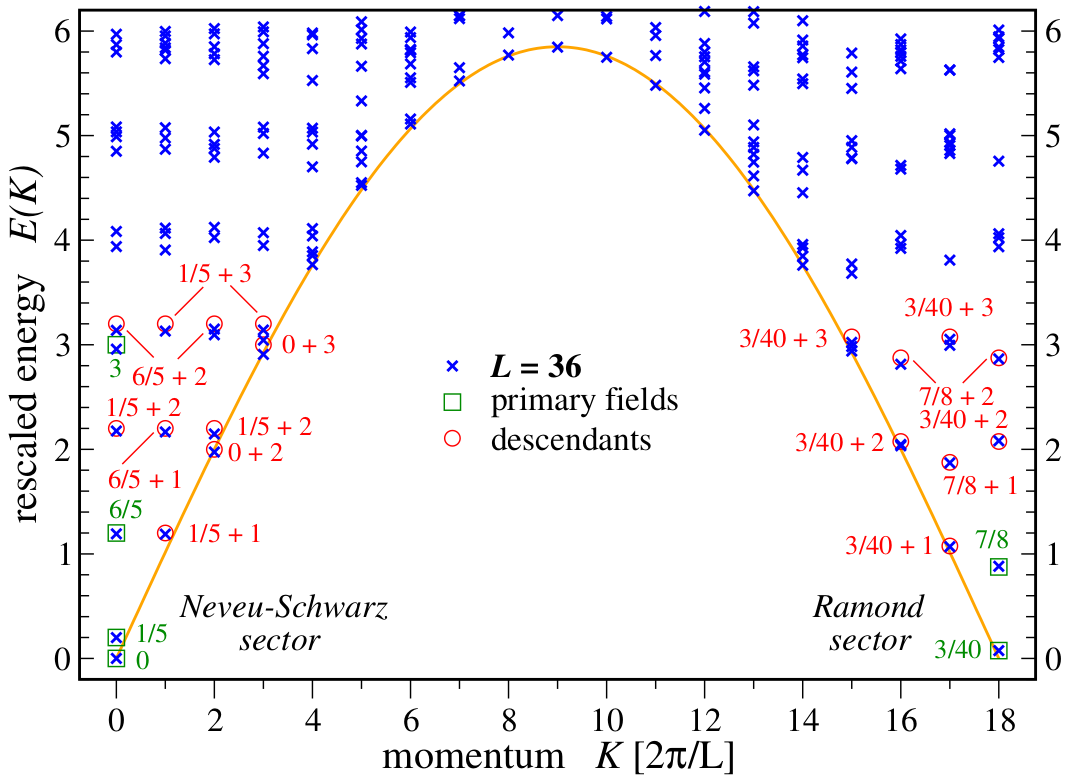} 
\end{center}

\caption{The energy-momentum spectrum  of the Fibonacci-anyon chain with $L$
anyons at $\te=0$, from \Rf{FFc0612341}.  } \label{GoldenChainMM5}
\end{figure}

The gapless states in Fibonacci-anyon chain were studied numerically in
\Rf{FFc0612341,TT08014602} (see Fig. \ref{GoldenChainPhases}). Indeed, a
$(c,\bar c)=(\frac7{10},\frac7{10})$ state was found, with the energy-momentum
spectrum given in Fig.  \ref{GoldenChainMM5}. We see that the states in the 
modular invariant partition function of tricritical Ising CFT \eqref{ZafFibMM5}
matches those read off from the numerically computed energy-momentum 
spectrum.  

The low energy spectrum contains two sectors $\onebb$ and $\phi\bar\phi$,
described by $Z_{\onebb\text{-cnd};\onebb}^{\eM_\text{dFib}}$ and $
Z_{\onebb\text{-cnd};\phi \bar\phi}^{\eM_\text{dFib}} $.  We see that the low
energy states, $\frac15=\frac{1}{10}+\frac{1}{10}$ and
$\frac65=\frac35+\frac35$ at $k=0$, come from $ \chi^{m5 \times
\overline{m5}}_{2,\frac{1}{10}; 2,-\frac{1}{10}}$ and $ \chi^{m5 \times
\overline{m5}}_{3,\frac{3}{5}; 3,-\frac{3}{5}}$ in the sector-$\phi\bar\phi$.
In other words, these low energy states carry a non-trivial charge of the
$\tl\cR_\text{Fib}$-symmetry.  The operators associated with these low energy
states (via the operator-state correspondence) have scaling dimensions $h+\bar
h = \frac15$ and $ \frac65$ which are less than 2, meaning that these operators are
RG-relevant. However, they carry the
charge $\phi\bar\phi$ under the $\tl\cR_\text{Fib}$-symmetry, so they cannot be
added to the Hamiltonian without breaking the $\tl\cR_\text{Fib}$-symmetry.

On the other hand, the operator from the lowest energy state in $ \chi^{m5
\times \overline{m5}}_{5,\frac{7}{16}; 5,-\frac{7}{16}} $ of the sector-$\onebb$
has a scaling dimension $h+\bar h = \frac78 < 2$ and is a symmetric operator
for the $\tl\cR_\text{Fib}$-symmetry.  The presence of such a symmetric
relevant operator implies that the gapless state is an unstable critical point.
This seemingly contradicts the numerical observation, which indicates that the
gapless state is a stable phase.  

This apparent contradiction can be resolved by noticing that the $h+\bar h =
\frac78$ state carries a large crystal momentum $k=\pi$ (see Fig.
\ref{GoldenChainMM5}).  Due to the translation symmetry of the Fibonacci-anyon
chain, the low energy states carry an effective $\Z_2$ quantum number $k\approx
0$ or $k \approx \pi$.  Such a $\Z_2$ quantum number corresponds to a $\Z_2$
grading of the tricritical Ising CFT $m5\times \overline{m5}$ that describes
the low energy states.  We know that both right- and left-moving
$(5,4)$-minimal models, $m5$ and $\overline{m5}$, have $\Z_2$ grading (see
Appendix \ref{grading}).  It turns out that the $k\approx 0,\pi$ crystal
momenta correspond to the $\Z_2$ grading of the left-moving $(5,4)$-minimal
model $\overline{m5}$.  This allows us to conclude that the states described by
the conformal characters $ \chi^{m5 \times \overline{m5}}_{1,0; 1,0} $ and $
\chi^{m5 \times \overline{m5}}_{4,\frac{3}{2}; 4,-\frac{3}{2}} $ carry
$k\approx 0$, while the states described by the conformal character $\chi^{m5
\times \overline{m5}}_{5,\frac{7}{16}; 5,-\frac{7}{16}} $ carry $k\approx \pi$.
This exactly matches the numerical result in Fig. \ref{GoldenChainMM5}.

Thus we conclude that the $h+\bar h = \frac78$ relevant operator cannot be 
added to the Hamiltonian if we preserve the translation
symmetry of the Fibonacci-anyon chain.  \Rf{FFc0612341} mentioned that the
stable gapless state is protected by the topological symmetry (\ie the
$\tl\cR_\text{Fib}$-symmetry).  From the above discussion, we see that, in fact,
the $\tl\cR_\text{Fib}$-symmetry alone is not enough.  We also need the
translation symmetry of the Fibonacci-anyon chain to have a stable gapless
state.

\begin{figure}[t]
\begin{center}
\includegraphics[height=2.2in]{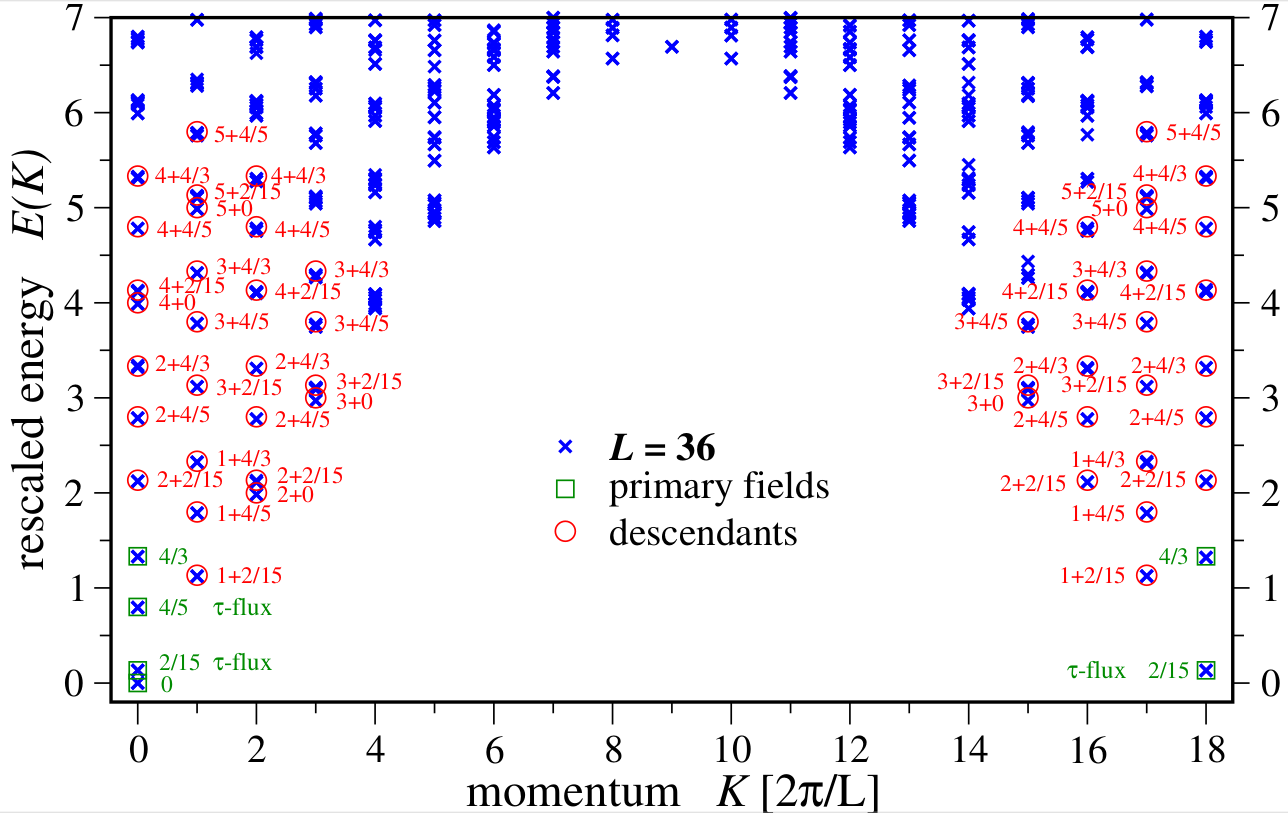} 
\end{center}

\caption{ The energy-momentum spectrum of the Fibonacci-anyon chain with $L$
anyons at $\tan \te= 1/\vphi$ and $\cos \te >0$, from \Rf{TT08014602}.  }
\label{GoldenChainMM6}
\end{figure}

The symTO $\eM_\text{dFib}$ allows other gapless states, corresponding
to other $\onebb$-condensed boundaries of $\eM_\text{dFib}$.  One of them
is given by the following vector-valued partition function
\begingroup
\allowdisplaybreaks
\begin{align}
\label{FibMM6Potts}
Z_{\onebb\text{-cnd};\onebb }^{\eM_\text{dFib}} &=
  \chi^{m6 \times \overline{m6}}_{1,0; 1,0} +
  \chi^{m6 \times \overline{m6}}_{1,0; 5,-3} +
  2\chi^{m6 \times \overline{m6}}_{3,\frac{2}{3}; 3,-\frac{2}{3}} 
\nonumber\\
&\ \ \ \ 
  +\chi^{m6 \times \overline{m6}}_{5,3; 1,0} +
  \chi^{m6 \times \overline{m6}}_{5,3; 5,-3} 
 \nonumber \\ 
Z_{\onebb\text{-cnd};\bar \phi }^{\eM_\text{dFib}} &=
  \chi^{m6 \times \overline{m6}}_{1,0; 6,-\frac{2}{5}} +
  \chi^{m6 \times \overline{m6}}_{1,0; 10,-\frac{7}{5}} +
  2\chi^{m6 \times \overline{m6}}_{3,\frac{2}{3}; 8,-\frac{1}{15}} 
\nonumber\\
&\ \ \ \ 
  +\chi^{m6 \times \overline{m6}}_{5,3; 6,-\frac{2}{5}} +
  \chi^{m6 \times \overline{m6}}_{5,3; 10,-\frac{7}{5}} 
 \nonumber \\ 
Z_{\onebb\text{-cnd};\phi }^{\eM_\text{dFib}} &=
  \chi^{m6 \times \overline{m6}}_{6,\frac{2}{5}; 1,0} +
  \chi^{m6 \times \overline{m6}}_{6,\frac{2}{5}; 5,-3} +
  2\chi^{m6 \times \overline{m6}}_{8,\frac{1}{15}; 3,-\frac{2}{3}} 
\nonumber\\
&\ \ \ \ 
  +\chi^{m6 \times \overline{m6}}_{10,\frac{7}{5}; 1,0} +
  \chi^{m6 \times \overline{m6}}_{10,\frac{7}{5}; 5,-3} 
 \nonumber \\ 
Z_{\onebb\text{-cnd};\phi \bar\phi}^{\eM_\text{dFib}} &=
  \chi^{m6 \times \overline{m6}}_{6,\frac{2}{5}; 6,-\frac{2}{5}} +
  \chi^{m6 \times \overline{m6}}_{6,\frac{2}{5}; 10,-\frac{7}{5}} +
  2\chi^{m6 \times \overline{m6}}_{8,\frac{1}{15}; 8,-\frac{1}{15}} 
\nonumber\\
&\ \ \ \ 
  +\chi^{m6 \times \overline{m6}}_{10,\frac{7}{5}; 6,-\frac{2}{5}} +
  \chi^{m6 \times \overline{m6}}_{10,\frac{7}{5}; 10,-\frac{7}{5}} 
\end{align}
\endgroup
This gapless state is described by the $(6,5)$-minimal model $m6\times \overline{m6}$
for the right- and left-movers. It has a central charge
$(c,\bar c)=(\frac45,\frac45)$. Its modular invariant partition function is
given by
\begin{align}
Z_{af} &= 
\sum_\al
Z^{\eM_\text{dFib}}_{\onebb\text{-cnd};\al}(\tau,\bar\tau)
\left (Z^{\eM_\text{dFib}}_{\tl\cR_\text{Fib};\al}\right )^*  
\\
 &=  |\chi^{m6}_{0} +  \chi^{m6}_{3}|^2 +  |\chi^{m6}_{\frac{2}{5}} +  \chi^{m6}_{\frac{7}{5}}|^2 
+
2 |\chi^{m6}_{\frac{2}{3}}|^2 + 2 |\chi^{m6}_{\frac{1}{15}}|^2 
,
\nonumber
\end{align}
which is the partition function of 3-state Potts critical point; we will refer 
to this gapless state as the 3-state Potts CFT.

We note that the $(6,5)$-minimal model has a $\Z_2$ grading (see Appendix
\ref{grading}). Only the $\Z_2$ trivial primary fields appear in the above
partition functions.  The Fibonacci-anyon chain has a critical point at $\tan
\te =1/\vphi$, whose energy-momentum spectrum is given by Fig.
\ref{GoldenChainMM6} which matches that of \eqn{FibMM6Potts} (\ie only the
$\Z_2$ trivial sectors are present).

From the trivial component of the partition function,
$Z_{\onebb\text{-cnd};\onebb }^{\eM_\text{dFib}} $, we see that the
symmetric sub-Hilbert space $\cV_\text{symmetric}$ contains states with
energy-momentum $\left(E,(k \text{ mod } \pi) \frac{L}{2\pi} \right) = $ $(0,0)$,
$(3,\pm 3)$, $(\frac43,0)$, $(6,0)$, \etc.  These states all appear in the
spectrum in Fig.  \ref{GoldenChainMM6} and correspond to primary fields. We notice that some of these states carry $k\approx 0$ while others carry
$k\approx \pi$.  To understand this effective $\Z_2$ quantum number within the
$m6\times \overline{m6}$ CFT, we note that the primary field $V_{\bar h = 3}$
in $\overline{m6}$ has a $\Z_2$ fusion $ V_{\bar h = 3} V_{\bar h = 3} \sim
V_{\bar h =0} $ (see Appendix \ref{grading}).  Thus we may regard $V_{\bar h =
3}$ as the operator that boosts the momentum by $\Del k= \pi$.  This assignment
is compatible with the fusion rule of $\overline{m6}$ given in Appendix
\ref{grading}.  From the fusion rule $V_{\bar h = 3} V_{\bar h = \frac23} \sim
V_{\bar h =\frac23} $, we see that the primary field $ V_{\bar h =\frac23}$
contains a part with $k\approx 0$ and another part with $k\approx \pi$.  As a
result, the states described by the character $2\chi^{m6 \times
\overline{m6}}_{3,\frac{2}{3}; 3,-\frac{2}{3}}$ include both $k\approx 0$ and
$k\approx \pi$ states.  This agrees with the calculated spectrum in Fig.
\ref{GoldenChainMM6}, which includes states with 
$\left(E,k \frac{L}{2\pi} \right)$
= $\left (\frac43, 0\right )$ and $\left (\frac43, \frac{L}{2}\right )$.

The state with $\left(E,k \frac{L}{2\pi} \right) = $ $(\frac43, 0)$ corresponds to
a symmetric operator that preserves the translation symmetry.  Thus, the
gapless state described by the 3-state Potts CFT \eqn{FibMM6Potts} is a
critical point with one relevant direction. It corresponds to the critical
point of the Fibonacci-anyon chain at $\tan \te =1/\vphi$ in Fig.
\ref{GoldenChainMM6} \cite{TT08014602}.

The third $\onebb$-condensed boundary of $\eM_\text{dFib}$ is given by 
\begingroup
\allowdisplaybreaks
\begin{align}
\label{FibMM6Tetra}
Z_{\onebb\text{-cnd};\onebb }^{\eM_\text{dFib}} &=
  \chi^{m6 \times \overline{m6}}_{1,0; 1,0} +
  \chi^{m6 \times \overline{m6}}_{2,\frac{1}{8}; 2,-\frac{1}{8}} +
  \chi^{m6 \times \overline{m6}}_{3,\frac{2}{3}; 3,-\frac{2}{3}} 
\nonumber\\ &\ \ \ \
  +\chi^{m6 \times \overline{m6}}_{4,\frac{13}{8}; 4,-\frac{13}{8}} +
  \chi^{m6 \times \overline{m6}}_{5,3; 5,-3} 
 \nonumber \\ 
Z_{\onebb\text{-cnd};\bar\phi }^{\eM_\text{dFib}} &=
  \chi^{m6 \times \overline{m6}}_{1,0; 10,-\frac{7}{5}} +
  \chi^{m6 \times \overline{m6}}_{2,\frac{1}{8}; 9,-\frac{21}{40}} +
  \chi^{m6 \times \overline{m6}}_{3,\frac{2}{3}; 8,-\frac{1}{15}} 
\nonumber\\ &\ \ \ \
  +\chi^{m6 \times \overline{m6}}_{4,\frac{13}{8}; 7,-\frac{1}{40}} +
  \chi^{m6 \times \overline{m6}}_{5,3; 6,-\frac{2}{5}} 
 \nonumber \\ 
Z_{\onebb\text{-cnd};\phi }^{\eM_\text{dFib}} &=
  \chi^{m6 \times \overline{m6}}_{6,\frac{2}{5}; 5,-3} +
  \chi^{m6 \times \overline{m6}}_{7,\frac{1}{40}; 4,-\frac{13}{8}} +
  \chi^{m6 \times \overline{m6}}_{8,\frac{1}{15}; 3,-\frac{2}{3}} 
\nonumber\\ &\ \ \ \
  +\chi^{m6 \times \overline{m6}}_{9,\frac{21}{40}; 2,-\frac{1}{8}} +
  \chi^{m6 \times \overline{m6}}_{10,\frac{7}{5}; 1,0} 
 \nonumber \\ 
Z_{\onebb\text{-cnd};\phi \bar\phi}^{\eM_\text{dFib}} &=
  \chi^{m6 \times \overline{m6}}_{6,\frac{2}{5}; 6,-\frac{2}{5}} +
  \chi^{m6 \times \overline{m6}}_{7,\frac{1}{40}; 7,-\frac{1}{40}} +
  \chi^{m6 \times \overline{m6}}_{8,\frac{1}{15}; 8,-\frac{1}{15}} 
\nonumber\\ &\ \ \ \
  +\chi^{m6 \times \overline{m6}}_{9,\frac{21}{40}; 9,-\frac{21}{40}} +
  \chi^{m6 \times \overline{m6}}_{10,\frac{7}{5}; 10,-\frac{7}{5}} 
\end{align}
\endgroup
The corresponding modular invariant partition function is given by
\begin{align}
Z_{af} &= 
\sum_\al
Z^{\eM_\text{dFib}}_{\onebb\text{-cnd};\al}(\tau,\bar\tau)
\left (Z^{\eM_\text{dFib}}_{\tl\cR_\text{Fib};\al}\right )^*
\\
 &=  |\chi^{m6}_{0}|^2 +  |\chi^{m6}_{3}|^2 
+  |\chi^{m6}_{\frac{2}{5}}|^2 +  |\chi^{m6}_{\frac{7}{5}}|^2 
+ |\chi^{m6}_{\frac{2}{3}}|^2 
\nonumber\\
&\ \
+  |\chi^{m6}_{\frac{1}{15}}|^2 + |\chi^{m6}_{\frac{1}{8}}|^2 +  |\chi^{m6}_{\frac{13}{8}}|^2 
+ |\chi^{m6}_{\frac{1}{40}}|^2 +  |\chi^{m6}_{\frac{21}{40}}|^2 
,
\nonumber
\end{align}
which is the partition function of Ising tetracritical point.  So we will refer 
to this gapless state as tetracritical Ising CFT.  If we regard the $\Z_2$ grading of
$\overline{m6}$ (see Appendix \ref{grading}) as the effective $\Z_2$ quantum
number for critical momenta $k\approx 0$ and $k\approx \pi$, then only the
primary field $V_{h,-\bar h}=V_{\frac23,-\frac23}$ corresponds to symmetric
relevant operator that preserves translation symmetry.  The
gapless state \eqref{FibMM6Tetra} describes the critical point of the
Fibonacci-anyon chain at $ \te \approx 1.528 \pi$ in Fig.  \ref{GoldenChainMM6}
\cite{TT08014602}.

In fact,  the 3-state Potts CFT \eq{FibMM6Potts} and the tetra-critical Ising
CFT \eq{FibMM6Potts} are the only two $\onebb$-condensed boundaries with
central charge $(c\,\bar c)=(\frac45, \frac45)$ of $\eM_\text{dFib}$.  Thus,
the gapless state of the Fibonacci-anyon chain around $\te=\pi$ in Fig.
\ref{GoldenChainPhases} is likely described by the 3-state Potts CFT
\eq{FibMM6Potts}.  Without lattice symmetry, the 3-state Potts CFT has relevant
operators of dimension $h+\bar h = \frac43$ that respect the the
$\tl\cR_\text{Fib}$-symmetry.  So we need to figure out how lattice translation
and reflection symmetries are represented in the 3-state Potts CFT to decide if
the gapless state around $\te=\pi$ is stable or not.

\section{Computing \lowercase{sym}TO using symmetry twists: an example}\label{SymmTwist}
It is often possible to identify emergent symmetries in a gapless theory. Once 
these symmetries are identified, it is possible to project down to symmetry 
charge sectors and into symmetry-twisted Hilbert spaces, to further resolve the 
theory into sub-sectors of the Hilbert space. This information is concisely 
captured by a symTO,  \ie a topological order in one higher dimension 
associated with the (emergent) symmetries of the gapless theory. In this 
section, we explore a calculation that uses symmetry charges and symmetry 
twists to identify the symTO of the gapless Ising critical point.

\subsection{Symmetry, dual symmetry, and patch operators}

We have mentioned that symmetry (or low energy emergent symmetry) is more fully
described by a symTO.  In this section, we will compute the symTO via 
multi-component partition functions associated with symmetry
twists (\ie topological defect lines).\cite{JW191213492,CW220303596} To
understand symmetry twists, we first discuss patch operators.  In
\Rf{JW191213492}, it was pointed out that the symmetries in a local system 
can be described by patch symmetry transformation
operators.  It turns out that the patch symmetry transformation operators carry
more information about the symmetry than the global symmetry transformation
operators and allow us to compute the symTO.

As an example, let us consider the 1+1D Ising model of size $L$ with a periodic
boundary condition:
\begin{align}
	\label{HIs}
	&H=-\sum_{I=1}^L \left(BX_I+JZ_{I}Z_{I+1}\right),
\end{align}
where $X,Y,Z$ are the Pauli matrices.
This theory has a $\Z_2$ symmetry generated by
\begin{align}
	&U_{\Z_2}= \prod_{I=1,2,\cdots L}X_I .
\end{align}
The patch operator associated with this $\Z_2$ symmetry is
\begin{align}
	W^m(I,J) = \prod_{I+\frac12<K<J+\frac12} X_K.
\end{align}

We can use the patch operators to select the so-called \emph{local symmetric
	operators} $O_K$ via
\begin{align}
	W^m(I,J) O_K &= O_K W^m(I,J),
	\nonumber\\
	\text{ for } &K \text{ far away from } I,J,
\end{align}
where $O_K$ acts on sites near the site $K$.  The Hamiltonian is a sum of local
symmetric operators.  This is how the patch operators
$W^m(I,J)$ impose the $\Z_2$ symmetry.
We also have a trivial operator $W^\onebb(I,J)$ which is a
product of identity operators
\begin{align}
	W^\onebb (I,J) = \prod_{I+\frac12<K<J+\frac12} \id_K.
\end{align}

To see the dual symmetry $\tl\Z_2$ explicitly, we make a Kramers-Wannier duality
transformation
\begin{align}
	X_I\rightarrow \tilde{X}_{I-\frac12}\tilde{X}_{I+\frac12},~~~
	Z_{I}Z_{I+1}\rightarrow \tilde{Z}_{I+\frac12},
\end{align}
which transforms the Ising model to the dual Ising model with
dual spins living on the links labeled by $I+\frac12$:
\begin{align}
	\label{HDW}
	&H=-\sum_{I} \left(B\tilde X_{I-\frac12}\tilde{X}_{I+\frac12}+J\tilde{Z}_{I+\frac12}\right).
\end{align}
We see a dual $\tl\Z_2$ symmetry generated by
\begin{align}
	&U_{\tl\Z_2}= \prod_{I}\tl Z_{I+\frac12} .
\end{align}
This gives us patch operators for the dual $\tl\Z_2$ symmetry
\begin{align}
	W^e(I,J) = \prod_{I<K+\frac12<J}\tl Z_{K+\frac12} .
\end{align}
In the original basis,  the patch operators for $\tl\Z_2$ are given by
\begin{align}
	W^e(I,J) = Z_I Z_J .
\end{align}
From the patch operators of the two symmetries, $\Z_2$ and $\tl \Z_2$, we can 
construct the patch operators of a third
symmetry $\Z_2^f$, given by
\begin{align}
	W^f(I,J)=W^m(I,J) W^e(I,J). 
\end{align}
This allows us to find all three sets of patch
operators, generating the $\Z_2$, $\tl\Z_2$, and $\Z_2^f$ symmetries:
\begin{align}
	W^m(I,J) &= \hskip -1em \prod_{I+\frac12<K<J+\frac12 } \hskip -1em  X_K, \ \ \  
	W^e(I,J) = Z_I Z_J , 
	\nonumber\\
	W^f(I,J) &= Z_I\left(\prod_{I+\frac12<K<J+\frac12 } \hskip -1em X_K \right) 
	Z_J.
\end{align}
With these explicit formulas for the patch operators, we can compute the associated symTO\, as outlined in \Rf{CW220303596}. Here we will treat these operators from a slightly different point of view. Since these operators implement their corresponding symmetry on a finite patch, their boundaries realize symmetry twists. From the form of the patch operators, we can identify how the symmetry twists can be implemented in concrete lattice models. In order to study the Ising critical point, as mentioned above, we would like to first transform the above discussion into the language of a Majorana fermion model. At the same time, it is instructive to obtain the form of the patch operators in terms of the Majorana variables.

To achieve this, we use the Jordan-Wigner transformation. Our goal is to obtain a Majorana representation of the
patch operators that create pairs of $ \Z_2 $ domain walls and $ \Z_2 $ charges. As a starting point for this transformation, we
work with the \emph{dual} Ising variables, as in \eqn{HDW}. The JW transformation on these variables is implemented as 
\begin{equation}\label{JW}
	\begin{split}
		\tilde{Z}_{J+\frac{1}{2}} &=1-2f_{J+\frac{1}{2}}^\dagger f_{J+\frac{1}{2}}\\
		\tilde{\sigma}^+_{J+\frac{1}{2}} &=f_{J+\frac{1}{2}}^\dagger \prod_{I<J}(1-2f_{I+\frac{1}{2}}^\dagger f_{I+\frac{1}{2}})\\
		\tilde{\sigma}^-_{J+\frac{1}{2}} &=f_{J+\frac{1}{2}} \prod_{I<J}(1-2f_{I+\frac{1}{2}}^\dagger f_{I+\frac{1}{2}})
	\end{split}
\end{equation}
where the $f$ operators satisfy canonical fermionic anti-commutation relations $\{f_{I+\frac{1}{2}},f_{J+\frac{1}{2}}^\dagger\}=\delta_{IJ}$, $\{f_{I+\frac{1}{2}},f_{J+\frac{1}{2}}\}=0=\{f_{I+\frac{1}{2}}^\dagger,f_{J+\frac{1}{2}}^\dagger\} $. Let us define Majorana fermions,
\begin{equation}\label{Maj}
	\lambda_{J}=f_{J+\frac{1}{2}}^\dagger+f_{J+\frac{1}{2}}, \quad \lambda_{J+\frac{1}{2}}=\ii(f_{J+\frac{1}{2}}^\dagger-f_{J+\frac{1}{2}})
\end{equation}
which satisfy $ \{\lambda_i,\lambda_j\}=2\delta_{ij}, \lambda_i^\dagger=\lambda_i $. In the Majorana representation, we have
\begin{equation}\label{Num}
	\begin{split}
		f_{J+\frac{1}{2}}^\dagger f_{J+\frac{1}{2}} &= \frac{1}{2}(\lambda_{J}- \ii \lambda_{J+\frac{1}{2}})\cdot \frac{1}{2}(\lambda_{J}+ \ii \lambda_{J+\frac{1}{2}})\\
		&= \frac{1}{2}(1+\ii \lambda_{J}\lambda_{J+\frac{1}{2}})
	\end{split}
\end{equation}
Under the JW transformation (\eqn{JW}), the Pauli operators $ \tilde{X}_I $ and $ \tilde{Z}_I $ transform as follows
\begin{align}
	\tilde{X}_{I+\frac{1}{2}} \equiv \tilde{\sigma}_{I+\frac{1}{2}}^+ + \tilde{\sigma}_{I+\frac{1}{2}}^- &\xrightarrow{\text{JW}} \lambda_{I} \prod_{J<I}\left(-\ii \lambda_{J}\lambda_{J+\frac{1}{2}}\right)\\
	\tilde{Z}_{I+\frac{1}{2}} & \xrightarrow{\text{JW}} -\ii \lambda_{I}\lambda_{I+\frac{1}{2}}
\end{align}
We can now transform the patch symmetry operators to the Majorana representation. The patch operator corresponding to $\Z_2^f$-symmetry transforms as
\begin{align}\label{WfMaj}
	W^f(I,J) =& \tilde{X}_{I+\frac12}  \left( \prod_{I<K+\frac12<J}\tl Z_{K+\frac12}\right) \tilde{X}_{J+\frac12} \notag\\
	\xrightarrow{\text{JW}} &\lambda_{I} \prod_{K<I}\left(-\ii \lambda_{K}\lambda_{K+\frac{1}{2}}\right) \prod_{I<K+\frac12<J}\left(-\ii  \lambda_{K}\lambda_{K+\frac12}\right)  \notag\\ 
	& \ \ \  \prod_{K<J}\left(-\ii \lambda_{K}\lambda_{K+\frac{1}{2}}\right) \lambda_{J} 
	= \lambda_{I} \lambda_{J} 
\end{align}
The patch operator for the $ \Z_2 $ symmetry transforms as
\begin{align}\label{WmMaj}
	W^m(I,J) &= \tilde{X}_{I+\frac12} \tilde{X}_{J+\frac12} \xrightarrow{\text{JW}} 
	\la_I \prod_{I<K+\frac12 < J}\left(-\ii  \lambda_{K}\lambda_{K+\frac12}\right) \lambda_{J}\notag\\
	&=\prod_{I<K-\frac12<J } \left(-\ii \la_{K-\frac12}\la_K\right)
\end{align}
Lastly, the patch operator for the dual $ \tilde{\Z}_2 $ symmetry transforms as
\begin{align}\label{WeMaj}
	W^e(I,J)  &=\prod_{I<K+\frac12<J}\tl Z_{K+\frac12} \xrightarrow{\text{JW}} \prod_{K=I+1}^{J}\left(-\ii  \lambda_{2K-1}\lambda_{2K}\right)\notag\\
	&= \prod_{I<K+\frac12<J } \left(-\ii \la_K\la_{K+\frac12}\right)
\end{align}
For completeness, let us transform the Hamiltonian \eqn{HDW} to the Majorana representation as well,
\begin{equation}\label{HMaj}
	H_\text{Maj}=\sum_{I}\ii(B \lambda_{I-\frac{1}{2}}\lambda_{I} + J \lambda_{I}\lambda_{I+\frac{1}{2}})
\end{equation}
which at the Ising critical point $ B=J=1 $ becomes
\begin{equation}
	H_\text{Maj}=\sum_{j\in \frac12 \Z} \ii\lambda_{j}\lambda_{j+\frac12}
\end{equation}
This is the Majorana model describing a $ \Z_2 $-symmetry-breaking critical point, defined on an infinite chain. For convenience of notation, we may equivalently write this as
\begin{equation}
	\label{HMajC}
	H_\text{Maj}=\sum_{j\in \Z} \ii\lambda_{j}\lambda_{j+1}
\end{equation}

With this concrete model of the $ \Z_2^m $-symmetry-breaking critical point (a.k.a. Ising critical point) in hand, we now proceed with computing its partition function in the presence of various symmetry twists. Particularly, we want to uncover the maximal symTO of this theory.
From \eqn{WmMaj} and \eqn{WeMaj}, we see that the lattice translation $j \to j+1$ in
\eqn{HMajC} exchanges $W^e$ and $W^m$. Thus the emergent $ e $-$m $ exchange symmetry at the critical point, $\Z_2^{em}$ is
realized by the translation $j\to j+1$. On the other hand, the $ \Z_2^m $ symmetry of the Ising model translates into the fermion parity of the Majorana model.

\subsection{Symmetry twists}
\label{symmtw}
The patch operators discussed above are very closely related to the notion of a
disorder operator.\cite{KC7118} When a symmetry transformation is restricted to
a finite patch in 1 spatial dimension, each of the two endpoints represents a
disorder operator, which implements a symmetry twist. The disorder operator has
associated fusion rules, which are particularly simple when the symmetry
involved is $ \Z_2 $. In this case, two symmetry twists become equivalent to no
twist. From this discussion, we see that the patch operators discussed in
previous sections can tell us how to implement spatial symmetry twists on the
Hilbert space. In particular, each endpoint of a patch symmetry operator
describes a symmetry twist in the space direction. On the other hand, symmetry
twists in the time direction are implemented by applying the global symmetry
transformation on the entire Hilbert space of states. In an operational sense,
this is implemented by inserting the symmetry transformation operator in the
partition function.

For the $ \Z_2^m $ symmetry, a non-trivial spatial symmetry twist amounts to introducing antiperiodic boundary conditions for the Majorana degrees of freedom,
\[ \la_{j+N}=-\la_j \]
In the time direction, a non-trivial $ \Z_2^m $ twist corresponds to \emph{periodic} temporal boundary conditions for the Majorana fermions. In other words, the untwisted case corresponds to antiperiodic boundary conditions along the time direction of the spacetime torus. Time antiperiodicity is automatic from the definition of fermion path integrals, which is why the non-trivial symmetry twist corresponds to periodic and not anti-periodic temporal boundary condition. This is in contrast to the bosonic $ \Z_2 $ boundary conditions discussed above in \eqn{Z2bc} (cf. \Rf{CFT12}, pp.346-347).

For the $ \Z_2^{em} $ symmetry, a non-trivial spatial symmetry twist corresponds to considering a Majorana chain with an odd number of sites. A non-trivial temporal symmetry twist is obtained by inserting into the partition function an operator that translates the system by a single lattice site. We will find it useful to represent this operator in terms of momentum space variables.

\subsection{Multi-component partition function from symmetry twists of $\Z_2^m$
	and $\Z_2^{em}$ symmetries}\label{Z2mZ2emZ}

In this section, we compute the partition functions of the 1+1D critical Ising 
theory in the presence of various symmetry twists of $ \Z_2^m $ and $ 
\Z_2^{em} $. Since there are 4 possible combinations along the space and the 
time directions each, we should expect a total of 16 possible symmetry twist 
combinations. 

Recall the Majorana representation of the critical Ising Hamiltonian, defined on a lattice of size N,
\begin{equation}\tag{\ref{HMajC}}
	H_\text{Maj}=\sum_{j=1}^N \ii\lambda_{j}\lambda_{j+1}
\end{equation}
where we have left the boundary conditions unspecified for now. We can define 
Fourier-transformed Majorana operators as
\begin{equation}\label{MajFT}
	\lambda_j = \sqrt{\frac{2}{N}}\sum_k \tilde{\lambda}_k \ee^{2\pi \ii kj/N}
\end{equation}
In terms of these momentum space variables, we have $ k\in \Z_N $ for periodic boundary conditions and $ k\in
\frac{1}{2}+\Z_N $ for antiperiodic boundary conditions. The inverse Fourier
transformation reads
\begin{equation}\label{invFT}
	\tilde{\lambda}_k = \frac{1}{\sqrt{2N}}\sum_{j=1}^N \lambda_j \ee^{-2\pi \ii kj/N}
\end{equation}
The $k$-space Majorana operators satisfy the following properties:
\begin{equation}\label{MajK}
	\tilde{\lambda}_k^\dagger = \tilde{\lambda}_{N-k},\quad \{\tilde{\lambda}_k,\tilde{\lambda}_q^\dagger\} =\delta_{k,q}
\end{equation}
where the Kronecker delta is to be understood in a modulo $ N $ sense.
In terms of the $k$-space Majorana modes, the Hamiltonian becomes
\begin{equation}\label{HMajFT}
	H=\sum_{\om_k>0} \omega_k \tilde{\lambda}_k^\dagger \tilde{\lambda}_k + E_0
\end{equation}
where the ``zero energy" is given by 
\begin{equation}\label{E0}
	E_0= \frac{1}{2}\sum_{\om_k<0}\omega_k
\end{equation} 
and $ \omega_k = -v\sin{\frac{2\pi k}{N}} $ and $ v=4 $. 

\subsubsection*{Even $N$ with periodic boundary conditions} 

First, let us consider $N=$ even cases. We choose the set of independent
$k$-states as $ F_{E,P}=\{k\in \Z |-\frac{N}{4}\leq k\leq 0 \text{ or }
\frac{N}{2}\leq k<\frac{3N}{4} \} $. The subscripts indicate $ E $ for \textit{even} $ N $ and $ P $ for \textit{periodic} b.c.

Non-zero $k$-modes are described by
canonical fermion operators. In addition to these, we find two zero mode
operators which do not appear in the Hamiltonian in \eqn{HMajFT}, $
\tilde{\lambda}_0 $ and $ \tilde{\lambda}_N $, which satisfy $
\tilde{\lambda}_0^2=\frac{1}{2}=\tilde{\lambda}_N^2 $, $
\tilde{\lambda}_0^\dagger=\tilde{\lambda}_0  $, and $
\tilde{\lambda}_N^\dagger=\tilde{\lambda}_N  $. 

\textit{Hilbert space}--- We can combine the above-mentioned zero modes into a
single fermionic operator
\begin{equation}\label{EPZM}
	c=\frac{1}{\sqrt{2}}( \tilde{\lambda}_0+ \ii \tilde{\lambda}_N), \quad c^\dagger=\frac{1}{\sqrt{2}}( \tilde{\lambda}_0-\ii \tilde{\lambda}_N)
\end{equation}
Then the ground state $ \ket{0} $ is defined by 
\begin{equation}\label{EPgrnd}
	\tilde{\lambda}_k\ket{0} = 0 \quad \forall k \in F_{E,P}', \text{ and } c\ket{0}=0
\end{equation}
where $ F_{E,P}'=F_{E,P}\setminus \{0,\frac{N}{2}\} $. Excited states are
created by the action of $ \tilde{\lambda}_k^\dagger $ ($ \forall k\in F_{E,P}'
$) and $ c^\dagger $ on the ground state.

\textit{Fermion number operator}--- We define
\begin{equation}\label{EPFNum}
	F =c^\dagger c+ \sum_{k\in F_{E,P}'} \tilde{\lambda}_k^\dagger  \tilde{\lambda}_k
\end{equation}
which counts the non-zero Majorana modes as well as the fermion created from
the two zero modes. Note that the Hilbert space states mentioned above are all
eigenstates of the fermion number parity operator $ (-1)^F $.

\textit{Translation operator}--- In real space, lattice translation is defined
by
\begin{equation}\label{EPTreal}
	T \lambda_j T^\dagger = \lambda_{j+1}
\end{equation}
which leads to the momentum space relation
\begin{equation}\label{EPTk}
	T \tilde{\lambda}_k T^\dagger = \ee^{\frac{2\pi \ii k}{N}} \tilde{\lambda}_k
\end{equation}
to be satisfied for all $ k\in F_{E,P} $. It can be checked that the following definition works
\begin{equation}\label{EPTkDef}
	T=\ii\sqrt{2}\tilde{\lambda}_0 \exp\left[\ii K_0+\sum_{k\in F_{E,P}'}\ii\left(\frac{2\pi k}{N}+\pi \right)  \tilde{\lambda}_k^\dagger  \tilde{\lambda}_k  \right]
\end{equation}
where $ K_0 $ is a yet-undetermined real number which we interpret as ``ground state momentum". The translation operator $ T $ is related to the momentum operator $ K $ as $ T=\ee^{\ii K} $.

\textit{Partition function}--- The partition function is defined as
\begin{equation}\label{Zdef}
	Z(\beta) = \text{Tr}\ee^{-\beta H}
\end{equation}
We can introduce a $ \Z_2^{em} $ twist in the time direction by inserting the operator $ T=\ee^{\ii K} $ in the partition function above. To that end, we define
\begin{equation}\label{genZ}
	Z(\beta,X) = \text{Tr}\left[\ee^{-\beta H+\ii X K} \right]
\end{equation}
where setting $ X $ to be even or odd corresponds to trivial and non-trivial insertion of the $ \Z_2^{em} $ symmetry transformation respectively. In particular, odd $ X $ implements a non-trivial $ \Z_2^{em} $ twist in the time direction.

For \underline{odd $ X $}, we can see that $ T^X \equiv  \ee^{\ii X K}$ and $ (-1)^F $ anticommute. This means that this choice of boundary conditions (odd $ X $, even $ N $, periodic) is not consistent with the notion of independent temporal and spatial symmetry twists, so this partition function is not allowed. Also because of this anticommutation, a brute force calculation of the partition function yields 0 anyway, so we can consistently drop it from consideration.

For \underline{even $ X $}, $ T^X $ and $ (-1)^F $ commute, so the above subtlety disappears. By linearizing the Hamiltonian near $ k=0 $ and $ k=N/2 $, and taking $ N\to\infty $ we find the following partition function:
\begin{align}\label{EPZ}
	Z(\beta,X) &= 2\ee^{-\beta E_0+\ii X\left(K_0+\frac{\pi}{2}\right)} \sum_{\{n_k\}} \ee^{ \sum_k \left(-\beta\omega_k + \ii X \frac{2\pi k}{N} \right)n_k}\notag \\
	&\approx 2 \ee^{\beta \frac{2N}{\pi}-\beta\frac{2\pi v}{12N}} \prod_{k\in\N} \left(1+ \ee^{ -\beta v\frac{2\pi  k}{N} - \ii X \frac{2\pi k}{N}} \right) \notag \\
	&\qquad   \prod_{k\in\N} \left(1+ \ee^{ -\beta v\frac{2\pi  k}{N} + \ii X \frac{2\pi}{N}\left(\frac{N}{2}+k\right)} \right)
\end{align}
where $ \N $ denotes the set of all positive integers. In the second expression above, we have used \eqn{wLin} and $ E_0 \approx -\frac{2N}{\pi}+\frac{2\pi v}{12N} +\mathcal{O}\left(\frac{1}{N^3}\right)$, with $ v=4 $. This expression for $ E_0 $ is obtained by computing the sum in \eqn{E0} in the limit of $ N\to \infty $. We have also made the choice of $ K_0 = -\frac{\pi}{2} $ which will ensure good modular transformation properties. After some algebra, we find 
\begin{equation}\label{EPZfin}
	Z_{EP}^{++} = \ee^{\frac{2N\beta}{\pi}} \left|\frac{\theta_2(\tau)}{\eta(\tau)}\right|
\end{equation}
where $ \tau = \frac{X+\ii\beta v}{N} $ is the modular parameter. The first
sign in the superscript indicates even $ X $ (untwisted $ \Z_2^{em} $) and the second one stands for
untwisted $ \Z_2^m $ (anti-periodic) temporal boundary conditions. 

Due to state-operator correspondence, the total energy and the total momentum
of the ground state on a ring is related to the total central charge $c+\bar c$ and
total scaling dimension $h+\bar h$:
\begin{align}
	E_0 &= \# N + \left (-\frac{c+\bar c}{24} +h+\bar h\right ) v\frac{2\pi}{N} + 
	o(N^{-1} ),
	\nonumber\\
	K_0 &= \# N + \left (-\frac{c-\bar c}{24} +h-\bar h\right ) \frac{2\pi}{N} + 
	o(N^{-1}).
\end{align}
where $v$ is the velocity and $N$ is the length of the ring, so that $2\pi/N$ is
the momentum quantum.  The sector with the lowest energy has $h=\bar h=0$,
whose $E_0$ and $K_0$ allow us to determine central charge $c$ and $\bar c$.
From the $E_0$ and $K_0$ of other sectors, we can determine the scaling
dimensions $h,\bar h$ of the operator that maps the ground state sector to the
other sectors.

\textit{Fermion twisted sector}--- Inserting $ (-1)^F $ into the partition functions above, we get their fermion parity twisted versions. 
Since $ Z_{EP}^{-+} $ is ill-defined as discussed above, its fermion parity twisted partner $ Z_{EP}^{--} $ is similarly afflicted. 
On the other hand, the fermion parity twisted partner of $ Z_{EP}^{++} $ is well-defined but evaluates to zero because of the presence of a zero mode, i.e. $ Z_{EP}^{+-}=0 $.

\subsubsection*{Even $N$ with antiperiodic boundary conditions}
For antiperiodic b.c., we choose the set of independent $ k $-states as $ F_{E,A}=\{k\in \Z
+\frac{1}{2} |-\frac{N}{4}\leq k\leq 0 \text{ or } \frac{N}{2}\leq
k<\frac{3N}{4} \}  $.  The subscripts indicate $ E $ for
\textit{even} $ N $ and $ A $ for \textit{antiperiodic} b.c. Note that $ F_{E,A} $ does not contain any zero modes; there are exactly $ N/2 $ dynamical modes.

\textit{Hilbert space}--- 
The ground state $ \ket{0} $ is defined by 
\begin{equation}\label{EAgrnd}
	\tilde{\lambda}_k\ket{0} = 0 \quad \forall k \in F_{E,A}
\end{equation}
Excited states are created by the action of $ \tilde{\lambda}_k^\dagger $ ($ \forall k\in F_{E,A} $) on the ground state.

\textit{Fermion number operator}--- 
\begin{equation}\label{EAFNum}
	F =\sum_{k\in F_{E,A}} \tilde{\lambda}_k^\dagger  \tilde{\lambda}_k
\end{equation}
Unlike in the periodic case, here there is no zero mode contribution to the fermion number.

\textit{Translation operator}--- In real space, lattice translation is defined as in the periodic case by
\begin{equation}\label{EATreal}
	T \la_j T^\dagger = \la_{j+1}
\end{equation}
with the understanding that $ \lambda_{N+1} =-\lambda_1$  due to the boundary condition. This leads to the momentum space relation as before:
\begin{equation}\label{EATk}
	T \tilde{\lambda}_k T^\dagger = \ee^{\frac{2\pi \ii k}{N}} \tilde{\lambda}_k
\end{equation}
for all $ k\in F_{E,A} $. It can be checked that the following definition works
\begin{equation}\label{EATkDef}
	T = \exp\left[\ii K_0+ \sum_{k\in F_{E,A}}\ii \frac{2\pi k}{N} \tilde{\lambda}_k^\dagger  \tilde{\lambda}_k  \right]
\end{equation}

\textit{Partition function}--- 
\begin{align}\label{EAZ}
	Z(\beta,X) &= \ee^{-\beta E_0+\ii X K_0} \sum_{\{n_k\}} \ee^{ \sum_k \left(-\beta\omega_k + \ii X \frac{2\pi k}{N} \right)n_k}\notag \\
	&\approx \ee^{\beta \frac{2N}{\pi}+\beta\frac{2\pi v}{24N}} \prod_{k\in \N -\frac{1}{2}} \left(1+ \ee^{ -\beta v\frac{2\pi  k}{N} - \ii X \frac{2\pi k}{N}} \right) \notag \\
	&\qquad   \prod_{k\in \N -\frac{1}{2}} \left(1+ \ee^{ -\beta v\frac{2\pi  k}{N} + \ii X \frac{2\pi}{N}\left(\frac{N}{2}+k\right)} \right)
\end{align}
where $ \N $ denotes the set of all positive integers. In the second expression above, we have used 
\begin{equation}\label{wLin}
	\omega_k \approx 
	\begin{cases}
		-\frac{2\pi vk}{N} &\text{ for } k \lesssim 0\\
		\frac{2\pi v}{N}\left(k-\frac{N}{2}\right) &\text{ for } k \gtrsim \frac{N}{2}
	\end{cases}
\end{equation}
and $ E_0 \approx -\frac{2N}{\pi}-\frac{2\pi v}{24N} +\mathcal{O}\left(\frac{1}{N^3}\right)$, with $ v=4 $. This expression for $ E_0 $ is obtained by computing the sum in \eqn{E0} in the limit of $ N\to \infty $. 
In \eqn{EAZ}, we also chose $ K_0 = 0$. 
Simplifying this expression, we find
\begin{align}\label{EAZfin}
	Z_{EA}^{++}&=\ee^{\frac{2N\beta}{\pi}}\left|\frac{\theta_3(\tau)}{\eta(\tau)}\right|\\
	Z_{EA}^{-+}&=\ee^{\frac{2N\beta}{\pi}} \frac{\sqrt{\overline{\theta_3(\tau)} \theta_4(\tau)}}{|\eta(\tau)|}
\end{align}
for even $ X $ and odd $ X $ respectively (reflected by the first sign in the
superscript).

\textit{Fermion twisted sector}--- Inserting $ (-1)^F $ in the partition functions above, we get the following
\begin{align}\label{EAZF}
	Z_{EA}^{+-}&=\ee^{\frac{2N\beta}{\pi}}\left|\frac{\theta_4(\tau)}{\eta(\tau)}\right|\\
	Z_{EA}^{--}&=\ee^{\frac{2N\beta}{\pi}} \frac{\sqrt{\overline{\theta_4(\tau)} \theta_3(\tau)}}{|\eta(\tau)|}
\end{align}

\subsubsection*{Odd $N$ with periodic boundary conditions}

Now, let us consider $N=$ odd cases.  As before, we choose the set of
independent $k$-states as $ F_{O,P}=\{k\in \Z |-\frac{N}{4}\leq k\leq 0 \text{
	or } \frac{N}{2}\leq k<\frac{3N}{4} \} $, for periodic b.c. The subscripts
indicate $ O $ for \textit{odd} $ N $ and $ P $ for \textit{periodic} b.c. $ F_{O,P} 
$ contains one zero mode, corresponding to $ \tilde{\lambda}_0 $ which does 
not appear in the Hamiltonian (\eqn{HMajFT}). The remaining modes can be 
described by canonical fermion operators.

\textit{Hilbert space}--- An odd number of Majorana modes is unphysical in and
of itself. To define the Hilbert space, we need to introduce an extra ``ghost"
Majorana fermion $ \tilde{\lambda}_{gh} $. This can be interpreted as the
Majorana mode present in the bulk in a topologically non-trivial superselection
sector of the 2+1D theory underlying this discussion of the gapless boundary
theory. We define a new zero mode operator using this ghost mode and the zero
mode $ \tilde{\lambda}_0 $,
\begin{equation}\label{OPZM}
	c=\frac{1}{\sqrt{2}}( \tilde{\lambda}_0+ \ii\tilde{\lambda}_{gh}), \quad c^\dagger=\frac{1}{\sqrt{2}}( \tilde{\lambda}_0-\ii \tilde{\lambda}_{gh})
\end{equation}
where $ \tilde{\lambda}_{gh} $ satisfies $ \{ \tilde{\lambda}_{gh},  \tilde{\lambda}_k \}=0 \quad \forall k\in F_{O,P} $, $  \tilde{\lambda}_{gh}^\dagger =  \tilde{\lambda}_{gh} $, and $  \tilde{\lambda}_{gh}^2=\frac{1}{2} $. Then $ c $ and $ c^\dagger $ behave like canonical fermion operators. The ground state $ \ket{0} $ is defined by 
\begin{equation}\label{OPgrnd}
	\tilde{\lambda}_k\ket{0} = 0 \quad \forall k \in F_{O,P}', \text{ and } c\ket{0}=0
\end{equation}
where $ F_{O,P}'=F_{O,P}\setminus \{0\} $. Excited states are created by the action of $ \tilde{\lambda}_k^\dagger $ ($ \forall k\in F_{O,P}' $) and $ c^\dagger $ on the ground state.

\textit{Fermion number operator}--- We define the fermion number operator to include the zero mode operator,
\begin{equation}\label{OPFNum}
	F =\sum_{k\in F_{O,P}'} \tilde{\lambda}_k^\dagger  \tilde{\lambda}_k+c^\dagger c
\end{equation}

\textit{Translation operator}--- Using the real space definition, lattice translation is defined in the momentum space by
\begin{equation}\label{OPTk}
	T \tilde{\lambda}_k T^\dagger = \ee^{\frac{2\pi \ii k}{N}} \tilde{\lambda}_k
\end{equation}
for all $ k\in F_{O,P} $. Additionally, we postulate for the ghost Majorana,
\begin{equation}\label{OPTkZM}
	T \tilde{\lambda}_{gh} T^\dagger = \tilde{\lambda}_{gh}
\end{equation}
It can be checked that the following definition satisfies the above properties
\begin{equation}\label{OPTkDef}
	T = \exp\left[\ii K_0+ \sum_{k\in F_{O,P}'}\ii \frac{2\pi k}{N} \tilde{\lambda}_k^\dagger  \tilde{\lambda}_k  \right]
\end{equation}

\textit{Partition function}--- 
\begin{align}\label{OPZ}
	Z(\beta,X) &= 2 \ee^{-\beta E_0+\ii X K_0} \sum_{\{n_k\}} \ee^{ \sum_k \left(-\beta\omega_k + \ii X \frac{2\pi k}{N} \right)n_k}\notag \\
	&\approx 2 \ee^{\beta \frac{2N}{\pi}-\beta\frac{2\pi v}{48N}+\ii X K_0} \prod_{k\in \N} \left(1+ \ee^{ -\beta v\frac{2\pi  k}{N} - \ii X \frac{2\pi k}{N}} \right) \notag \\
	&\qquad   \prod_{k\in \N -\frac{1}{2}} \left(1+ \ee^{ -\beta v\frac{2\pi  k}{N} + \ii X \frac{2\pi}{N}\left(\frac{N}{2}+k\right)} \right)
\end{align}
where the factor of 2 is due to the zero mode degeneracy. Setting $ K_0 = -\frac{\pi}{8N}$ and using $ E_0 \approx -\frac{2N}{\pi}+\frac{2\pi v}{48N}+\mathcal{O}\left(\frac{1}{N^3}\right)$, we find
\begin{align}\label{OPZfin}
	Z_{OP}^{++}&=\ee^{\frac{2N\beta}{\pi}}\frac{\sqrt{2\overline{\theta_2(\tau)} \theta_3(\tau)}}{|\eta(\tau)|}\\
	Z_{OP}^{-+}&=\ee^{\frac{2N\beta}{\pi}} \frac{\sqrt{2\overline{\theta_2(\tau)} \theta_4(\tau)}}{|\eta(\tau)|}
\end{align}
for even $ X $ and odd $ X $ respectively.

\textit{Fermion twisted sector}--- Inserting $ (-1)^F $ in the partition functions above, we get 0 because of the zero mode, i.e. $ Z_{OP}^{+-}=Z_{OP}^{--}=0$.

\subsubsection*{Odd $N$ with antiperiodic boundary conditions}
In this case, we have the set of independent $k$-states given by 
$F_{O,A}=\{k\in \Z +\frac{1}{2} |-\frac{N}{4}\leq k\leq 0 \text{ or }
\frac{N}{2}\leq k<\frac{3N}{4} \}  $. The subscripts
indicate $ O $ for \textit{odd} $ N $ and $ A $ for \textit{antiperiodic} b.c. $ 
F_{O,A} $ contains one zero mode, corresponding to $ \tilde{\lambda}_{N/2} $ 
which does not appear in the Hamiltonian (\eqn{HMajFT}). The remaining 
modes can be described by canonical fermion operators.

\textit{Hilbert space}--- 
As in the periodic case, to define the Hilbert space, we need to introduce an extra ``ghost" Majorana fermion $ \tilde{\lambda}_{gh} $. We define a new zero mode operator using this ghost mode and the zero mode $ \tilde{\lambda}_0 $,
\begin{equation}\label{OAZM}
	c=\frac{1}{\sqrt{2}}(\tilde{\lambda}_{gh} + \ii \tilde{\lambda}_{N/2}), \quad c^\dagger=\frac{1}{\sqrt{2}}( \tilde{\lambda}_{gh}-\ii \tilde{\lambda}_{N/2})
\end{equation}
where $ \tilde{\lambda}_{gh} $ satisfies $ \{ \tilde{\lambda}_{gh},  \tilde{\lambda}_k \}=0 \quad \forall k\in F_{O,P} $, $  \tilde{\lambda}_{gh}^\dagger =  \tilde{\lambda}_{gh} $, and $  \tilde{\lambda}_{gh}^2=\frac{1}{2} $. Then $ c $ and $ c^\dagger $ behave like canonical fermion operators. The ground state $ \ket{0} $ is defined by 
\begin{equation}\label{OAgrnd}
	\tilde{\lambda}_k\ket{0} = 0 \quad \forall k \in F_{O,A}', \text{ and } c\ket{0}=0
\end{equation}
where $ F_{O,A}'=F_{O,A}\setminus \{\frac{N}{2}\} $. Excited states are created by the action of $ \tilde{\lambda}_k^\dagger $ ($ \forall k\in F_{O,A}' $) and $ c^\dagger $ on the ground state.

\textit{Fermion number operator}--- Similar to the periodic case, we define
\begin{equation}\label{OAFNum}
	F =\sum_{k\in F_{O,A}'} \tilde{\lambda}_k^\dagger  \tilde{\lambda}_k +c^\dagger c
\end{equation}

\textit{Translation operator}--- We need to satisfy \eqn{OPTk} for all $ k\in 
F_{O,A} $. Additionally, we postulate for the ghost Majorana,
\begin{equation}\label{OATkZM}
	T \tilde{\lambda}_{gh} T^\dagger =- \tilde{\lambda}_{gh}
\end{equation}
It can be checked that the following definition satisfies the above properties
\begin{align}\label{OATkDef}
	T &= 2\ii  \tilde{\lambda}_{N/2} \tilde{\lambda}_{gh} \exp\left[\ii K_0+ \sum_{k\in F_{O,A}'}\ii \frac{2\pi k}{N} \tilde{\lambda}_k^\dagger  \tilde{\lambda}_k  \right]\notag \\
	&=(-1)^{c^\dagger c} \exp\left[\ii K_0+ \sum_{k\in F_{O,A}'}\ii \frac{2\pi k}{N} \tilde{\lambda}_k^\dagger  \tilde{\lambda}_k  \right]
\end{align}

\textit{Partition function}--- 
For \underline{odd $ X $}, due to the $ (-1)^{c^\dagger c} $ factor in $ T\equiv \ee^{\ii K} $, the partition function simply evaluates to 0. 

For \underline{even $ X $}, the zero mode gives a factor of 2 instead of 0, and the partition function is given by
\begin{align}\label{OAZ}
	Z(\beta,X) &= 2\ee^{-\beta E_0+\ii X K_0} \sum_{\{n_k\}} \ee^{ \sum_k \left(-\beta\omega_k + \ii X \frac{2\pi k}{N} \right)n_k}\notag \\
	&\approx 2 \ee^{\beta \frac{2N}{\pi}-\beta\frac{2\pi v}{48N}+\ii X K_0} \prod_{k\in \N -\frac{1}{2}} \left(1+ \ee^{ -\beta v\frac{2\pi  k}{N} - \ii X \frac{2\pi k}{N}} \right) \notag \\
	&\qquad   \prod_{k\in \N } \left(1+ \ee^{ -\beta v\frac{2\pi  k}{N} + \ii X \frac{2\pi}{N}\left(\frac{N}{2}+k\right)} \right)
\end{align}
Setting $ K_0 = \frac{\pi}{8N}$ and using $ E_0 \approx -\frac{2N}{\pi}+\frac{2\pi v}{48N}+\mathcal{O}\left(\frac{1}{N^3}\right)$, we find
\begin{align}\label{OAZfin}
	Z_{OA}^{++}&=\ee^{\frac{2N\beta}{\pi}}\frac{\sqrt{2\overline{\theta_3(\tau)} \theta_2(\tau)}}{|\eta(\tau)|}\\
	Z_{OA}^{-+}&=0
\end{align}
for even $ X $ and odd $ X $ respectively.

\textit{Fermion twisted sector}--- Inserting $ (-1)^F $ in the partition functions above compensates for the $ (-1)^{c^\dagger c} $ factor for the odd $ X $ case, while it produces a factor of 0 in the even $ X $ case due to the new factor of $ -1 $ from the fermion twist operator. Therefore we have 
\begin{align}\label{OAZF}
	Z_{OA}^{+-}&=0\\
	Z_{OA}^{--}&=\ee^{\frac{2N\beta}{\pi}} \frac{\sqrt{2\overline{\theta_4(\tau)} \theta_2(\tau)}}{|\eta(\tau)|}
\end{align}
for even and odd $ X $ respectively. 

In the above calculation, we made some ad hoc choices for the way the translation operator acts on the momentum space Majorana modes and consequently for the values of $ K_0$ (ground state momentum) in the various Hilbert space sectors. In a more systematic calculation, we would start with a real space translation operator and derive its form in momentum space. Our only explanation for these choices at the moment is post-hoc, \ie these choices give us nice modular transformation properties of the multi-component partition function.

\subsection{Modular transformation properties of the multi-component partition function}

\subsubsection*{Symmetry Twist Basis}
Let's summarize the 16-component partition function obtained above,
\begingroup
\allowdisplaybreaks
\begin{equation}\label{Zst}
	\begin{split}
		Z_{EP}^{++} &=  2|\chi^\text{Is}_{\frac{1}{16}}|^2  \\ 
		Z_{EP}^{-+} &=  {\color{violet} N/A}  \\ 
		Z_{EP}^{+-} &=  0 \\ 
		Z_{EP}^{--} &= {\color{violet} N/A}	\\
		Z_{EA}^{++} &=  |\chi^\text{Is}_0+\chi^\text{Is}_{\frac{1}{2}}|^2 \\ 
		Z_{EA}^{-+} &= (\chi^\text{Is}_0-\chi^\text{Is}_{\frac{1}{2}})(\bar\chi^\text{Is}_0+\bar\chi^\text{Is}_{\frac{1}{2}})  \\ 
		Z_{EA}^{+-} &=  |\chi^\text{Is}_0-\chi^\text{Is}_{\frac{1}{2}}|^2 \\ 
		Z_{EA}^{--} &= (\chi^\text{Is}_0+\chi^\text{Is}_{\frac{1}{2}})(\bar\chi^\text{Is}_0-\bar\chi^\text{Is}_{\frac{1}{2}})	\\ 
		Z_{OP}^{++} &=  2\bar{\chi}^\text{Is}_{\frac{1}{16}}(\chi^\text{Is}_0+\chi^\text{Is}_{\frac{1}{2}})  \\ 
		Z_{OP}^{-+} &=  2\bar\chi^\text{Is}_{\frac{1}{16}}(\chi^\text{Is}_0-\chi^\text{Is}_{\frac{1}{2}})   \\ 
		Z_{OP}^{+-} &=  0 \\ 
		Z_{OP}^{--} &= 0\\ 
		Z_{OA}^{++} &=  2(\bar{\chi}^\text{Is}_0+\bar{\chi}^\text{Is}_{\frac{1}{2}}) \chi^\text{Is}_{\frac{1}{16}}\\ 
		Z_{OA}^{-+} &=  0\\ 
		Z_{OA}^{+-} &=  0 \\ 
		Z_{OA}^{--} &=  2(\bar\chi^\text{Is}_0-\bar\chi^\text{Is}_{\frac{1}{2}})\chi^\text{Is}_{\frac{1}{16}} 
	\end{split}
\end{equation}
\endgroup
The 16-component partition function is expressed in terms of Ising CFT
characters. The subscripts include $ E $ and $ O $ for even and odd number of
lattice sites, and $ A $ and $ P $ for antiperiodic and periodic boundary
conditions respectively. The superscripts have two $ \pm $ signs, the first of
which indicates whether $ X $ is even or odd by $ + $ and $ - $ respectively,
while the second indicates periodic or antiperiodic temporal b.c. by $ - $ and
$ + $ respectively. ``N/A" stands for ``not allowed", indicating that the
corresponding spatial and temporal boundary conditions are incompatible. In the
following, we will sometimes also refer to even and odd $ X $ by $ E^X $ and $
O^X $ respectively. Similarly, we will also refer to $ \Z_2^m $ untwisted i.e.
antiperiodic temporal b.c. by $ A^f $ and $ \Z_2^m $ twisted i.e. periodic
temporal b.c. by $ P^f $.  In \eqn{Zst}, we have dropped the $
\mathcal{O}(\ee^N) $ factor from each of the partition function components.

\Eqn{Zst} describes the multi-component partition function of the Ising
critical point in the so-called symmetry twist basis. Using the known modular
transformation properties of the Ising characters, we find that the nine
non-zero components transform into each other under modular transformations,
but the $ S $ matrix is not unitary. To get a unitary $ S $ matrix in the
symmetry twist basis, we need to strip off a factor of $ \sqrt{2} $ from the
partition function components corresponding to odd $ N $. This can be
interpreted as the quantum dimension of the ghost Majorana degree of freedom;
we put in the ghost fermion by hand so it only makes sense to take off the
extra factor from the partition function. One can understand this in the same
spirit as regulators used in quantum field theory calculations. This issue was
also discussed in \Rf{DG210102218} where the authors found that an odd number
of Majorana fermions do not admit a well-defined graded Hilbert space. We
approach this issue differently --- we add in a ghost Majorana fermion so that
the Hilbert space may be well-defined, with the ``ghost" being interpreted as
an insertion of a quasiparticle in the bulk 2+1D topological order.  We dub
this basis, in which $S$ and $T$ matrices are unitary, the ``unitary symmetry
twist" (UST) basis. In this basis, the $ 9\times 9 $ modular $S$ and $T$
matrices are found to be given by 
\begingroup
\allowdisplaybreaks
\begin{align}
		\label{MajST1a}
		S &=\begin{pmatrix}
			0 & 0 & 1   & 0 & 0 & 0   & 0 & 0 & 0 \\
			0 & 1 & 0   & 0 & 0 & 0   & 0 & 0 & 0 \\
			1 & 0 & 0   & 0 & 0 & 0   & 0 & 0 & 0 \\
			0 & 0 & 0   & 0 & 0 & 0   & 0 & 1 & 0 \\
			0 & 0 & 0   & 0 & 0 & 1   & 0 & 0 & 0 \\
			0 & 0 & 0   & 0 & 1 & 0   & 0 & 0 & 0 \\
			0 & 0 & 0   & 0 & 0 & 0   & 0 & 0 & 1 \\
			0 & 0 & 0   & 1 & 0 & 0   & 0 & 0 & 0 \\
			0 & 0 & 0   & 0 & 0 & 0   & 1 & 0 & 0
		\end{pmatrix} \\
\label{MajST1b}
		T &=\begin{pmatrix}
			1 & 0 & 0   & 0 & 0 & 0   & 0 & 0 & 0 \\
			0 & 0 & 1   & 0 & 0 & 0   & 0 & 0 & 0 \\
			0 & 1 & 0   & 0 & 0 & 0   & 0 & 0 & 0 \\
			0 & 0 & 0   & 0 & 1 & 0   & 0 & 0 & 0 \\
			0 & 0 & 0   & 1 & 0 & 0   & 0 & 0 & 0 \\
			0 & 0 & 0   & 0 & 0 & 0   & \ee^{-\ii \frac{\pi}{8}} & 0 & 0 \\
			0 & 0 & 0   & 0 & 0 & \ee^{-\ii \frac{\pi}{8}}   & 0 & 0 & 0 \\
			0 & 0 & 0   & 0 & 0 & 0   & 0 & 0 & \ee^{\ii \frac{\pi}{8}} \\
			0 & 0 & 0   & 0 & 0 & 0   & 0 & \ee^{\ii \frac{\pi}{8}} & 0
		\end{pmatrix}
\end{align}
\endgroup

\subsubsection*{Quasiparticle Basis}

The partition functions in \eqn{Zst}, when expanded in terms of $ q\equiv \ee^{2\pi\ii\tau} $ and $ \bar{q} \equiv \ee^{-2\pi\ii\bar\tau} $, do not all have positive integer coefficients. In the so-called ``quasiparticle basis", however, these coefficients indicate the degeneracy of different excited states in the spectrum, and hence must be positive integer valued. In order to convert from the symmetry twist to the quasiparticle basis, we must take suitable linear combinations of the different temporal boundary conditions so as to project to different symmetry charge sectors. Each of the partition function components in the symmetry-twist basis has the general form of 
$ Z = Z_{00} - Z_{10} - Z_{01} + Z_{11} $
where $ Z_{00}, Z_{01}, Z_{10}, Z_{11} $ are polynomials in $ q,\bar q $ with positive integer coefficients. $ Z_{00} $ collects the terms without $ \Z_2^{em} $ twist and $ (-1)^F $ insertion, $ Z_{10} $ those with only $ \Z_2^{em} $ twist, $ Z_{01} $ those with a negative contribution due to only $ (-1)^F $ insertion, and $ Z_{11} $ collects the remaining terms getting a negative sign from both (hence has a positive sign). The four new $ Z $'s can be interpreted as the components of the partition function in the quasiparticle basis. The general prescription to extract them is given by the following formulas (cf. excitation basis of $ \Z_2 $ topological order in \Rf{JW190513279})
\begin{align}\label{STtoQP}
	\begin{split}
		Z^{00} &= \frac{Z^{++}+Z^{-+}+Z^{+-}+Z^{--}}{4}\\
		Z^{10} &= \frac{Z^{++}-Z^{-+}+Z^{+-}-Z^{--}}{4}\\
		Z^{01} &= \frac{Z^{++}+Z^{-+}-Z^{+-}-Z^{--}}{4}\\
		Z^{11} &= \frac{Z^{++}-Z^{-+}-Z^{+-}+Z^{--}}{4}
	\end{split}
\end{align}
where the superscripts on the r.h.s. indicate the symmetry twist in the time direction for the $ \Z_2^{em} $ and $ \Z_2^m $ symmetries, as in \eqn{Zst}.
The subscript labels are suppressed since we apply this formula separately for each of the four spatial symmetry twists. There is, however, a subtlety with applying this definition to the $ EP $ sector of \eqn{Zst}. Since two of the components in this sector, labeled ``N/A", correspond to disallowed boundary conditions, we define $ Z^{00} = \frac{1}{2}(Z^{++}+Z^{+-}) $ and $ Z^{01} = \frac{1}{2}(Z^{++}-Z^{+-}) $ for this column, while leaving ``N/A" labels for $ Z_{10}, Z_{11} $. The 16-component partition function in this new basis is given by
\begin{alignat}{4}\label{Zqp}
			& Z_{EP}^{00} =  |\chi^\text{Is}_{\frac{1}{16}}|^2 , \ 
			&& Z_{EP}^{10} =  {\color{violet} N/A} , \ 
			&& Z_{EP}^{01} =  {\color{violet} |\chi^\text{Is}_{\frac{1}{16}}|^2 } , \ 
			&& Z_{EP}^{11} = {\color{violet} N/A}	\nonumber
			\\
			& Z_{EA}^{00} =  |\chi^\text{Is}_0|^2 , \ 
			&& Z_{EA}^{10} =|\chi^\text{Is}_{\frac{1}{2}}|^2 , \ 
			&& Z_{EA}^{01} = \bar\chi^\text{Is}_{\frac{1}{2}} \chi^\text{Is}_0  , \ 
			&& Z_{EA}^{11} =\chi^\text{Is}_{\frac{1}{2}}\bar\chi^\text{Is}_0	
			\nonumber
			\\ 
			& Z_{OP}^{00} =  \bar{\chi}^\text{Is}_{\frac{1}{16}}\chi^\text{Is}_0 , \ 
			&& Z_{OP}^{10} = 
			\bar\chi^\text{Is}_{\frac{1}{16}}\chi^\text{Is}_{\frac{1}{2}}  
			, \ 
			&& Z_{OP}^{01} =  {\color{violet} 
			\bar{\chi}^\text{Is}_{\frac{1}{16}}\chi^\text{Is}_0 } , \ 
			&& Z_{OP}^{11} = {\color{violet} 
			\bar{\chi}^\text{Is}_{\frac{1}{16}}\chi^\text{Is}_{\frac12} } \nonumber
			\\ 
			& Z_{OA}^{00} =   \bar{\chi}^\text{Is}_0 \chi^\text{Is}_{\frac{1}{16}} , \ 
			&& Z_{OA}^{10} =   
			\bar\chi^\text{Is}_{\frac{1}{2}}\chi^\text{Is}_{\frac{1}{16}} , \ 
			&& Z_{OA}^{01} =  {\color{violet} 
			\bar\chi^\text{Is}_{\frac{1}{2}}\chi^\text{Is}_{\frac{1}{16}}} , \ 
			&& Z_{OA}^{11} = {\color{violet} 
			\bar\chi^\text{Is}_0\chi^\text{Is}_{\frac{1}{16}}}
\end{alignat}
We note that the nine distinct partition functions seen here can be interpreted as anomalous partition functions corresponding to the appropriate defect lines inserted into the bulk double Ising topological order corresponding to fusion of chiral $ h=0,\frac{1}{2},\frac{1}{16} $ and anti-chiral $ \bar h=0,\frac{1}{2},\frac{1}{16} $ excitations.\cite{JW190513279}

Turns out, the modular $S$ and $T$ matrices in this basis are unitary if we 
\emph{don't} strip off the factor of $ \sqrt{2} $, unlike in the UST basis above. 
We can interpret this peculiarity as follows. In the symmetry twist basis, we 
focused on the 1+1D CFT without considering the bulk topological order, hence 
the bulk/ghost Majorana should not be included in the partition function 
calculation. However, in the quasiparticle basis, we are computing the partition 
function for the boundary along with the 2+1D bulk, \ie with the insertion of 
defect 
lines in the bulk topological order. For consistency with that description, the 
bulk/ghost Majorana must not be factored out if we are to retain a unitary 
description of the noninvertible gravitational anomaly of the 1+1D CFT. The 
explicit expressions of the $S$ and $T$ matrices in the quasiparticle basis are 
given by
	\begin{equation}\label{MajST2}
	\begin{split}
		&S=\begin{pmatrix}
			0 & \frac{1}{2} & \frac{1}{2}   & -\frac{1}{2} & -\frac{1}{2} & 0   & 0 & 0 & 0 \\
			\frac{1}{2} & \frac{1}{4} & \frac{1}{4}   & \frac{1}{4} & \frac{1}{4} & \frac{1}{\sqrt{8}}   & \frac{1}{\sqrt{8}} & \frac{1}{\sqrt{8}} & \frac{1}{\sqrt{8}} \\
			\frac{1}{2} & \frac{1}{4} & \frac{1}{4}   & \frac{1}{4} & \frac{1}{4}   & -\frac{1}{\sqrt{8}}   & -\frac{1}{\sqrt{8}} & -\frac{1}{\sqrt{8}} & -\frac{1}{\sqrt{8}} \\
			-\frac{1}{2} & \frac{1}{4} & \frac{1}{4}   & \frac{1}{4} & \frac{1}{4}   & -\frac{1}{\sqrt{8}}   & -\frac{1}{\sqrt{8}} & \frac{1}{\sqrt{8}} & \frac{1}{\sqrt{8}} \\
			-\frac{1}{2} & \frac{1}{4} & \frac{1}{4}   & \frac{1}{4} & \frac{1}{4}   & \frac{1}{\sqrt{8}}   & \frac{1}{\sqrt{8}} & -\frac{1}{\sqrt{8}} & -\frac{1}{\sqrt{8}} \\
			0 & \frac{1}{\sqrt{8}}   & -\frac{1}{\sqrt{8}} & -\frac{1}{\sqrt{8}} & \frac{1}{\sqrt{8}} & 0   & 0 & \frac{1}{2} & -\frac{1}{2} \\
			0 & \frac{1}{\sqrt{8}}   & -\frac{1}{\sqrt{8}} & -\frac{1}{\sqrt{8}} & \frac{1}{\sqrt{8}} & 0   & 0 & -\frac{1}{2} & \frac{1}{2} \\
			0 & \frac{1}{\sqrt{8}}   & -\frac{1}{\sqrt{8}} & \frac{1}{\sqrt{8}} & -\frac{1}{\sqrt{8}} & \frac{1}{2} & -\frac{1}{2}   & 0 & 0 \\
			0 & \frac{1}{\sqrt{8}}   & -\frac{1}{\sqrt{8}} & \frac{1}{\sqrt{8}} & -\frac{1}{\sqrt{8}} & -\frac{1}{2} & \frac{1}{2}   & 0 & 0 
		\end{pmatrix} \\
		&T= \mathrm{diag} \left (1,1,1,-1,-1, \ee^{-\frac{\ii \pi}{8}} ,  -\ee^{-\frac{\ii 
		\pi}{8}} 
		,  \ee^{\frac{\ii \pi}{8}}, - \ee^{\frac{\ii \pi}{8}}   \right  )
	\end{split}
	\end{equation}
Here, we leave out the disallowed
(labeled by N/A) components of \eqn{Zqp} and average over the redundant ones, so that we have nine distinct components of the
partition function,
\begin{multline}
	\mathbf{Z}^{QP}= \Bigg(\frac{Z_{EP}^{00}+Z_{EP}^{01}}{2},
	Z_{EA}^{00},
	Z_{EA}^{10},
	Z_{EA}^{01},
	Z_{EA}^{11} ,\\
	\frac{Z_{OP}^{00}+Z_{OP}^{01}}{2}, 
	\frac{Z_{OP}^{10}+Z_{OP}^{11}}{2}, 
	\frac{Z_{OA}^{00}+Z_{OA}^{11}}{2}, 
	\frac{Z_{OA}^{10} +Z_{OA}^{01}}{2}  \Bigg) 
\end{multline}
The $T$ matrix is diagonal in the quasiparticle basis, with the diagonal elements indicating the topological spins of the corresponding
excitations. We also note that both $S$ and $T$ matrices are unitary and
symmetric, as expected from the properties of minimal models. In particular,
\eqn{MajST2} exactly matches the modular transformation matrices of a theory
defined by the direct product of left and right moving Ising characters.

\subsubsection*{Relating the Different Bases}
To make the above basis changes and projections more systematic, we look for linear transformations between the symmetry-twist (ST) basis, the unitary symmetry twist (UST) basis, and
the quasiparticle (QP) basis. $S$ and $T$ matrices in the symmetry-twist basis have
the general form given in \eqn{ZpropG}. For $ \Z_2 \times \Z_2 $ symmetry
twists, this gives us $ 16\times 16 $ $S$ and $T$ matrices. To connect
these to the $ 9\times 9 $ matrices in the UST basis displayed in \eqns{MajST1a} and \eqref{MajST1b},
we project onto the relevant subspace of non-zero components of the
partition function. Moreover, the $T$ matrix gets some complex phase factors,
which can be interpreted as anomalies
of the partition function. The $ 16\times 16 $ $S$ and $T$ matrices (with the
appropriate complex phase factors plugged into the $T$ matrix) can also be transformed into the
quasiparticle basis directly by a change of basis combined with a projection.
These two transformations, therefore, take us from the appropriately modified
\eqn{ZpropG} to \eqns{MajST1a}-\eqref{MajST1b} and \eqn{MajST2} directly.

In the Majorana model discussed here, the symmetry $ \Z_2^m \times \Z_2^{em} $ can also be interpreted as a product of two fermion parity $ \Z_2 $ groups, one each for left and right movers. We denote this group as $\Z_2^L\times\Z_2^R= \{++,+-,-+,--\} $, where $ + $ is for periodic and $ - $ for antiperiodic. As explained 
below \eqn{EPZfin},
the presence and absence of fermion parity operator in the partition function corresponds respectively to periodic and antiperiodic boundary conditions in the time direction. In the space direction, periodic and antiperiodic b.c. correspond to integer and half-integer momenta, as explained above. With this in mind, let's map the superscript and subscript labels on the l.h.s. of \eqn{Zst} to $ \Z_2^L\times\Z_2^R $ elements:
\begin{equation}\label{STinZ2Z2}
	\begin{alignedat}{2}
		EP &\rightarrow ++ \qquad && E^XA^f \rightarrow --\\
		EA &\rightarrow -- \qquad && O^XA^f \rightarrow -+\\
		OP &\rightarrow +- \qquad && E^XP^f \rightarrow ++\\
		OA &\rightarrow -+ \qquad && O^XP^f \rightarrow +-
	\end{alignedat}
\end{equation}
In light of these new symmetry labels, let us re-write \eqn{Zst} and also 
incorporate the division by $ \sqrt{2} $ for the cases with an odd number of 
lattice sites, as explained above. The result of these operations is shown in 
\eqn{Zst2}. 
\begingroup
\allowdisplaybreaks
\begin{align}
		Z_{++}^{--} &=  2|\chi^\text{Is}_{\frac{1}{16}}|^2  \nonumber\\ 
		Z_{++}^{-+} &=  {\color{violet} N/A}  \nonumber\\ 
		Z_{++}^{++} &=  0 \nonumber\\ 
		Z_{++}^{+-} &= {\color{violet} N/A}	\nonumber
		\\
		Z_{--}^{--} &=  |\chi^\text{Is}_0+\chi^\text{Is}_{\frac{1}{2}}|^2 \label{Zst2} \\ 
		Z_{--}^{-+} &= (\chi^\text{Is}_0-\chi^\text{Is}_{\frac{1}{2}})(\bar\chi^\text{Is}_0+\bar\chi^\text{Is}_{\frac{1}{2}})  \nonumber\\ 
		Z_{--}^{++} &=  |\chi^\text{Is}_0-\chi^\text{Is}_{\frac{1}{2}}|^2 \nonumber\\ 
		Z_{--}^{+-} &= (\chi^\text{Is}_0+\chi^\text{Is}_{\frac{1}{2}})(\bar\chi^\text{Is}_0-\bar\chi^\text{Is}_{\frac{1}{2}})	\nonumber
		\\ 
		Z_{+-}^{--} &=  2\bar{\chi}^\text{Is}_{\frac{1}{16}}(\chi^\text{Is}_0+\chi^\text{Is}_{\frac{1}{2}})  \nonumber\\ 
		Z_{+-}^{-+} &=  2\bar\chi^\text{Is}_{\frac{1}{16}}(\chi^\text{Is}_0-\chi^\text{Is}_{\frac{1}{2}})   \nonumber\\ 
		Z_{+-}^{++} &=  0 \nonumber\\ 
		Z_{+-}^{+-} &= 0 		\nonumber
		\\ 
		Z_{-+}^{--} &=  2(\bar{\chi}^\text{Is}_0+\bar{\chi}^\text{Is}_{\frac{1}{2}}) \chi^\text{Is}_{\frac{1}{16}} 		\nonumber\\ 
		Z_{-+}^{-+} &=  0 		\nonumber\\ 
		Z_{-+}^{++} &=  0 		\nonumber\\ 
		Z_{-+}^{+-} &=  2(\bar\chi^\text{Is}_0-\bar\chi^\text{Is}_{\frac{1}{2}})\chi^\text{Is}_{\frac{1}{16}} 		\nonumber
\end{align}
\endgroup

The appropriately modified form of \eqn{ZpropG},
incorporating the phase factors in the modular $ T $ transformation, is
\begin{equation}\label{ZstST}
	\begin{split}
		Z_{g',h'}(-1/\tau) &=  S_{(g',h'),(g,h)} Z_{g,h}(\tau),\\
		Z_{g',h'}(\tau+1) &=  T_{(g',h'),(g,h)} Z_{g,h}(\tau),\\
		Z_{g',h'}(\tau) &=  R_{(g',h'),(g,h)}(u) Z_{g,h}(\tau),\\
		S_{(g',h'),(g,h)} &= \del_{(g',h'),(h,g)},\\
		T_{(g',h'),(g,h)} &= 
		\begin{cases}
			\ee^{-\frac{\ii \pi}{8}} \del_{(g',h'),(g,hg)} &\text{ for } g = +-\\
			\ee^{\frac{\ii \pi}{8}} \del_{(g',h'),(g,hg)} &\text{ for } g=-+\\
			\del_{(g',h'),(g,hg)} &\text{ otherwise }
		\end{cases},\\
		R &= 1
		,
	\end{split}
\end{equation}
where $ g,h\in \Z_2\times\Z_2 $. These $ S $ and $ T $ matrices seemingly also act on the disallowed components of the partition function. However, this is not really an issue because we are aiming to extract the physically meaningful components of the 16-component partition function by suitable projection and/or change of basis. In this spirit, the 9-component partition function in the UST basis can be obtained by a projection to the subspace of the 9 non-zero components of \eqn{Zst2}, described by the $ 9\times 16 $ matrix,
\begin{align}
	\label{STtoUST}
	M=\begin{pmatrix}
		0 & 0 & 0 & 1   & 0 & 0 & 0 & 0   & 0 & 0 & 0 & 0   & 0 & 0 & 0 & 0\\
		0 & 0 & 0 & 0   & 0 & 0 & 1 & 0   & 0 & 0 & 0 & 0   & 0 & 0 & 0 & 0\\
		0 & 0 & 0 & 0   & 0 & 0 & 0 & 1   & 0 & 0 & 0 & 0   & 0 & 0 & 0 & 0\\
		0 & 0 & 0 & 0   & 0 & 0 & 0 & 0   & 0 & 1 & 0 & 0   & 0 & 0 & 0 & 0\\
		0 & 0 & 0 & 0   & 0 & 0 & 0 & 0   & 0 & 0 & 0 & 1   & 0 & 0 & 0 & 0\\
		0 & 0 & 0 & 0   & 0 & 0 & 0 & 0   & 0 & 0 & 0 & 0   & 1 & 0 & 0 & 0\\
		0 & 0 & 0 & 0   & 0 & 0 & 0 & 0   & 0 & 0 & 0 & 0   & 0 & 1 & 0 & 0\\
		0 & 0 & 0 & 0   & 0 & 0 & 0 & 0   & 0 & 0 & 0 & 0   & 0 & 0 & 1 & 0\\
		0 & 0 & 0 & 0   & 0 & 0 & 0 & 0   & 0 & 0 & 0 & 0   & 0 & 0 & 0 & 1
	\end{pmatrix}
	,
\end{align}
which satisfies $ M M^\dag =I_{9\times 9} $ (\ie it is an isometry). Acting on the 16$ \times $16  $S$ and $T$ matrices described by \eqn{ZstST} with $ M $ (by conjugation), we derive the $ 9\times 9 $ $S$ and $T$ matrices in UST basis given in \eqns{MajST1a} and \eqref{MajST1b},
\begin{equation}\label{USTproj}
	S^{UST}=MS^{ST}M^\dag, \qquad T^{UST}=MT^{ST}M^\dag
\end{equation}
where the $S$, $T$ matrices on the l.h.s. stand for those in \eqns{MajST1a} and \eqref{MajST1b} and those on the r.h.s. stands for the ones in \eqn{ZstST}. 

The next task is to find a linear transformation to go from \eqn{ZstST} to \eqn{MajST2}. We do this in two parts, first a change of basis going from the partition function in \eqn{Zst2} to that in \eqn{Zqp}, and then an appropriate projection onto the 9 independent components of the quasiparticle basis. The change of basis is described by the following block-diagonal matrix,
\begin{align}
	\label{STtoQPmat}
	N=\begin{pmatrix}
		A & \mathrm O & \mathrm O & \mathrm O  \\
		\mathrm O & \sqrt{2}B & \mathrm O & \mathrm O  \\
		\mathrm O & \mathrm O & \sqrt{2} B & \mathrm O \\
		\mathrm O & \mathrm O & \mathrm O & B
	\end{pmatrix},
\end{align}
where the $ 4\times 4 $ matrices $ A $ and $ B $ are
\begin{equation}\label{Nsubmat}
	A=\begin{pmatrix}
		\frac{1}{2} & 0 & 0 & \frac{1}{2}   \\
		0 & 1 & 0 & 0   \\
		-\frac{1}{2} & 0 & 0 & \frac{1}{2}   \\
		0 & 0 & 1 & 0
	\end{pmatrix},
	\ \
	B=\frac{1}{4}\begin{pmatrix}
		1 & 1 & 1 & 1\\
		1 & -1 & -1 & 1\\
		-1 & -1 & 1 & 1\\
		-1 & 1 & -1 & 1
	\end{pmatrix},
\end{equation}
and $\mathrm O$ is the $ 4\times 4 $ null matrix. This change of basis is 
essentially identical to \eqn{STtoQP}, with the extra factors of $ \sqrt{2} $ 
simply accounting for the fact that we removed a factor of the quantum 
dimension of the ghost Majorana mode in defining the partition function in the 
symmetry-twist basis which we must restore when we go to the quasiparticle 
basis (as argued above). Also, note that we have an identity matrix in the 
subspace of the two components that are not allowed because of the 
incompatible space and time direction symmetry twists (labeled ``N/A" in 
\eqns{Zst} and \eqref{Zst2}). The action of matrix $ N $ on the 16-component 
partition function in the symmetry-twist basis produces the 16-component 
partition function in \eqn{Zqp},
\begin{equation}\label{STtoQP16}
	\mathbf Z^{QP16} = N \mathbf Z^{ST},
\end{equation} 
where $\mathbf Z^{ST} $ is the 16-component partition function displayed in \eqn{Zst2} and $\mathbf Z^{QP16} $ is the 16-component partition function shown in \eqn{Zqp}. From $ \mathbf Z^{QP16} $, we project out the independent components to form the 9-component partition function $\mathbf Z^{QP} $ corresponding to the double Ising quasiparticle basis, i.e. labeled by $ h,\bar h\in\{0,\frac12,\frac1{16}\} $. This is achieved by the action of the $ 9\times 16 $ matrix,
\begin{align}
	P=\begin{pmatrix}
		\frac{1}{2} & 0 & \frac{1}{2} & 0   & 0 & 0 & 0 & 0   & 0 & 0 & 0 & 0   & 0 & 0 & 0 & 0\\
		0 & 0 & 0 & 0   & \frac{1}{2} & 0 & \frac{1}{2} & 0   & 0 & 0 & 0 & 0   & 0 & 0 & 0 & 0\\
		0 & 0 & 0 & 0   & 0 & \frac{1}{2} & 0 & \frac{1}{2}   & 0 & 0 & 0 & 0   & 0 & 0 & 0 & 0\\
		0 & 0 & 0 & 0   & 0 & 0 & 0 & 0   & \frac{1}{2} & 0 & 0 & \frac{1}{2}   & 0 & 0 & 0 & 0\\
		0 & 0 & 0 & 0   & 0 & 0 & 0 & 0   & 0 & \frac{1}{2} & \frac{1}{2} & 0   & 0 & 0 & 0 & 0\\
		0 & 0 & 0 & 0   & 0 & 0 & 0 & 0   & 0 & 0 & 0 & 0   & 1 & 0 & 0 & 0\\
		0 & 0 & 0 & 0   & 0 & 0 & 0 & 0   & 0 & 0 & 0 & 0   & 0 & 1 & 0 & 0\\
		0 & 0 & 0 & 0   & 0 & 0 & 0 & 0   & 0 & 0 & 0 & 0   & 0 & 0 & 1 & 0\\
		0 & 0 & 0 & 0   & 0 & 0 & 0 & 0   & 0 & 0 & 0 & 0   & 0 & 0 & 0 & 1
	\end{pmatrix}
	,
\end{align}
where we have taken averages over the duplicate components in $\mathbf Z^{QP16} $ (cf. \eqn{Zqp}) so as to put them on an equal footing. In equations, we can express the above transformations from 16-component $\mathbf Z^{ST} $ to 16-component $\mathbf Z^{QP16} $ to 9-component $\mathbf Z^{QP} $ is summarized as
\begin{equation}\label{QP16toQP}
	\mathbf Z^{QP}=P \mathbf Z^{QP16}=PN \mathbf Z^{ST}
\end{equation} 
It turns out that $ 4 (PN)^\dag $ is the right inverse of the matrix $ PN $, i.e. $ 4 P N (PN)^\dag =I_{9\times 9} $. In terms of these matrices, we can define a transformation from the $S$, $T$ matrices in \eqn{ZstST} to those in \eqn{MajST2},
\begin{equation}\label{QPproj}
	S^{QP} = 4 PN S^{ST} N^\dag P^\dag, \qquad T^{QP} = 4 PN T^{ST} N^\dag P^\dag
\end{equation}
where the $S$, $T$ matrices on the l.h.s. stand for those in \eqn{MajST2} and those on the r.h.s. stand for the ones in \eqn{ZstST}. \Eqn{USTproj} and \eqn{QPproj} are thus the desired transformations that convert $ S $ and $ T $ matrices from the $ \Z_2 \times \Z_2 $ symmetry twist (ST) basis to the UST and QP bases respectively.

Therefore, we may view the critical point of $\Z_2$-symmetry breaking
transition as a $ \onebb $-condensed boundary of the 2+1D double-Ising topological order
$\eM_\text{dIs}$.  It is described by the nine-component partition
function labeled by a pair $(h,\bar h)$
\begin{equation}
	\label{IsIs}
	Z_{h,\bar h}^{QP}(\tau,\bar \tau) =
	\chi^\text{Is}_h(\tau)
	\bar\chi^\text{Is}_{\bar h}(\bar\tau),\ \ \ \
	h,\bar h =0,\frac12,\frac1{16}.
\end{equation}
A modular invariant partition function is obtained by stacking on the $\eM_\text{dIs}$ bulk and a gapped boundary obtained by condensing $ \onebb \oplus\si\bar\si\oplus\psi\bar\psi $, as shown in Fig. \ref{CFTdIs}.
\begin{align*}\label{IsModInv}
Z^{af}_{Is} (\tau,\bar\tau) &=
Z^{\eM_\text{dIs}}_{\onebb\text{-cnd};(\onebb,\bar \onebb)} (\tau,\bar\tau) 
+ Z^{\eM_\text{dIs}}_{\onebb\text{-cnd};(\si,\bar \si)} (\tau,\bar\tau) \\
& \qquad \qquad \qquad \qquad + Z^{\eM_\text{dIs}}_{\onebb\text{-cnd};(\psi,\bar \psi)} (\tau,\bar\tau) \\
&=|\chi^\text{Is}_0(\tau)|^2 + |\chi^\text{Is}_\frac12(\tau)|^2 + |\chi^\text{Is}_\frac{1}{16}(\tau)|^2
\end{align*}
\smallskip

\begin{table*}[t]
\caption{
Fusion of anyons in $(6,5)$-minimal model.
}
\label{fusion65}
\begin{tabular}{ |c||c|c|c|c|c|c|c|c|c|c|}
 \hline 
$\otimes$ & $\onebb$  & $a$  & $b$  & $c$  & $\vc d$  & $\vc e$  & $f$  & $\vc g$  & $\vc h$  & $i$ \\ 
\hline 
 \hline 
$\onebb$  & $ \onebb$  & $ a$  & $ b$  & $ c$  & $ \vc d$  & $ \vc e$  & $ f$  & $ \vc g$  & $ \vc h$  & $ i$  \\ 
 \hline 
$a$  & $ a$  & $ \onebb$  & $ c$  & $ b$  & $ \vc e$  & $ \vc d$  & $ f$  & $ \vc h$  & $ \vc g$  & $ i$  \\ 
 \hline 
$b$  & $ b$  & $ c$  & $ \onebb \oplus c$  & $ a \oplus b$  & $ \vc g$  & $ \vc h$  & $ i$  & $ \vc d \oplus \vc h$  & $ \vc e \oplus \vc g$  & $ f \oplus i$  \\ 
 \hline 
$c$  & $ c$  & $ b$  & $ a \oplus b$  & $ \onebb \oplus c$  & $ \vc h$  & $ \vc g$  & $ i$  & $ \vc e \oplus \vc g$  & $ \vc d \oplus \vc h$  & $ f \oplus i$  \\ 
 \hline 
$\vc d$  & $ \vc d$  & $ \vc e$  & $ \vc g$  & $ \vc h$  & $ \onebb \oplus f$  & $ a \oplus f$  & $ \vc d \oplus \vc e$  & $ b \oplus i$  & $ c \oplus i$  & $ \vc g \oplus \vc h$  \\ 
 \hline 
$\vc e$  & $ \vc e$  & $ \vc d$  & $ \vc h$  & $ \vc g$  & $ a \oplus f$  & $ \onebb \oplus f$  & $ \vc d \oplus \vc e$  & $ c \oplus i$  & $ b \oplus i$  & $ \vc g \oplus \vc h$  \\ 
 \hline 
$f$  & $ f$  & $ f$  & $ i$  & $ i$  & $ \vc d \oplus \vc e$  & $ \vc d \oplus \vc e$  & $ \onebb \oplus a \oplus f$  & $ \vc g \oplus \vc h$  & $ \vc g \oplus \vc h$  & $ b \oplus c \oplus i$  \\ 
 \hline 
$\vc g$  & $ \vc g$  & $ \vc h$  & $ \vc d \oplus \vc h$  & $ \vc e \oplus \vc g$  & $ b \oplus i$  & $ c \oplus i$  & $ \vc g \oplus \vc h$  & $ \onebb \oplus c \oplus f \oplus i$  & $ a \oplus b \oplus f \oplus i$  & $ \vc d \oplus \vc e \oplus \vc g \oplus \vc h$  \\ 
 \hline 
$\vc h$  & $ \vc h$  & $ \vc g$  & $ \vc e \oplus \vc g$  & $ \vc d \oplus \vc h$  & $ c \oplus i$  & $ b \oplus i$  & $ \vc g \oplus \vc h$  & $ a \oplus b \oplus f \oplus i$  & $ \onebb \oplus c \oplus f \oplus i$  & $ \vc d \oplus \vc e \oplus \vc g \oplus \vc h$  \\ 
 \hline 
$i$  & $ i$  & $ i$  & $ f \oplus i$  & $ f \oplus i$  & $ \vc g \oplus \vc h$  & $ \vc g \oplus \vc h$  & $ b \oplus c \oplus i$  & $ \vc d \oplus \vc e \oplus \vc g \oplus \vc h$  & $ \vc d \oplus \vc e \oplus \vc g \oplus \vc h$  & $ \onebb \oplus a \oplus b \oplus c \oplus f \oplus i$  \\ 
 \hline 
\end{tabular}
\end{table*}

\section{Conclusion}

In this paper, we have used the isomorphic holographic decomposition
\cite{KZ150201690} to reveal the emergent symmetry in a quantum field theory:
$QFT_{af} = QFT_{ano} \boxtimes_{\eM} \tl\cR$ (see Figs.  \ref{CDiso} and
\ref{QFTR}).  This decomposition
means that the partition function of gravitational anomaly-free $QFT_{af}$ is 
reproduced by the
composite system $QFT_{ano} \boxtimes_{\eM} \tl\cR$, where the bulk $\eM$ 
and
the boundary $\tl\cR$ are assumed to have infinite energy gap.  The
decomposition makes explicit the emergent symTO $\eM$ and the emergent 
symmetry
$\tl\cR$, where $\tl\cR$ describes the fusion of symmetry defects.

Using such a decomposition picture, we define the notion of \emph{maximal 
symTO}.
We believe that the maximal symTO is a very detailed characterization of a
gapless state.  We argue that it largely characterizes and determines the
gapless state (up to holo-equivalence).  In other words, just knowing maximal
symTO may allow us to determine local low energy dynamical properties, with
just a few ambiguities.  This may open up a new direction to study gapless
states.

~~~~~~~

We acknowledge many helpful discussions with Michael DeMarco, Liang Kong, Ho
Tat Lam, Ryan Lanzetta, Salvatore Pace, Shu-Heng Shao, and Carolyn Zhang.  This
work is partially supported by NSF DMR-2022428 and by the Simons Collaboration
on Ultra-Quantum Matter, which is a grant from the Simons Foundation (651446,
XGW).

\appendix

\section{Minimal models: fusion rules and $\Z_2$ grading}
\label{grading}

The topological order described by the $(5,4)$-minimal model has the following
set of anyons, which correspond to the primary fields of the minimal model
CFT:\\[3mm] 
\centerline{
\begin{tabular}{|c|c|c|c|c|c|c|}
\hline
anyon& $\onebb$ 
& $a$ 
& $\vc b$ 
& $c$ 
& $d$ 
& $\vc e$ 
\\ 
\hline
$s$: & $0$
& $\frac{3}{2}$
& $\frac{7}{16}$
& $\frac{3}{5}$
& $\frac{1}{10}$
& $\frac{3}{80}$
\\ 
\hline
$d$: & $1$
& $1$
& $\sqrt{2}$
& $\frac{1+\sqrt{5}}{2}$
& $\frac{1+\sqrt{5}}{2}$
& $\frac{5+\sqrt{5}}{\sqrt{10}}$
\\ 
\hline
\end{tabular}
}
\vskip 3mm 
These anyons have the following fusion rules.

\vspace*{3mm}
\centerline{
\begin{tabular}{ |c||c|c|c|c|c|c|}
 \hline 
$\otimes$ & $\onebb$  & $a$  & $\vc b$  & $c$  & $d$  & $\vc e$ \\ 
\hline 
 \hline 
$\onebb$  & $ \onebb$  & $ a$  & $ \vc b$  & $ c$  & $ d$  & $ \vc e$  \\ 
 \hline 
$a$  & $ a$  & $ \onebb$  & $ \vc b$  & $ d$  & $ c$  & $ \vc e$  \\ 
 \hline 
$\vc b$  & $ \vc b$  & $ \vc b$  & $ \onebb \oplus a$  & $ \vc e$  & $ \vc e$  & $ c \oplus d$  \\ 
 \hline 
$c$  & $ c$  & $ d$  & $ \vc e$  & $ \onebb \oplus c$  & $ a \oplus d$  & $ \vc b \oplus \vc e$  \\ 
 \hline 
$d$  & $ d$  & $ c$  & $ \vc e$  & $ a \oplus d$  & $ \onebb \oplus c$  & $ \vc b \oplus \vc e$  \\ 
 \hline 
$\vc e$  & $ \vc e$  & $ \vc e$  & $ c \oplus d$  & $ \vc b \oplus \vc e$  & $ \vc b \oplus \vc e$  & $ \onebb \oplus a \oplus c \oplus d$  \\ 
 \hline 
\end{tabular}
}
\vspace*{3mm}

From the fusion rules, we see a $\Z_2$ grading, where the non-trivial $\Z_2$
sector is indicated by bold labels.
The fusion rules also allow us to regard the Abelian anyon $a$ as carrying a
$\Z_2^a$ charge.  From the fusion rule, we see that $\onebb$ and $c$ carry no
$\Z_2^a$ charge, while $a$ and $d$ carry a non-trivial $\Z_2^a$ charge.  On the
other hand $\vc b$ and $\vc e$ carry uncertain $\Z_2^a$ charges, \ie both trivial
$\Z_2^a$ charge and non-trivial $\Z_2^a$ charge. 

The $(6,5)$-minimal model has the following set of anyons, which correspond to
the primary fields of the minimal model CFT:\\[3mm] \centerline{
\begin{tabular}{|c|c|c|c|c|c|c|c|c|c|c|}
\hline
anyon& $\onebb$ 
& $a$ 
& $b$ 
& $c$ 
& $\vc d$ 
& $\vc e$ 
& $f$ 
& $\vc g$ 
& $\vc h$ 
& $i$ 
\\ 
\hline
$s$: & $0$
& $3$
& $\frac{2}{5}$
& $\frac{7}{5}$
& $\frac{1}{8}$
& $\frac{13}{8}$
& $\frac{2}{3}$
& $\frac{1}{40}$
& $\frac{21}{40}$
& $\frac{1}{15}$
\\ 
\hline
$d$: & $1$
& $1$
& $\frac{1+\sqrt{5}}{2}$
& $\frac{1+\sqrt{5}}{2}$
& $\sqrt{3}$
& $\sqrt{3}$
& $2$
& $\frac{15+3\sqrt{5}}{2\sqrt{15}}$
& $\frac{15+3\sqrt{5}}{2\sqrt{15}}$
& $1+\sqrt{5}$
\\ 
\hline
\end{tabular}
}
\vskip 3mm 
These anyons have the fusion rules given in Table \ref{fusion65}.  Again, from
the fusion rules, we see a $\Z_2$ grading, where the non-trivial $\Z_2$ sector
is indicated by bold labels. Similarly, these fusion rules also allow us to
regard the Abelian anyon $a$ as carrying a $\Z_2^a$ charge.  From the fusion
rules, we see that $\onebb$ and $c$ carry no $\Z_2^a$ charge, while $a$ and $b$
carry a non-trivial $\Z_2^a$ charge.  On the other hand $f$, $i$, $\vc d$, $\vc
e$, $\vc g$, and $\vc h$ carry uncertain $\Z_2^a$ charges, \ie both trivial
$\Z_2^a$ charge and non-trivial $\Z_2^a$ charge.

\bibliography{./local,../../bib/all,../../bib/allnew,../../bib/publst,../../bib/publstnew}

\end{document}